\documentclass[a4paper,12pt]{article}

\usepackage{graphicx} 
\usepackage[margin=1in]{geometry}
\usepackage{setspace}
\usepackage[authoryear]{natbib}
\usepackage{adjustbox}
\usepackage{threeparttable}
\usepackage{indentfirst}

\usepackage{authblk}

\usepackage{rotating} 

\usepackage{multirow}
\usepackage{amsmath}
\usepackage{amsthm}
\usepackage{bbm}
\usepackage{mathrsfs}
\usepackage{etoolbox}
\AtBeginEnvironment{quote}{\itshape}
\usepackage[usestackEOL]{stackengine}
\usepackage{yhmath}
\usepackage{caption}
\usepackage{graphicx}
\usepackage{algorithm}
\usepackage{algpseudocode}
\usepackage{subfigure}
\usepackage[table]{xcolor}
\usepackage{xcolor}

\usepackage[T1]{fontenc}
\usepackage{array,booktabs}
\usepackage{tikz}           
\usepackage{cleveref}
\usepackage{chngcntr}

\counterwithin{figure}{section}
\renewcommand{\thefigure}{\thesection.\arabic{figure}}

\crefname{appendix}{Online Appendix}{Online Appendices}
\Crefname{appendix}{Online Appendix}{Online Appendices}

\crefname{subfigure}{Figure}{Figures}
\Crefname{subfigure}{Figure}{Figures}

\makeatletter
\renewcommand{\p@subfigure}{\thefigure}
\makeatother

\newtheorem{theorem}{Theorem}[section]
\newtheorem{lemma}{Lemma}[section]
\newtheorem{proposition}{Proposition}[section]

\numberwithin{equation}{section}
\usepackage{pifont}

\crefname{assumption}{Assumption}{Assumptions}  
\Crefname{assumption}{Assumption}{Assumptions} 

\usepackage{enumitem}
\usepackage{url}
\usepackage{mathtools}

\DeclareMathOperator*{\argmax}{argmax}

\usepackage{bbding}
\usepackage{pifont}
\usepackage{wasysym}
\usepackage{amssymb}
\usepackage{makecell}
\usepackage{ragged2e} 

\theoremstyle{definition}
\newtheorem*{assumption*}{Assumption}
\newtheorem{assumption}{Assumption}[section]

\newtheorem{example}{Example}[section]
\AtBeginEnvironment{example}{%
  \pushQED{\qed}%
}
\AtEndEnvironment{example}{\popQED\endexample}
\newcolumntype{L}[1]{>{\raggedright\arraybackslash}m{#1}}
\newcolumntype{C}[1]{>{\centering\arraybackslash}m{#1}}
\newcolumntype{R}[1]{>{\raggedleft\arraybackslash}m{#1}}

\makeatletter
\newcommand*{\rom}[1]{\expandafter\@slowromancap\romannumeral #1@}

\title{\Large Identifying Inattention and Active Choice in Incomplete Social Assistance Takeup: The Case of WIC}

\author[1]{Lei Bill Wang\thanks{Email: \texttt{lei.wang@austin.utexas.edu}}}
\author[2]{Sooa Ahn\thanks{Email: \texttt{sahn.econ@gmail.com}}}

\affil[1]{The University of Texas at Austin}
\affil[2]{Bank of Korea}

\date{}
\begin{document}

\maketitle
\pagenumbering{gobble}

\begin{abstract}

Existing literature on incomplete social assistance takeup points to two distinct behavioral mechanisms: inattention and active choice. We develop an econometric framework that semiparametrically identifies these mechanisms by exploiting institutional features common to public programs. Applying the framework to WIC, we compare two interventions: choice-nudging messages (CNM) and attention-boosting messages (ABM). While CNM increases takeup more because of its dynamic attention channel, ABM generates higher marginal value of public funds because its marginal participants are positively selected on willingness to pay. These results demonstrate that distinguishing inattention from active choice is crucial for both decomposing policy effects and evaluating welfare.

\vspace{1cm}


\end{abstract}

\vspace{1cm}

\noindent\textbf{Acknowledgements}: This research received no external financial support. The authors have no financial relationships or other potential conflicts of interest to disclose.




\clearpage


\clearpage










\pagenumbering{arabic}
\section{Introduction}\label{sec introduction}

Why do millions of eligible families not participate in government assistance programs? 
Existing literature on incomplete social assistance takeup points to two fundamentally different behavioral mechanisms: inattention and active choice \citep{finkelstein2019take,bergman2024creating,ko2024take}. 
\begin{itemize}
    \item[] Some households are \textit{inattentive}. They may be unaware of the program altogether or aware of it but fail to consider participation because of limited mental bandwidth or competing demands.
    \item[] Others are attentive to the program but decide that the paperwork, hassle, or perceived stigma cost \text{outweighs the perceived benefits} and therefore \textit{actively choose} not to participate.
\end{itemize} 
Although these two behaviors are conceptually distinct, they are observationally indistinguishable in data. An eligible household may never participate in a program because it is never attentive, or because it never thinks it is worthwhile, or a mix of both frictions (i.e., the household is inattentive in some months while deeming participation not worthwhile in the months they are attentive). Consequently, policymakers do not know whether low takeup primarily reflects a lack of attention or an unwillingness to participate. This information problem makes it difficult to evaluate the takeup effect and welfare implications of potential interventions that target either behavioral frictions.

This paper develops a new econometric framework that separately identifies inattention and active choice using panel participation data. We model program participation as a two-stage decision process. In the first stage, households become attentive to the program with some probability. Conditional on paying attention, they then choose whether to participate by comparing the perceived benefits of participation with its associated costs in the second stage. A household participates in the program if and only if it is attentive at the attention stage and deems participation worthwhile at the choice stage. Unlike the existing attention-choice models (see \cite{crawford2021survey} for a comprehensive review of inattentive consumer models), our framework does not require an exclusive attention shifter that is independent of the choice stage decision, while allowing lagged participation status and unobserved heterogeneity to simultaneously enter attention and choice stages, accommodating a rich set of specifications.

The main identification challenge is that unobserved attention and choice jointly determine observed participation, making their effects difficult to disentangle. We tackle this challenge by considering two institutional features common to many means-tested programs that enable separate identification of attention and choice parameters. First, current participants remain actively engaged with the program through benefit redemption and ongoing use, which prompt them to be attentive in the \emph{following} period. The second is the periodic eligibility recertification process, which introduces variation in participation hassle costs. 

The first institutional feature implies that two observable dimensions of participation histories—probability of continuing participation and probability of starting participation—contain distinct identifying information. Because current participants are fully attentive, we use their continuation probabilities to isolate the choice parameters unrelated to hassle costs. Next, variation generated by recertification recovers the choice parameter associated with hassle costs. Once the choice parameters are pinned down, we compare the starting and continuation probabilities to identify the attention parameters. We formalize these
intuitions as three sets of semiparametric identification propositions and theorems, which impose linear utility functions for attention and choice stages, but leave the utility shocks nonparametrically specified. 
We also briefly discuss how our identification strategy can extend to a fully nonparametric setup. 

Our identification strategy hinges on the assumption that participating households are fully attentive next period. We motivate this assumption using institutional features and empirical evidence. Institutionally, social assistance program participation often entails active engagement from the households, prompting these households to be likely attentive in the following period. Empirically, \Cref{sec persistent take up} shows that for our application to the Special Supplemental Nutrition Program for Women, Infants, and Children (WIC), existing participants continue WIC enrollment in the following period at a rate of 96.1\%. Even those who need to recertify their eligibility have a continuation probability of 94.4\%. These two takeup rates suggest that period $t$ participants are extremely likely to be attentive in period $t+1$, as being attentive is necessary for participation. In contrast, those who are not enrolled in period $t$ have a 4.0\% probability of starting participation in period $t+1$. This more than 90\% gap in takeup probability strongly motivates our two-stage framework in which current participants skip the attention stage. 

To make the framework empirically tractable, we develop a parametric estimator suitable for large administrative or survey datasets. We apply the proposed parametric estimator to the National Longitudinal Survey of Youth 1997 (NLSY97) data, which records 2,991 eligible households' WIC takeup decisions and other characteristics at the monthly level. The estimated parameters are consistent with existing evidence on WIC participation while providing a decomposition that existing approaches cannot recover. For example, it is well-documented that families with infants have the highest participation rate among all eligible groups \citep{bitler2003wic}. We show that conditional on benefit amount, such a high participation rate can be mostly attributed to attention. The attention-triggering impact of having an infant is equivalent to \$53.68 in benefit, and whether the family has an infant has a much smaller (equivalent to \$11.38 in benefit) and only marginally significant impact on choice. 

Because the structural estimation relies on the full attention among participants assumption and random-effect design of household-level heterogeneity, we implement two sets of robustness checks. First, we allow the attention of current enrollees to fall below one during recertification periods to test the sensitivity of the models to the full attention setup. Second, we introduce flexible household-type and regional-economic-condition fixed effects to address potential endogeneity concerns. The main parameter estimates remain stable across these alternative specifications.

Next, we study two interventions tested by Vermont's WIC2Five quasi-experimental project in 2016-2017.  One intervention reminded \textit{recently exited} households (stochastically attentive) that they were likely still eligible for WIC, representing an attention-boosting policy. The other sent messages to \textit{existing participants} (fully attentive) emphasizing the benefits of continued participation, representing a choice-nudging policy. We call these two interventions attention-boosting messages (ABM) and choice-nudging messages (CNM), respectively. Analyzing the WIC2Five quasi-experimental data with various reduced-form methods yields a clear pattern: ABM has little effect on site-level retention rate, whereas CNM increases site-level retention rate by approximately 8--10 percentage points (pp).

We simulate ABM and CNM with the attention-choice structural model. 
The counterfactual simulations reveal an interesting trade-off between overall takeup effect and marginal value of public funds (MVPF). 
\begin{itemize}
    \item[] Consistent with the WIC2Five reduced-form results, CNM increases overall takeup substantially more than ABM. Furthermore, the counterfactual analysis decomposes the effect of CNM into two channels. First, increasing the probability of continued participation directly raises current-month retention. Second, retaining households in the program keeps them actively engaged with WIC and therefore fully attentive in the following month. This dynamic link between current participation and future attention generates a large indirect attention channel that accounts for 77\% of CNM's policy effect on takeup.
    \item[] Despite its larger takeup effect, CNM has substantially lower MVPF than ABM across all the policy intensities we consider. ABM changes whether households reach the choice stage but leaves their participation choice unchanged, so the marginal participants induced by ABM are positively selected on choice-stage surplus, which is proportional to willingness to pay (WTP). CNM instead directly shifts the choice margin. As the intervention becomes more intensive, marginal participants have progressively lower WTP. The opposite directions of targeting on WTP drive the difference in MVPF of ABM and CNM. 
\end{itemize}
Consequently, the policymaker may select different policies depending on whether the objective is to maximize takeup or to maximize takeup among those with higher WTP.

This paper makes two contributions. 
\emph{Methodologically}, we develop a novel econometric framework that sequentially models attention and choice. By exploiting institutional features common to many public programs, we establish a set of sufficient identifying conditions for this two-stage model. Among them, the key identifying condition is the existence of an observed fully attentive subgroup. Our framework shows how the behavior of this subgroup can be combined with panel participation data to semiparametrically identify attention and choice parameters without exclusive attention shifters while allowing households to be heterogeneous. Importantly, the existence of a fully attentive subgroup is not unique to WIC. Related models, including \cite{hortaccsu2017power,heiss2021inattention,einav2025selling}, likewise feature a subset of agents who are fully attentive.\footnote{
These papers also assume that a subset of agents is fully attentive, although
the attentive subgroup is generated by exogenous events in their settings,
whereas lagged takeup, which determines full attention in our full model, is
endogenous. The fully attentive subgroup is not essential to the identification
strategies in \cite{hortaccsu2017power} and \cite{heiss2021inattention}, but is
central to \cite{einav2025selling}.
}
To demonstrate the broad implications of our theoretical results, we show how our identification argument provides a complementary source of identification for these related empirical models.

\emph{Empirically}, we combine structural estimates from NLSY97 with reduced-form evidence from the WIC2Five quasi-experiment to compare interventions targeting attention and choice. Both approaches show that CNM generates larger participation effects than ABM. Yet the ranking of the two policies reverses when evaluated using the MVPF framework from \cite{hendren2020unified}: ABM generates substantially higher welfare per dollar of government spending. This ranking reversal echoes a broader lesson from \cite{finkelstein2019take}: policies generating large takeup effects need not generate correspondingly large welfare gains. Our framework goes further by uncovering the behavioral mechanisms behind this ranking reversal. CNM generates larger takeup effects because of a large dynamic attention channel, whereas ABM generates a higher MVPF because its marginal participants are positively selected on WTP. Thus, distinguishing attention from choice allows us not only to rank policies under different objectives, but also to explain why those rankings differ.

The remainder of the paper proceeds as follows. 
The following subsection relates our contribution to relevant literature. 
Section~\ref{sec preliminary model} introduces a preliminary model to illustrate the key intuition behind our identification strategy. Section~\ref{sec full model} introduces the full model, identification propositions, and estimation procedure. Section~\ref{sec data} describes the WIC program and the NLSY97 data. Sections~\ref{sec estimation results} and \ref{sec policy comparison} present the empirical application to WIC. Section~\ref{sec conclusion} concludes.

\subsection{Related literature} \label{sec related literature}
This paper builds on a structural literature that uses attention-choice models to evaluate counterfactual policies and welfare \citep{honka2017advertising,hortaccsu2017power,heiss2021inattention,crawford2021survey,einav2025selling}. We extend this approach to incomplete social assistance takeup. This adaptation contributes to three related literatures.

First, compared with the existing literature on attention-choice models, we provide a novel identification strategy tailored to institutional features common to social assistance programs. This differs from commonly used approaches based on a large-support exclusive attention shifter \citep{agarwal2025demand} or Slutsky symmetry at the choice stage \citep{abaluck2021consumers}. \Cref{appendix related literature and contribution} explains why these alternative identifying conditions may be difficult to satisfy in the social assistance context.

Second, much of the social assistance takeup literature estimates the causal effects of program features or interventions on participation \citep{rossin2013wic,ganong2018decline,ko2024take}, while some field experiments are specifically designed to shed light on the behavioral mechanisms underlying these effects \citep{finkelstein2019take,bergman2024creating}. We complement this work by providing a structural framework that decomposes policy effects into attention and choice margins. The decomposition allows us to explore a dynamic attention channel that, to the best of our knowledge, has not been discussed in the literature. 

Third, the attention-choice decomposition has direct implications for targeting and welfare evaluation \citep{finkelstein2020welfare,hendren2020unified,shepard2025ordeals}. Policies targeting different behavioral margins may select different households into participation and therefore induce marginal participants with different WTP. The structural decomposition therefore links the behavioral source of incomplete takeup to both targeting and welfare evaluation.
\section{Preliminary model} \label{sec preliminary model}
We first set up a preliminary model to illustrate the key intuition behind the identification strategy for the full model.\footnote{The preliminary model does not allow for unobserved heterogeneity and imposes a fully parametric specification. The full model in \Cref{sec full model} relaxes both restrictions.}

\subsection{Model setup} \label{sec preliminary model setup}

In the panel data, we observe household $i$'s previous-period takeup decision
$D_{it-1}$, current takeup decision $D_{it}$, household characteristics
$X_{it}\in\mathcal{X}\subseteq\mathbb{R}^K$, and a recertification indicator
$Z_{it}\in\{0,1\}$. The vector $X_{it}$ includes program benefit amounts,
demographic characteristics, and a constant term.
$Z_{it}$ indicates whether the household must recertify its eligibility in
period $t$ in order to remain enrolled and is administratively determined by
program rules. For example, WIC requires households to recertify when a child
reaches ages 1 and 13 months, when the benefit structure changes substantially.

\subsubsection{Attention stage}
Let $A_{it}\in\{0,1\}$ denote household $i$'s attention to the program in
period $t$. We assume that a household that participated in the previous
period is fully attentive in the current period. For a previous
nonparticipant, attention is determined by household characteristics
$X_{it}$ and an idiosyncratic attention shock $\mathcal{E}_{it}$:
\begin{align*}
A_{it}
=
D_{it-1}
+
(1-D_{it-1})
\mathbbm{1}\{X_{it}\gamma>\mathcal{E}_{it}\}.
\end{align*}
Thus, previous-period participants satisfy $A_{it}=1$ with certainty, whereas the
attention of previous-perid nonparticipants depends on $X_{it}\gamma$ relative to
the attention shock.

We additionally impose the exclusion restriction
$A_{it}\perp Z_{it}\mid X_{it},D_{it-1}$.
That is, conditional on $X_{it}$ and previous participation, the
recertification requirement $Z_{it}$ does not directly affect attention.
Instead, $Z_{it}$ affects participation through the choice-stage hassle cost
defined below. This restriction is motivated by the fact that $Z_{it}$ is
administratively determined by program rules rather than chosen by the
household.

\subsubsection{Choice stage}
Let $C_{it}\in\{0,1\}$ denote the latent participation choice household $i$
would make if it reached the choice stage. We assume that the household makes
a static utility-maximizing decision subject to an idiosyncratic choice shock $\Xi_{it}$:
\begin{align*}
    C_{it}
    =
    \mathbbm{1}\left\{
    X_{it}\beta + S_{it}\alpha > \Xi_{it}
    \right\},
\end{align*}
where $S_{it}
=
\max\{1-D_{it-1},\,D_{it-1}Z_{it}\}$ indicates whether participation in period $t$ requires the household to incur
an enrollment or recertification hassle cost.
$S_{it} = 1$ in two cases. If the household did not participate in period $t-1$, it must pay a hassle cost to enroll in the program in period $t$. If the household participated in the previous
period, it incurs the hassle cost only when it is required to recertify,
i.e., when $Z_{it}=1$.

$X_{it}\beta$ captures the monthly recurring utility (e.g., from the monetary savings) and disutility (e.g., from stigma cost when redeeming the benefits) of participating in the program. 
$\alpha$ captures the disutility from sign-up hassle or recertification hassle, so we expect that $\alpha$ to be negative.

For exposition, we impose two simplifying restrictions on hassle costs. First, we assume that enrollment and recertification generate the same disutility, so that the common parameter $\alpha$ captures both types of hassle. This assumption is plausible in our WIC application because enrollment and recertification involve similar administrative procedures. Furthermore, \Cref{sec model extension} discusses extension models that allow enrollment and recertification hassle costs to differ. Second, we assume that hassle costs do not vary with observed household characteristics. This restriction is not critical to our identification argument. A more flexible hassle cost specification can be obtained by interacting $S_{it}$ with $X_{it}$ and our identification strategy would still apply.

\subsubsection{Participation status}
An econometrician observes that the household $i$ participates in the program if and only if household $i$ is attentive and chooses to participate. Therefore, $D_{it} = A_{it} C_{it}$. 

\subsection{Identification}

Following \cite{abaluck2021consumers}, we treat the conditional takeup probability
$P(D_{it}=1\mid D_{it-1},X_{it},Z_{it})$
as known on the support of $(D_{it-1},X_{it},Z_{it})$. We show that,
under the assumptions below and sufficient variation in the observed
covariates, these conditional takeup probabilities identify
$(\gamma,\beta,\alpha)$.

\begin{assumption}\label{assumption on shocks preliminary model}
\leavevmode
\begin{enumerate}[label=(\alph*), ref=(\alph*)]

    \item \label{assumption preliminary attention shock distribution}
    $\mathcal{E}_{it}
    \mid D_{it-1},X_{it},Z_{it}
    \sim G$. $G$ is a known, strictly increasing CDF with support
    $\mathbb{R}$.

    \item \label{assumption preliminary choice shock distribution}
    $\Xi_{it}
    \mid D_{it-1},X_{it},Z_{it}
    \sim H$. $H$ is a known, strictly increasing CDF with support
    $\mathbb{R}$.

    \item \label{assumption preliminary conditional independence}
    $\mathcal{E}_{it}\perp\Xi_{it}
    \mid D_{it-1},X_{it},Z_{it}$.

\end{enumerate}
\end{assumption}
Under \Cref{assumption on shocks preliminary model} (a) and (b), the latent decision
rules imply the following attention and choice probabilities.
\begin{align*}
P(A_{it}=1\mid D_{it-1},X_{it},Z_{it})=&~
G(X_{it}\gamma)^{1-D_{it-1}}.\\
P(C_{it}=1\mid D_{it-1},X_{it},Z_{it})=&~
H(X_{it}\beta+S_{it}\alpha).
\end{align*}
Given that $D_{it} = A_{it}C_{it}$, the conditional independence between the two shocks that influence $A_{it}$ and $C_{it}$, i.e., \Cref{assumption on shocks preliminary model} (c), implies the following equation.
\begin{align}
    P(D_{it} = 1 \mid D_{it-1},X_{it},Z_{it}) = G(X_{it}\gamma)^{1-D_{it-1}} H(X_{it}\beta + S_{it}\alpha) \label{eq decomposing observed prob}
\end{align}

The proof of \Cref{eq decomposing observed prob} can be found in \Cref{proof of eq decomposing observed prob}. By setting $D_{it-1} = 1$ and varying $(X_{it},Z_{it})$ to different values, we can identify $\beta$ and $\alpha$. When $D_{it-1} = 1$, a household is fully attentive. Hence, the observed probability of \textit{continuing} participation is a function of choice parameters $(\beta,\alpha)$ only. 
\begin{align*}
    &P(D_{it} = 1 \mid D_{it-1} = 1, X_{it}, Z_{it}) \overset{\Cref{eq decomposing observed prob}}{=} H(X_{it}\beta + Z_{it}\alpha). \\
    & \quad \quad \overset{\text{strict monotonicity of $H$}}{\implies} H^{-1}(P(D_{it} = 1 \mid D_{it-1} = 1, X_{it}, Z_{it})) = X_{it}\beta + Z_{it}\alpha.
\end{align*}


Provided that $(X_{it},Z_{it})$ has full rank on its support among households with $D_{it-1} = 1$, the preceding equation identifies $\beta$ and $\alpha$. Because $X_{it}$ contains a constant, the full rank condition requires variation in $Z_{it}$ among previous participants. This highlights the role of the recertification process in our identification strategy.

After identifying the choice parameters, we set $D_{it-1} = 0$ to identify $\gamma$.
The observed probability of \textit{starting} participation is
\begin{align*}
    &P(D_{it} = 1 \mid D_{it-1} = 0, X_{it}, Z_{it}) \overset{\Cref{eq decomposing observed prob}}{=} G(X_{it}\gamma)H(X_{it}\beta + \alpha). \\
    &\quad \quad \overset{\text{Strict monotonicity of $G$}}{\implies} G^{-1}\left(\frac{P(D_{it} = 1 \mid D_{it-1} = 0, X_{it}, Z_{it})}{H(X_{it}\beta + \alpha)}\right) = X_{it}\gamma.
\end{align*}
Provided that $X_{it}$ has full rank on its support among households with $D_{it-1} = 0$, the preceding equation identifies $\gamma$. The following proposition spells out the ``full rank'' conditions for identifying all the parameters.

\begin{proposition}\label{prop identification preliminary}
Under \Cref{assumption on shocks preliminary model} and $P(D_{it} = 1 \mid D_{it-1} = 1, X_{it}, Z_{it})$ is observed on the support of $(D_{it-1}, X_{it}, Z_{it})$,

\begin{enumerate}
\item[(IC)] 
$(\beta,\alpha)$ are identified under the  support condition
\text{FR1:} The support of $(X_{it},Z_{it})$ conditional on
$D_{it-1}=1$ contains $\dim(\beta)+1$ linearly independent vectors.

\item[(IA)] 
$\gamma$ is identified under FR1 and the support condition
\text{FR2:} The support of $X_{it}$ conditional on $D_{it-1}=0$
contains $\dim(\gamma)$ linearly independent vectors.
\end{enumerate}
\end{proposition}
\Cref{prop identification preliminary} contains three intuitions: choice parameters can be identified using continuation probabilities, $D_{it-1} = 1$, alone; choice parameters associated with hassle cost can be identified using variations in recertification status, $Z_{it}$; attention parameters can be identified by comparing continuation and starting probabilities, $D_{it-1} = 1$ versus $D_{it-1} = 0$. We leverage these intuitions for our full model identification in \Cref{sec the full model semiparametric identification} and explain how they extend to other empirical cases in \Cref{appendix extension to binary exclusive attention shifter}.

\section{Full model} \label{sec full model}

The preliminary model rules out persistent unobserved heterogeneity and
assumes that the conditional distributions of the attention and choice
shocks, $G$ and $H$, are known. The full model relaxes both restrictions.
First, we introduce a time-invariant random effect $Q_i$ that generates
persistent heterogeneity in both the attention and choice stages.
Intuitively, $Q_i$ captures latent household traits, such as thriftiness
and financial awareness, that may affect both attention and
participation choices. Second, we leave the distributions of the
idiosyncratic attention and choice shocks nonparametrically specified.

For notational convenience, let
$W_{it}:=(X_{it},Z_{it})$ denote the current observed covariates and
$R_{it}:=(D_{it-1},W_{it})$ denote the current observed state.
Define the corresponding histories as
$W_{i,1:s}:=(W_{i1},\ldots,W_{is})$ and
$R_{i,1:s}:=(R_{i1},\ldots,R_{is})$.

\begin{assumption}\label{assumption on shocks full model}
\leavevmode

\begin{enumerate}[label=(\alph*), ref=(\alph*)]

    \item
    $\mathcal{E}_{it}=\epsilon_{it}-\sigma_1Q_i$ and
    $\Xi_{it}=\xi_{it}-\sigma_2Q_i$.

    \item
    $Q_i\sim F$ for all $i$, where $F$ has full support on
    $\mathbb{R}$, and
    $Q_i\perp W_{i,1:T}$.

\end{enumerate}

\begin{enumerate}[label=(c\arabic*), ref=(c\arabic*)]

    \item
    For every $t\in\{2,\ldots,T\}$,
    $(\epsilon_{it},\xi_{it})
    \perp
    R_{i,1:(t-1)}
    \mid
    R_{it},Q_i$.

    \item
    For every $t\in\{2,\ldots,T\}$,
    $(\epsilon_{i,1:(t-1)},\xi_{i,1:(t-1)})
    \perp
    W_{it}
    \mid
    W_{i,1:(t-1)},Q_i$.

    \item
    For every $t\in\{1,\ldots,T\}$,
    $\epsilon_{it}\perp\xi_{it}\mid R_{it},Q_i$;
    $\epsilon_{it}\mid R_{it},Q_i\sim G$;
    and
    $\xi_{it}\mid R_{it},Q_i\sim H$.
    $G$ and $H$ are unknown, strictly increasing, differentiable CDFs
    with support $\mathbb{R}$.

\end{enumerate}

\end{assumption}

\Cref{assumption on shocks full model} (a) and (b) constitute the standard
random-effects structure. Conditions (c1)--(c3) jointly imply strict
exogeneity of $W_{i,1:T}$ with respect to the idiosyncratic shocks
conditional on $Q_i$. Together with (a) and (b), this yields strict
exogeneity with respect to the composite shocks. Moreover, conditions
(c1)--(c3) restrict serial dependence in $(\epsilon_{it},\xi_{it})$ by
requiring that the takeup history $D_{i,1:(t-1)}$ contain no information
about the period-$t$ idiosyncratic shocks.

Under this shock structure,\footnote{We further compare
\Cref{assumption on shocks full model} and
\Cref{assumption on shocks preliminary model} in
\Cref{appendix strength of the full model}. We also discuss several extensions of the full model in \Cref{sec model extension}, including allowing hassle costs to differ between first-time enrollment and recertification, allowing for stage-specific unobserved heterogeneity in attention and choice, and allowing for latent group effects that may be arbitrarily correlated with observed characteristics.} the transition from period $t-1$ to $t$ conditional on period $t$ covariates and $Q_i$ are characterized by the following two probabilities
\begin{align*}
P(A_{it}=1\mid R_{it},Q_i)
=&~
G(U_{it}^a(X_{it},Q_i))^{1-D_{it-1}}
\\
P(C_{it}=1\mid R_{it},Q_i)
=&~
H(U_{it}^c(R_{it},Q_i))
,
\end{align*}
where $U_{it}^a(X_{it},Q_i):=X_{it}\gamma+\sigma_1Q_i$
and
$U_{it}^c(R_{it},Q_i):=X_{it}\beta+S_{it}\alpha+\sigma_2Q_i$ denote the latent utilities
from the attention and choice stages, respectively. As in the preliminary
model, current-period takeup satisfies $D_{it}=A_{it}C_{it}$.

\subsection{Initial Condition}
So far, we have characterized the transition from period $t-1$ to $t$. To complete the model, we must specify the initial condition, which is central in panel models with random effects \citep{wooldridge2005simple}. We assume the econometrician has access to a sufficiently long panel dataset that observes each household from the first period it becomes eligible for the program. We index the period immediately preceding eligibility as t=0 and set $D_{i0}=0$ by construction. This yields the fixed initial condition, which is non-random and therefore independent of the household-specific random effect $Q_i$, avoiding the initial conditions problem that typically arises in dynamic random-effects models.

Longitudinal datasets such as the NLSY, SIPP, and PSID can permit researchers to construct samples satisfying this condition by restricting attention to households whose observed eligibility spell begins within the panel.

\subsection{Identification} \label{sec the full model semiparametric identification}

The key difference between the identification strategies of the preliminary
and full models is that, instead of analyzing a single-period transition
probability, we analyze the probability of an entire takeup sequence.
For notational simplicity, suppose all households are eligible for the
program for $T$ periods.\footnote{This restriction can be relaxed by
replacing $T$ with household-specific eligibility duration $T_i$.}
Following the population-level identification approach in the preliminary
model, we treat
$
P(D_{i,1:T}=d\mid W_{i,1:T}=w)
$
as known for every $(d,w)$ in the relevant support.

Under \Cref{assumption on shocks full model}(b), the random effect can be
integrated out:
\begin{align}
P(D_{i,1:T}\mid W_{i,1:T})
=
\int
P(D_{i,1:T}\mid W_{i,1:T},Q_i=q)
\,dF(q). \label{eq observed prob integrate out q}
\end{align}
Conditional on $Q_i$,
the probability of a takeup sequence can be decomposed into the product
of its single-period transition probabilities.

\begin{lemma} \label{lemma multiple period likelihood}
Under \Cref{assumption on shocks full model} (b), (c1) and (c2),
\begin{align*} 
P(D_{i,1:T}\mid W_{i,1:T},Q_i)
=
\prod_{t=1}^{T}
P(D_{it}\mid R_{it},Q_i).
\end{align*}
\end{lemma}
The proof of \Cref{lemma multiple period likelihood} is provided in
\Cref{proof for multiple period likelihood}.

We next consider an infinitesimal change in a continuous covariate
$X_{i\tau}^{\omega}$ at a fixed period $\tau$, where $\omega$ indexes
the covariate being varied. We assume that the conditional sequence
probability is differentiable with respect to $X_{i\tau}^{\omega}$ on
the relevant interior of its support and treat the derivative
\begin{align*}
\frac{\partial}{\partial X_{i\tau}^{\omega}}
P(D_{i,1:T}\mid W_{i,1:T})
\end{align*}
as known in the relevant support.\footnote{Related identification
papers for attention-choice models that exploit derivatives of observed probabilities
include \cite{abaluck2021consumers,agarwal2025demand}.}
Denote this derivative by
$
FOD_{\tau}^{\omega}(D_{i,1:T},W_{i,1:T}).
$
Plugging \Cref{eq observed prob integrate out q} and \Cref{lemma multiple period likelihood} into the derivative,
we obtain the following decomposition.

\begin{lemma} \label{lemma decomposing fod}
Under \Cref{assumption on shocks full model} (b), (c1), and (c2), 
\begin{align} 
FOD_{\tau}^{\omega}(D_{i,1:T},W_{i,1:T})
=
\int
\underbrace{
\prod_{t=1}^{T}
P(D_{it}\mid R_{it},Q_i=q)
}_{CPL(D_{i,1:T},W_{i,1:T},q)}
\times
\underbrace{
\frac{
\frac{\partial}{\partial X_{i\tau}^{\omega}}
P(D_{i\tau}\mid R_{i\tau},Q_i=q)
}{
P(D_{i\tau}\mid R_{i\tau},Q_i=q)
}
}_{SE_{\tau}^{\omega}(D_{i\tau},R_{i\tau},q)}
\,dF(q),
\label{eq decomposing CPL and SE}
\end{align}
\end{lemma}
CPL denotes the conditional probability level of the takeup sequence given $Q_i=q$, and SE denotes the period-$\tau$ semi-elasticity.
Proof of \Cref{lemma decomposing fod} can be found in \Cref{proof decomposing fod}.

\subsubsection{Identification of $\beta$}

The preliminary model identifies the choice parameters from continuation
behavior, as shown in part (IC) of \Cref{prop identification preliminary}.
We exploit the same intuition in the full model, but now use a continuation
transition embedded within an entire takeup sequence.

For any $\tau\in\{2,\ldots,T\}$, define
\begin{align*}
\widetilde{\mathcal D}_\tau
:=
\left\{
d\in\{0,1\}^T:
d_{\tau-1}=d_\tau=1
\right\}.
\end{align*}
Fix $\tilde d\in\widetilde{\mathcal D}_\tau$ and a covariate sequence
$w$ in the relevant support. Because $\tilde d_{\tau-1}=1$, the household
is fully attentive at period $\tau$. Hence, the period-$\tau$
semi-elasticity depends only on the choice stage:
\begin{align*}
SE_\tau^\omega(1,w_\tau,q)
=
\frac{
h\left(U_{i\tau}^c((1,w_\tau),q)\right)
}{
H\left(U_{i\tau}^c((1,w_\tau),q)\right)
}
\beta^\omega,
\end{align*}
where $h(\cdot):=H'(\cdot)$.

Substituting this expression into \Cref{lemma decomposing fod} gives
\begin{align}
FOD_\tau^\omega(\tilde d,w)
=
\int
CPL(\tilde d,w,q)
\frac{
h\left(U_{i\tau}^c((1,w_\tau),q)\right)
}{
H\left(U_{i\tau}^c((1,w_\tau),q)\right)
}
dF(q) \beta^\omega,
\label{eq fod prob of continue}
\end{align}

Because the integral in \Cref{eq fod prob of continue} is common across
$\omega$, we can pin down the integral by
normalizing $\beta^B=1$, where $B$ denotes the benefit amount.\footnote{
\Cref{assumption on shocks full model} does not impose a scale
normalization. We normalize $|\beta^B|=1$ and use the sign of
$FOD_\tau^B(\tilde d,w)$ to determine the sign of $\beta^B$.
For notational simplicity, we present the identification argument under
$\beta^B=1$. The same argument applies under $\beta^B=-1$.}
Then
\begin{align*}
FOD_\tau^B(\tilde d,w)
=
\int
CPL(\tilde d,w,q)
\frac{
h\left(U_{i\tau}^c((1,w_\tau),q)\right)
}{
H\left(U_{i\tau}^c((1,w_\tau),q)\right)
}
dF(q),
\end{align*}
which yields the following identification result.

\begin{proposition} \label{prop beta overidentified}
Under \Cref{assumption on shocks full model}, normalize $\beta^B=1$.
For any $\tau\in\{2,\ldots,T\}$, $\tilde d\in\widetilde{\mathcal D}_\tau$,
and $w$ in the relevant support such that
$FOD_\tau^B(\tilde d,w)\neq0$,
\begin{align}
FOD_\tau^\omega(\tilde d,w)
= FOD_\tau^B(\tilde d,w) \beta^\omega.
\label{eq beta overidentified}
\end{align}
\end{proposition}

\Cref{prop beta overidentified} has a simple interpretation. For a
sequence containing a continuation transition at period $\tau$, the
effect of $X_{i\tau}^\omega$ on the sequence probability relative to the
effect of the benchmark covariate, benefit amount, identifies $\beta^\omega$. 

\Cref{prop beta overidentified} overidentifies $\beta$ for two reasons. One, a fixed sequence $(d,w)$ may contain multiple continuation behaviors. Two, by varying $(d,w)$, we can obtain infinitely many identifying equations in the form of \Cref{eq beta overidentified}. In \Cref{appendix theorem identifying beta}, we address these two dimensions of overidentification by formulating \Cref{theorem identifying beta}. The overidentification result also elucidates that some model restrictions can be relaxed. \Cref{proof for fully nonparametric model} shows that we can extend our identification argument to a fully nonparametric setup, where the linear component of the utility functions are replaced with varying-coefficient specifications.

\subsubsection{Identification of $\alpha$} \label{sec identifying alpha}

The preceding identification strategy for $\beta$ exploits derivatives
with respect to continuous covariates. It cannot be applied directly to
$\alpha$ because the recertification indicator $Z_{it}$ is binary.
Instead, we identify $\alpha$ using a compensating-variation argument
that compares two covariate histories differing only in the
recertification requirement and benefit amount at a continuation period.
The same argument can be applied to other discrete covariates.

Fix $\tau\in\{2,\ldots,T\}$ and
$\tilde d\in\widetilde{\mathcal D}_\tau$, so that
$\tilde d_{\tau-1}=\tilde d_\tau=1$.
Consider a base covariate history $\tilde w$ satisfying
$(\tilde B_\tau,\tilde Z_\tau)=(b,0)$.
For any scalar $c$, construct a comparison history $\ddot w(c)$ that is
identical to $\tilde w$ in every period except that $(\ddot B_\tau(c),\ddot Z_\tau)=(b+c,1)$.
For any takeup and covariate histories $(d,w)$, denote
$\pi(d,w):=P(D_{i,1:T}=d\mid W_{i,1:T}=w)$.
By \Cref{lemma multiple period likelihood},
\begin{align*}
\pi(\tilde d,\tilde w)
=&~
\int
\prod_{t\neq\tau}
P(\tilde d_t\mid \tilde R_t,Q_i=q)
\times
H\left(
b+X_{i\tau}^{-B}\beta^{-B}+\sigma_2q
\right)
\,dF(q),\\
\pi(\tilde d,\ddot w(c))
=&~
\int
\prod_{t\neq\tau}
P(\tilde d_t\mid \tilde R_t,Q_i=q)
\times
H\left(
b+c+X_{i\tau}^{-B}\beta^{-B}
+\alpha+\sigma_2q
\right)
\,dF(q),
\end{align*}

Under the normalization $\beta^B = 1$, the two period-$\tau$ choice utilities coincide for every $q$ when $c = -\alpha$. Since all remaining transition probabilities are identical across the two histories, $\pi(\tilde{d},\tilde{w}) = \pi(\tilde{d},\ddot{w}(-\alpha))$. 
Moreover, because $H$ is strictly increasing, $\pi(\tilde{d},\ddot{w}(c))$ is strictly increasing in $c$. Provided the required comparison history lies in the support, $c= -\alpha$ is the unique value that equalizes the two observed sequence probabilities.

\begin{proposition} \label{prop alpha overidentified}
Under \Cref{assumption on shocks full model}, normalize $\beta^B=1$.
Fix $\tau\in\{2,\ldots,T\}$,
$\tilde d\in\widetilde{\mathcal D}_\tau$, and a base covariate history
$\tilde w$ with $(\tilde B_\tau,\tilde Z_\tau)=(b,0)$.
Suppose that $\ddot w(c)$, defined by replacing
$(b,0)$ with $(b+c,1)$ at period $\tau$ while holding the remainder of
the covariate history fixed, lies in the relevant support for values of
$c$ in a neighborhood containing $-\alpha$.
Then $\alpha$ is constructively identified as
$\alpha=-c^\ast$,
where $c^\ast$ is the unique value satisfying
$\pi(\tilde d,\tilde w)
=
\pi(\tilde d,\ddot w(c^\ast))$.
\end{proposition}
The proposition has a simple compensating-variation interpretation. At a continuation period, requiring a household to recertify imposes the hassle cost $\alpha$, while increasing the benefit amount raises the utility of continued participation. We compare two otherwise identical participation histories: one without recertification and one with recertification but a different benefit amount. The unique benefit adjustment that makes the two sequence probabilities equal exactly offsets the utility loss from recertification.

Similar to \Cref{prop beta overidentified}, \Cref{prop alpha overidentified} overidentifies $\alpha$. In \Cref{appendix theorem identifying alpha}, we formalize \Cref{theorem identifying alpha} to aggregate the information contained in all admissible matched pairs of baseline history and comparison history.

\subsubsection{Identification of $\gamma$}

Following the intuition of part (IA) of
\Cref{prop identification preliminary}, we identify the attention
parameters by comparing continuation and starting behavior. In the full
model, however, these transitions are embedded within entire takeup
sequences. Fix a period $\tau\in\{2,\ldots,T\}$. Let
$(\tilde d,\tilde w)$ denote a base sequence satisfying
$\tilde d_{\tau-1}=\tilde d_\tau=1$, so that the period-$\tau$
transition is a continuation. Let $(\ddot d,\ddot w)$ denote a
comparison sequence satisfying
$\ddot d_{\tau-1}=0$ and $\ddot d_\tau=1$, so that the
period-$\tau$ transition is a start. We impose additional restrictions on the base and comparison sequences below so that their comparison isolates $\gamma$.

Let
$\tilde r_\tau=(1,\tilde w_\tau)$ and
$\ddot r_\tau=(0,\ddot w_\tau)$ denote the period-$\tau$ observed
states of the base and comparison sequences, respectively. Because the
base sequence involves a continuation, the household is fully attentive
at period $\tau$, and its semi-elasticity contains only the choice
component:
\begin{align*}
SE^\omega_\tau(\tilde d_\tau,\tilde r_\tau,q)
=
\underbrace{
\frac{
h\left(U_{i\tau}^c(\tilde r_\tau,q)\right)
}{
H\left(U_{i\tau}^c(\tilde r_\tau,q)\right)
}
\beta^\omega
}_{\text{Choice}}.
\end{align*}
By contrast, the comparison sequence involves a start, so its
semi-elasticity contains both attention and choice components:
\begin{align*}
SE^\omega_\tau(\ddot d_\tau,\ddot r_\tau,q)
=
\underbrace{
\frac{
g\left(U_{i\tau}^a(\ddot x_\tau,q)\right)
}{
G\left(U_{i\tau}^a(\ddot x_\tau,q)\right)
}
\gamma^\omega
}_{\text{Attention}}
+
\underbrace{
\frac{
h\left(U_{i\tau}^c(\ddot r_\tau,q)\right)
}{
H\left(U_{i\tau}^c(\ddot r_\tau,q)\right)
}
\beta^\omega
}_{\text{Choice}},
\end{align*}
where $g(\cdot):=G'(\cdot)$ and $h(\cdot):=H'(\cdot)$.

The difference between the two semi-elasticities does not yet isolate the
attention component. First, their choice components generally differ because
the period-$\tau$ choice utilities may differ across the base and comparison
sequences. Second, by \Cref{lemma decomposing fod}, the observed FOD integrates
the period-$\tau$ semi-elasticity weighted by the $q$-specific CPL of the entire takeup sequence. We therefore impose two
high-level matching conditions on the base and comparison sequences. H1 equates their
period-$\tau$ choice utilities, while H2
equates their CPLs:
\begin{align}
U_{i\tau}^c(\tilde r_\tau,q)
&=
U_{i\tau}^c(\ddot r_\tau,q)
\qquad \forall q\in\mathbb R.
\tag{H1}
\label{condition gamma H1}\\
CPL(\tilde d,\tilde w,q)
&=
CPL(\ddot d,\ddot w,q)
\qquad \forall q\in\mathbb R,
\tag{H2}
\label{condition gamma H2}
\end{align}
Under H1 and H2,
differencing the two FODs isolates the attention-stage semi-elasticity.
$$FOD^\omega_\tau(\ddot{d},\ddot{w}) - FOD^\omega_\tau(\tilde{d},\tilde{w}) = \int CPL(\ddot{d},\ddot{w},q) \frac{g(U_{i\tau}^a(\ddot{x}_\tau,q))}{G(U_{i\tau}^a(\ddot{x}_\tau,q))} dF(q) \gamma^\omega.$$

\begin{proposition} \label{prop gamma overidentified}
Under \Cref{assumption on shocks full model}, normalize $\gamma^B=1$. For any
$\tau\in\{2,\ldots,T\}$ and any base and comparison sequences
$(\tilde d,\tilde w)$ and $(\ddot d,\ddot w)$ satisfying H1 and H2,
\begin{align}
FOD^\omega_\tau(\ddot d,\ddot w)
-
FOD^\omega_\tau(\tilde d,\tilde w)
=
\left[
FOD^B_\tau(\ddot d,\ddot w)
-
FOD^B_\tau(\tilde d,\tilde w)
\right]
\gamma^\omega.
\label{eq gamma overidentified}
\end{align}
\end{proposition}
\Cref{prop gamma overidentified} has the same basic interpretation as the
identification argument in the preliminary model. The continuation sequence
isolates the choice response because a previous participant is fully attentive,
whereas the starting sequence reflects both the attention and choice responses.
Under H1 and H2, differencing the two FODs removes the common choice response and
leaves only the attention response. Relative to the corresponding difference
in FODs with respect to the benefit amount, this identifies the sign and scale
of $\gamma^\omega$.

Similar to \Cref{prop beta overidentified}, \Cref{prop gamma overidentified} overidentifies $\gamma$. We formalize \Cref{theorem identifying gamma} in \Cref{appendix theorem identifying gamma} to aggregate the information contained in all admissible matched sequence pairs.

\Cref{prop gamma overidentified} and \Cref{theorem identifying gamma} are conditional on the existence of
base and comparison sequences satisfying H1 and H2. We now show
constructively that such sequence pairs can exist. Consider a population
of households observed for four periods of eligibility. The following
base and comparison sequences provide one example:

\begin{table}[H]
    \centering
    \begin{minipage}{0.48\textwidth}
        \centering
        \begin{tabular}{c c c c c c}
            $t$ & 0 & 1 & 2 & 3 & 4   \\
            $\tilde{d}$ & 0 & 1 & 1 & 0 & 0  \\
            $\tilde{x}$ &  & $x^*$ & $x^*$ & $x^*$ & $x^*$ \\
            $\tilde{z}$ &  & 0 & 1 & 0 & 0
        \end{tabular}
    \end{minipage}
    \hfill
    \begin{minipage}{0.48\textwidth}
        \centering
        \begin{tabular}{c c c c c c}
            $t$ & 0 & 1 & 2 & 3 & 4   \\
            $\ddot{d}$ & 0 & 0 & 1 & 1 & 0   \\
            $\ddot{x}$ &  & $x^*$ & $x^*$ & $x^*$ & $x^*$ \\
            $\ddot{z}$ &  & 0 & 0 & 1 & 0
        \end{tabular}
    \end{minipage}
\end{table}

We set $X_{it}=x^*$ in every period for both sequences only to simplify
the construction; time-invariant covariates are not required for the
general identification argument. The base sequence contains, in order,
a signup, a continuation with recertification, an exit, and a period of
remaining out of the program. The comparison sequence contains exactly
the same four transition types, but in a different order: remaining out,
signup, continuation with recertification, and exit. Because the
associated covariates are the same, the two sequence probabilities
conditional on $Q_i=q$ consist of the same four transition-probability
factors, rearranged across periods. Hence,
\begin{align*}
\prod_{t=1}^4
P(D_{it}=\tilde d_t\mid R_{it} = \tilde r_t,Q_i=q)
=
\prod_{t=1}^4
P(D_{it}=\ddot d_t\mid R_{it} = \ddot r_t,Q_i=q)
\qquad \forall q\in\mathbb R.
\end{align*}
By \Cref{lemma multiple period likelihood}, this implies
\begin{align*}
CPL(\tilde d,\tilde w,q)
=
CPL(\ddot d,\ddot w,q)
\qquad \forall q\in\mathbb R,
\end{align*}
so H2 is satisfied.

We next verify H1 at $\tau=2$. In the base sequence,
$\tilde d_1=\tilde d_2=1$ and $\tilde z_2=1$, so period 2 is a
continuation with recertification. In the comparison sequence,
$\ddot d_1=0$ and $\ddot d_2=1$, so period 2 is a signup. Both
transitions involve the same hassle state, $S_{i2}=1$. Because the two
sequences also share $X_{i2}=x^*$,
\begin{align*}
U_{i2}^c(\tilde r_2,q)
=
U_{i2}^c(\ddot r_2,q)
=
x^*\beta+\alpha+\sigma_2q
\qquad \forall q\in\mathbb R.
\end{align*}
Thus H1 is also satisfied.

This construction shows that H1 and H2 are jointly attainable with a
four-period participation history. Combined with
\Cref{prop gamma overidentified}, it provides a constructive
identification argument for $\gamma$ whenever the required sequence and
covariate configurations lie in the relevant support. Together with the
preceding identification results for $\alpha$ and $\beta$, this implies
that, under the maintained assumptions and support conditions, a
four-period panel can be sufficient to identify
$(\alpha,\beta,\gamma)$. This short-panel requirement broadens the
potential applicability of the framework to settings such as college
financial aid, where participation histories may be observed for only a
limited number of periods.

\subsection{Estimation}
We propose an MLE for the model, which maximizes the log likelihood of observing the \textit{sequence} of welfare takeup decisions for all households in the sample, denoted as $\mathcal{I}$, 
\begin{align*}
    &\log\left(\prod_{i \in \mathcal{I}}P(D_{i,1:T_i} \mid  X_{i,1:T_i},Z_{i,1:T_i})\right) \\
    =&~ \sum_{i \in \mathcal{I}} \log \int P(D_{i,1:T_i} \mid  X_{i,1:T_i},Z_{i,1:T_i}, Q_i = q) dF(q) \qquad \text{Integrate out random effect} \\
    =&~ \sum_{i \in \mathcal{I}} \log \int \prod_{t=1}^{T_i} P(D_{it} \mid  D_{it-1}, X_{it}, Z_{it},Q_i = q) dF(q) \qquad \text{\Cref{lemma multiple period likelihood}} \\
    =&~ \sum_{i \in \mathcal{I}} \log \int \prod_{t=1}^{T_i} \left(P_a(D_{it-1},X_{it}, Q_i) P_c(D_{it-1}, X_{it}, Z_{it}, Q_i)\right)^{D_{it}} \qquad\text{Model setup}\\
    & \qquad\qquad\qquad \left(1-P_a(D_{it-1},X_{it}, Q_i) P_c(D_{it-1},X_{it}, Z_{it}, Q_i)\right)^{1-D_{it}} dF(q)
\end{align*}

For tractability, We set $\{F,G,H\}$ as independent standard normal. Under this specification the MLE is
    \begin{equation*}
    \begin{split}
        \{\hat{\gamma},\hat{\beta},\hat{\sigma}\} =& \argmax_{\gamma,\beta,
        \sigma} ~ \sum_{i \in \mathcal{I}} \log \int
        \prod_{t=1}^{T_i} \Bigg[\Phi(X_{it}\gamma + \sigma_1 q)^{1-D_{it-1}}  \Phi(X_{it}\beta +  S_{it}\alpha + \sigma_2 q)\Bigg]^{D_{it}} \\
        &\qquad \qquad \Bigg[1 - \Phi(X_{it}\gamma + \sigma_1 q)^{1-D_{it-1}} \Phi(X_{it}\beta +  S_{it}\alpha + \sigma_2 q)\Bigg]^{1 - D_{it}} \, d\Phi(q)
    \end{split}
    \end{equation*}
Following \cite{heiss2021inattention}, we use Gauss-Hermite quadrature to approximate the integral and solve the maximization problem numerically.


\section{Data and institutional details} \label{sec data}
We apply our framework to the National Longitudinal Survey of Youth 1997 (NLSY97) to study eligible households' takeup behavior of the Special Supplemental Nutrition Program for Women, Infants, and Children (WIC).\footnote{WIC is one of the largest nutrition assistance programs in the US, providing free healthy food, nutrition education, and healthcare support to low-income pregnant women, new mothers, and young children up to age of 48 months who are at nutritional risk. In 2002  (one of the years in our sample period), it accounted for almost 12 percent of total Federal spending on food and nutrition assistance.} 
We identify WIC-eligible households in the NLSY97 using two requirements: categorical eligibility and income eligibility.

\begin{itemize}
    \item[] \textit{Categorical requirement:} A household is categorically eligible if it includes a pregnant woman; a woman within six weeks after childbirth; a non-breastfeeding postpartum woman up to six months after childbirth or the end of pregnancy; a breastfeeding woman up to the infant's first birthday; an infant under age one; or a child under age five.

    \item[] \textit{Income requirement:} A household is income-eligible if its income is below 185 percent of the federal poverty level. Households participating in certain other means-tested programs, including SNAP (formerly the Food Stamp Program), Medicaid, and Temporary Assistance for Needy Families (formerly Aid to Families with Dependent Children, or AFDC), are also automatically income-eligible for WIC.
\end{itemize}

Using information on household composition and annual household income, we construct a sample of WIC-eligible households from 2000 through 2009. The sample period ends in 2009 because the NLSY97 stopped collecting information on welfare-program participation in 2010. 

On top of these categorical and income requirements, the WIC program additionally requires households to prove their nutritional-risk status. Following \cite{bitler2003wic}, we do not impose the nutritional-risk requirement. Instead, we subsume proving nutritional-risk eligibility as part of the hassle cost.

The resulting sample contains 2,991 unique households and 109,591 household-month observations. The overall WIC take-up rate in the sample is 47.3\%. Official WIC reports indicate that take-up during the 2000s ranged from 56\% to 62\%. Because WIC participation in the NLSY97 is self-reported, it is not surprising that the estimated take-up rate is lower than the official figures \citep{ko2024take}. Estimates from other self-reported survey data are similarly lower. For example, \cite{jackson2016child} estimates an average WIC take-up rate of 44.6\% using Survey of Income and Program Participation data from 2004 through 2008.

\subsection{Summary statistics}

\Cref{table sample summary} presents descriptive statistics for the estimation sample. Panel A shows how WIC take-up varies across demographic groups. Female household members account for approximately two-thirds of the household-month observations and have a substantially higher take-up rate than male household members, at 55.1\% compared with 31.6\%. Take-up also varies meaningfully across racial groups. Hispanic households have an average take-up rate of 52.6\%, approximately 5.3 pp above the overall sample rate, and their annual take-up rate exceeds the corresponding overall rate in all nine sample years. In contrast, Black households have an average take-up rate of 44.0\%, approximately 3.3 pp below the overall rate, and their annual rate remains below the overall rate in every sample year, as illustrated in \Cref{fig takeup by race}. Average take-up rates differ little across the two education groups, both of which have rates of slightly above 47\%.

\begin{table}[p]
\centering
\caption{Descriptive Statistics: WIC-Eligible Sample}
\label{table sample summary}

\setlength{\tabcolsep}{10pt}
\renewcommand{\arraystretch}{1}

\begin{adjustbox}{max width=\textwidth}
\begin{threeparttable}

\begin{tabular}{@{}lrrrr@{}}
\toprule
Subgroup
&
\makecell[r]{Household-month\\observations}
&
\makecell[r]{Sample\\share}
&
\makecell[r]{WIC take-up\\rate}
&
\makecell[r]{Above annual \\ rate (years/9)}
\\
\midrule

\addlinespace[5pt]
\textbf{Overall sample}
    & 109,591 & 100.0\% & 47.3\% & -- \\

\addlinespace[10pt]
\multicolumn{5}{@{}l}{\textbf{Panel A: Demographic characteristics}} \\

\addlinespace[5pt]
\multicolumn{5}{@{}l}{\textit{Gender}} \\[2pt]
\quad Male
    & 36,272 & 33.1\% & 31.6\% & 0/9 \\
\quad Female
    & 73,319 & 66.9\% & 55.1\% & 9/9 \\

\addlinespace[5pt]
\multicolumn{5}{@{}l}{\textit{Race}} \\[2pt]
\quad Black
    & 48,430 & 44.2\% & 44.0\% & 0/9 \\
\quad Hispanic
    & 26,372 & 24.1\% & 52.6\% & 9/9 \\
\quad Other
    & 34,789 & 31.7\% & 47.8\% & 5/9 \\

\addlinespace[5pt]
\multicolumn{5}{@{}l}{\textit{Education}} \\[2pt]
\quad Less than high school
    & 37,404 & 34.1\% & 47.4\% & 2/9 \\
\quad At least high school
    & 72,187 & 65.9\% & 47.2\% & 7/9 \\

\addlinespace[10pt]
\multicolumn{5}{@{}l}{\textbf{Panel B: Geographic and economic characteristics}} \\

\addlinespace[5pt]
\multicolumn{5}{@{}l}{\textit{Region}} \\[2pt]
\quad Northeast
    & 13,278 & 12.1\% & 48.7\% & 6/9 \\
\quad Midwest
    & 23,281 & 21.2\% & 45.2\% & 2/9 \\
\quad South
    & 53,031 & 48.4\% & 46.8\% & 2/9 \\
\quad West
    & 20,001 & 18.3\% & 50.0\% & 7/9 \\

\addlinespace[5pt]
\multicolumn{5}{@{}l}{\textit{Urban residence}} \\[2pt]
\quad Urban
    & 85,313 & 77.8\% & 46.4\% & 2/9 \\
\quad Rural
    & 21,271 & 19.4\% & 50.5\% & 7/9 \\

\addlinespace[5pt]
\multicolumn{5}{@{}l}{\textit{Household income}} \\[2pt]
\quad At or above sample median
    & 56,545 & 51.6\% & 45.6\% & 1/9 \\
\quad Below sample median
    & 53,046 & 48.4\% & 49.1\% & 8/9 \\

\addlinespace[5pt]
\multicolumn{5}{@{}l}{\textit{SNAP participation}} \\[2pt]
\quad Participating
    & 47,311 & 43.2\% & 60.3\% & 9/9 \\
\quad Not participating
    & 62,212 & 56.8\% & 37.4\% & 0/9 \\

\addlinespace[10pt]
\multicolumn{5}{@{}l}{\textbf{Panel C: Program-related characteristics}} \\

\addlinespace[5pt]
\multicolumn{5}{@{}l}{\textit{Categorical eligibility}} \\[2pt]
\quad Pregnant women
    & 29,019 & 26.5\% & 45.3\% & 1/9 \\
\quad Youngest child 1--12 months
    & 30,897 & 28.2\% & 67.3\% & 9/9 \\
\quad Youngest child 13 months or older
    & 49,675 & 45.3\% & 36.0\% & 0/9 \\

\addlinespace[5pt]
\multicolumn{5}{@{}l}{\textit{Imputed monthly WIC benefit}} \\[2pt]
\quad $\hat{B}_{it} < 100$
    & 45,851 & 41.8\% & 35.2\% & 0/9 \\
\quad $\hat{B}_{it} \in [100, 200)$
    & 58,789 & 53.6\% & 55.1\% & 9/9 \\
\quad $\hat{B}_{it} \geq 200$
    & 4,951 & 4.5\% & 66.0\% & 9/9 \\

\addlinespace[5pt]
\multicolumn{5}{@{}l}{\textit{Previous-period WIC status}} \\[2pt]
\quad $D_{i,t-1} = 1, Z_{it} = 0$
    & 49,762 & 45.4\% & 96.1\% & 9/9 \\
\quad $D_{i,t-1} = 1, Z_{it} = 1$
    & 4,012 & 3.7\% & 94.4\% & 9/9 \\
\quad $D_{i,t-1} = 0$
    & 58,808 & 53.7\% & 4.0\% & 0/9 \\

\bottomrule
\end{tabular}

\begin{tablenotes}[flushleft]
\footnotesize
\item
\textit{Notes:}
The unit of observation is a household-month. Sample shares are relative
to the 109,591 household-month observations in the estimation sample.
The WIC take-up rate is the subgroup mean of the WIC participation
indicator. The final column reports the number of years from 2001 through
2009 in which the subgroup take-up rate exceeded the overall take-up rate
in the corresponding year. 
\end{tablenotes}

\end{threeparttable}
\end{adjustbox}
\end{table}

Panel B shows how WIC take-up varies across geographic and economic characteristics. Differences across regions, urban status, and household income are relatively modest. The most salient pattern concerns SNAP participation. Households already participating in SNAP have a WIC take-up rate of 60.3\%, compared with 37.4\% among SNAP nonparticipants, a difference of approximately 23 pp. As shown in \Cref{fig takeup by SNAP}, the take-up rate among SNAP participants exceeds the overall rate in all nine sample years, whereas the rate among nonparticipants remains below the overall rate throughout the sample period.

Panel C begins by comparing households across categorical-eligibility groups. Among these groups, households with an infant aged 1--12 months have the highest WIC take-up rate, at 67.3\%. Their annual take-up rate also exceeds the corresponding overall rate in all nine sample years. This pattern is consistent with the well-documented high rate of WIC participation among families with infants \citep{bitler2003wic}. 

\subsection{Benefit imputation}

Monthly WIC benefit amounts are reported for most participating households but
are unobserved for nonparticipants. Because the structural model requires a
potential benefit amount for every eligible household-month, we construct an
imputed benefit measure for the entire sample. We apply the same imputation
procedure to participants and nonparticipants so that benefit amounts are
measured consistently across participation states.

Reported benefit amounts vary systematically with household composition,
calendar period, and geographic region. We therefore estimate a median
regression using observations with nonmissing reported benefits. The
household-composition vector $HC_{it}$ contains the number of pregnant women,
infants aged 1--12 months, and preschool-aged children aged 13--48 months in
household $i$ at month $t$. Let $Post_{it}$ be an indicator equal to one for
observations in 2007 or later. We allow the effects of household composition
to vary between the pre- and post-2007 periods and across the four Census
regions: the Midwest, Northeast, South, and West. The specification is
\begin{equation}
\label{eq benefit imputation}
\begin{aligned}
Q_{0.5}\left(
B_{it}^{\mathrm{reported}}
\mid HC_{it},Post_{it},Region_{it}
\right)
={}
\theta_0
+
HC_{it}\theta^{HC}
+
(HC_{it}\times Post_{it})\theta^{HP}& \\
+
(HC_{it}\times Region_{it})\theta^{HR}
+
(HC_{it}\times Post_{it}\times Region_{it})\theta^{HPR}&.
\end{aligned}
\end{equation}

We use median regression to limit the
influence of extreme reported benefit amounts. The fitted conditional median,
denoted by $\widehat B_{it}$, is assigned to every eligible household-month,
including observations for which a benefit amount is reported.
\Cref{appendix Benefit imputation variable details} presents the descriptive
evidence motivating the predictors and interaction terms, along with the
regression results.

\Cref{table median imputed benefit} summarizes the median imputed benefit for
three simple household-composition groups. Households with one infant have
the largest median benefit in both periods. Their median benefit increases
from \$120 before 2007 to \$150 during 2007--2009. The median benefit for
households with one preschool-aged child also increases, from \$75 to \$92.50,
while the median benefit for households with one pregnant woman remains at
\$100.

\begin{table}[htbp]
    \centering
    \caption{Median Imputed Monthly WIC Benefits by Household Composition}
    \label{table median imputed benefit}
    \begin{tabular}{lcc}
        \toprule
        Household composition & Before 2007 & 2007--2009 \\
        \midrule
        One pregnant woman  & \$100.00 & \$100.00 \\
        One infant          & \$120.00 & \$150.00 \\
        One preschooler     & \$75.00  & \$92.50  \\
        \bottomrule
    \end{tabular}
\end{table}

\begin{figure}[htbp]
    \centering
    \includegraphics[width=0.75\linewidth]{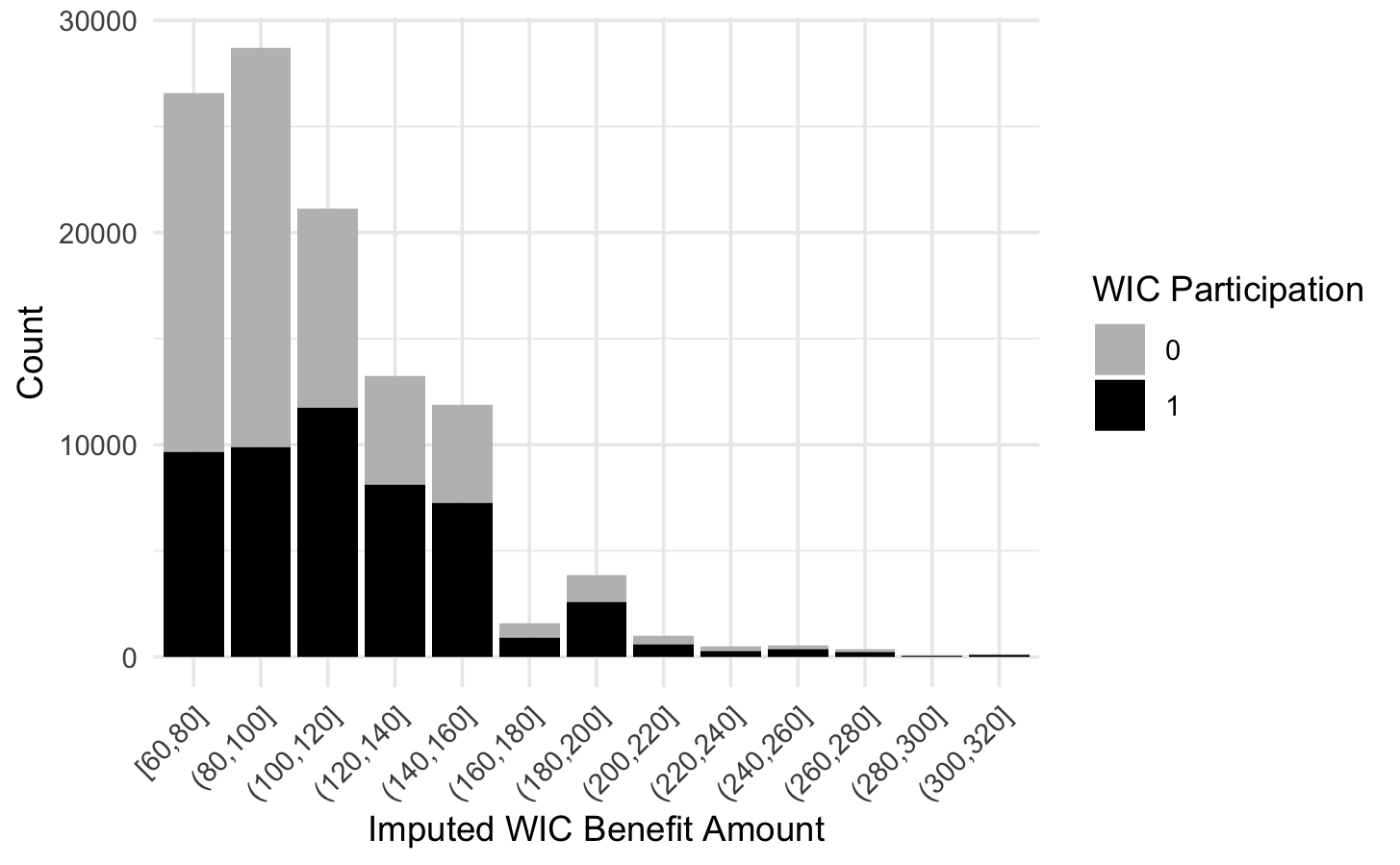}
    \caption{Distribution of Eligible Household-Months by Imputed Monthly
    WIC Benefit and Participation Status}
    \label{fig takeup by benefit}

    \caption*{\footnotesize
    \textit{Notes:} The figure groups eligible household-month observations
    into \$20 intervals based on the imputed monthly benefit
    $\widehat B_{it}$. Black and gray portions of each bar represent WIC
    participants and nonparticipants, respectively. The figure displays
    observations with imputed benefits between \$60 and \$320. Variation in
    household composition, calendar period, and region generates additional
    variation in $\widehat B_{it}$ beyond the six representative values
    reported in \Cref{table median imputed benefit}.
    }
\end{figure}

Panel C of
\Cref{table sample summary} shows a strong positive association between
imputed benefit amounts and WIC participation. Households with an imputed
monthly benefit below \$100 have a take-up rate of 35.2\%. The rate rises to
55.1\% among households with an imputed benefit between \$100 and \$200 and
to 66.0\% among households with an imputed benefit of at least \$200.
\Cref{fig takeup by benefit} illustrates this relationship across more
detailed benefit intervals.

\subsection{Highly persistent take-up} \label{sec persistent take up}

Panel C of \Cref{table sample summary} documents substantial persistence in
WIC participation. Among households participating in the previous month, the
probability of continuing participation is 96.1\% when recertification is not
required and 94.4\% when eligibility must be recertified. Thus, participation
remains highly persistent even in recertification months, although the
continuation rate is approximately 1.7 pp lower. Together with
the recurring interactions that current participants have with the program,
these high continuation rates support our assumption that participating
households are attentive in the following period.

In contrast, among households that did not participate in the previous month,
the probability of beginning participation is only 4.0\%. The gap of more than
90 pp between continuation and entry probabilities indicates
that participation histories contain markedly different information about
current participants and nonparticipants. This pattern motivates our
two-stage framework: current participants proceed directly to the choice
stage, whereas previous nonparticipants must first become attentive before
considering whether to participate.

\subsection{Definition of $Z_{it}$}
WIC participants must periodically recertify their eligibility. Recertification
generally occurs at 12-month intervals. In addition, a new certification event
is associated with changes in the benefit package when an infant begins
receiving infant benefits and when the child transitions out of the infant
benefit package at 13 months. These requirements generate predictable months
in which participating households face an additional administrative burden.

We do not directly observe the month in which each household completes
recertification. We therefore construct a recertification indicator using the
age of the household's youngest eligible child, $YA_{it}$. Specifically, we define
\begin{equation}
    Z_{it}
    =
    \mathbbm{1}
    \left\{
    YA_{it}
    \in \{1,13,25,37\}
    \right\}.
\end{equation}
The first two ages correspond to the beginning and end of the infant formula benefit
period, while ages 25 and 37 reflect subsequent 12-month certification
intervals. We refer to household-months with $Z_{it}=1$ as recertification
months.

To assess whether the age-based indicator captures recertification, we compare
continuation probabilities in months classified as recertification months with
those in the immediately following months. In particular, we compare households
whose youngest eligible child is 13 months old with those whose youngest child
is 14 months old, and households whose youngest child is 25 months old with
those whose youngest child is 26 months old. Within each comparison, the
applicable benefit category is unchanged, but ages 13 and 25 correspond to
predicted recertification months, whereas ages 14 and 26 do not.

As shown in \Cref{fig recertification}, the probability of continuing
participation is lower at age 13 than at age 14 and lower at age 25 than at age
26 in each of the nine sample years. The consistency of these differences
supports using $Z_{it}$ as a proxy for whether a participating household faces
recertification in month $t$.

\section{Estimation results} \label{sec estimation results}
We first report the preliminary and full model estimation results, and then compare the full model with other robustness check specifications.

\subsection{Preliminary model vs full model}
\Cref{table preliminary full estimates} reports estimates from the preliminary
and full models. The signs of most covariate coefficients are
stable across the two specifications, although several magnitudes change after
persistent heterogeneity is introduced.

\begin{table}[p]
    \centering
    \caption{Preliminary- and Full-Model Estimates}
    \label{table preliminary full estimates}

    \setlength{\tabcolsep}{6pt}
    \renewcommand{\arraystretch}{1}

    \footnotesize
    \begin{adjustbox}{max width=\textwidth}
    \begin{tabular}{lrrrrr}
        \toprule
        & \multicolumn{2}{c}{Preliminary model}
        & \multicolumn{3}{c}{Full model} \\
        \cmidrule(lr){2-3}
        \cmidrule(lr){4-6}
        Variable
        & \multicolumn{1}{c}{Estimate}
        & \multicolumn{1}{c}{SE}
        & \multicolumn{1}{c}{Estimate}
        & \multicolumn{1}{c}{SE}
        & \multicolumn{1}{c}{$\Delta P$ (pp)} \\
        \midrule

        \multicolumn{6}{l}{\textbf{Panel A: Attention parameters}} \\[1pt]

        Intercept
        & $-2.0198$ & $0.0604$
        & $-2.0720$ & $0.0829$
        & -- \\

        Imputed monthly benefit, $\widehat B_{it}$
        & $0.0041$ & $0.0003$
        & $0.0054$ & $0.0004$
        & $1.28$ \\

        Has infant
        & $0.1621$ & $0.0247$
        & $0.2899$ & $0.0302$
        & $4.53$ \\

        Black
        & $-0.1004$ & $0.0208$
        & $-0.1341$ & $0.0326$
        & $-1.54$ \\

        Hispanic
        & $0.1094$ & $0.0234$
        & $0.2057$ & $0.0347$
        & $3.03$ \\

        Female
        & $0.3426$ & $0.0197$
        & $0.5074$ & $0.0303$
        & $9.15$ \\

        Less than high school
        & $-0.0170$ & $0.0193$
        & $-0.0590$ & $0.0290$
        & $-0.72$ \\

        Urban
        & $-0.1051$ & $0.0212$
        & $-0.1434$ & $0.0289$
        & $-1.63$ \\

        SNAP participant
        & $0.2917$ & $0.0189$
        & $0.3925$ & $0.0254$
        & $6.57$ \\

        Income
        & $-0.0361$ & $0.0134$
        & $-0.0150$ & $0.0193$
        & $0.19$ \\

        Income squared
        & $0.0056$ & $0.0011$
        & $0.0041$ & $0.0016$
        & -- \\

        Year trend
        & $-0.0746$ & $0.0038$
        & $-0.1109$ & $0.0054$
        & $-1.29$ \\


        \addlinespace[10pt]
        \multicolumn{6}{l}{\textbf{Panel B: Choice parameters}} \\[1pt]

        Intercept
        & $1.5461$ & $0.0828$
        & $1.2866$ & $0.0932$
        & -- \\

        Imputed monthly benefit, $\widehat B_{it}$
        & $0.0033$ & $0.0004$
        & $0.0049$ & $0.0005$
        & $0.94$ \\

        Has infant
        & $0.0276$ & $0.0294$
        & $0.0558$ & $0.0314$
        & $0.62$ \\

        Black
        & $0.0034$ & $0.0262$
        & $-0.0251$ & $0.0330$
        & $-0.30$ \\

        Hispanic
        & $0.1034$ & $0.0298$
        & $0.1454$ & $0.0363$
        & $1.52$ \\

        Female
        & $0.1011$ & $0.0260$
        & $0.2386$ & $0.0323$
        & $2.31$ \\

        Less than high school
        & $-0.0431$ & $0.0236$
        & $-0.0749$ & $0.0296$
        & $-0.93$ \\

        Urban
        & $-0.0832$ & $0.0269$
        & $-0.0915$ & $0.0309$
        & $-1.15$ \\

        SNAP participant
        & $0.0041$ & $0.0222$
        & $0.0623$ & $0.0250$
        & $0.69$ \\

        Income
        & $-0.0083$ & $0.0180$
        & $0.0188$ & $0.0160$
        & $-0.07$ \\

        Income squared
        & $0.0001$ & $0.0014$
        & $-0.0023$ & $0.0012$
        & -- \\

        Eligibility (re)certification, $S_{it}$
        & $-0.3606$ & $0.0378$
        & $-0.3291$ & $0.0394$
        & $-4.92$ \\

        Year trend
        & $-0.0095$ & $0.0050$
        & $-0.0428$ & $0.0059$
        & $-0.52$ \\


 \addlinespace[10pt]
        \multicolumn{6}{l}{
            \textbf{Panel C: Persistent heterogeneity and model fit}
        } \\[1pt]

        Attention heterogeneity, $\sigma_1$
        & $0$ & --
        & $0.1553$ & $0.0045$
        & -- \\

        Choice heterogeneity, $\sigma_2$
        & $0$ & --
        & $0.1012$ & $0.0045$
        & -- \\

        Implied hassle cost
        & \multicolumn{2}{c}{\$109.27}
        & \multicolumn{2}{c}{\$67.16}
        & -- \\

        Log likelihood
        & \multicolumn{2}{c}{$-20{,}197.09$}
        & \multicolumn{2}{c}{$-19{,}707.82$}
        & -- \\

        AIC
        & \multicolumn{2}{c}{$40{,}444.18$}
        & \multicolumn{2}{c}{$39{,}469.64$}
        & -- \\

        BIC
        & \multicolumn{2}{c}{$40{,}684.29$}
        & \multicolumn{2}{c}{$39{,}728.96$}
        & -- \\

        Household-month observations
        & \multicolumn{2}{c}{$109{,}591$}
        & \multicolumn{2}{c}{$109{,}591$}
        & -- \\
        \bottomrule
    \end{tabular}
    \end{adjustbox}

    \vspace{3pt}

    \parbox{\textwidth}{
        \footnotesize
        \textit{Notes:}
        The table reports maximum-likelihood estimates and standard errors.
        The preliminary model restricts the persistent unobserved-heterogeneity
        parameters $\sigma_1$ and $\sigma_2$ to zero, whereas the full model
        estimates both parameters. The final column reports discrete changes
        in predicted probabilities under the full model, measured in percentage
        points, relative to a baseline household (explained in the main text). In Panel A, $\Delta P$ denotes the change in the attention
        probability, while in Panel B it denotes the change in the choice
        probability. 

        
        To compute $\Delta P$, binary characteristics are changed from zero to one while all
        other characteristics are held fixed. The benefit comparison increases
        $\widehat B_{it}$ from \$75 to \$92.50, the income comparison increases
        annual household income from \$20,000 to \$25,000, and the year
        comparison changes 2001 to 2002. The income probability changes jointly
        incorporate the linear and squared income terms. Random effects are
        integrated over their estimated distribution. 

        The benefit coefficients measure the effect of a one-dollar increase
        in the imputed monthly WIC benefit. The implied hassle cost is reported
        as a positive dollar amount and is calculated as
        $-\widehat{\beta}^{S}/\widehat{\beta}^{B}$.

        AIC and BIC are computed from the maximized log likelihood, with BIC using
the 109,591 household-month observations as the sample size.
        \par
    }
\end{table}

The preliminary model restricts the persistent unobserved heterogeneity parameters $\sigma_1$ and $\sigma_2$ to zero, whereas the full
model estimates both parameters. Both heterogeneity parameters are positive and precisely estimated in the full model. Allowing for persistent heterogeneity also substantially improves model fit: the log likelihood increases from -20,197.09 to -19,707.82, while both AIC and BIC decrease substantially. These results favor the full model over the restrictions imposed by the preliminary specification. Hence, for the remainder of this section, we focus on interpreting the full-model estimates.

\subsection{Full model estimates}

To facilitate interpretation of the nonlinear estimates, we compare the
predicted attention and choice probabilities of several comparison households
with those of a baseline household. The baseline household has a
monthly benefit of \$75, has no infant, is neither Black nor Hispanic, is male,
has at least a high-school education, resides in a rural area, does not
participate in SNAP, has annual household income of \$20,000, and is observed
in 2001. The baseline choice probability is evaluated in a
nonrecertification month, while the baseline attention probability applies to
a household that did not participate in the preceding month. After integrating
over the estimated distribution of persistent unobserved heterogeneity, the
predicted attention and choice probabilities for this household are 6.60\%
and 94.11\%, respectively.

For each covariate reported in Panels A and B of
\Cref{table preliminary full estimates}, we change the corresponding
characteristic while holding all other characteristics fixed. The final column
reports the resulting discrete change in the relevant predicted probability,
measured in percentage points. In Panel A, $\Delta P$ refers to the change in
the attention probability, whereas in Panel B it refers to the change in the
choice probability.

\subsubsection{Benefit amount}
The imputed benefit amount, $\widehat B_{it}$, has a positive and statistically
significant coefficient in both stages. Increasing the monthly benefit from
\$75 to \$92.50 raises the predicted attention probability by 1.28 pp and the predicted choice probability by 0.94 pp. Thus, within the model, increases in WIC benefit operate through both margins: they make eligible households more likely to become attentive to the program and, conditional on attention, more likely to choose participation. This mechanism suggests that benefit expansions, such as those associated with revisions to the WIC food package, can affect take-up through both attention and choice.

\subsubsection{Has infant}
The estimated effect of having an infant is substantially larger in the
attention stage than in the choice stage. Holding the imputed benefit fixed at
\$75, changing the household from having no infant to having an infant raises
the predicted attention probability by 4.53 pp. The
corresponding increase in the choice probability is 0.62 pp.
The attention coefficient is statistically significant at conventional
levels, while the choice coefficient is significant at the 10\% level but not
at the 5\% level. 

Previous research documents substantially higher WIC participation among
households with infants \citep{bitler2003wic}. Our estimates provide a
mechanism for this pattern: conditional on the imputed benefit amount and
other observed characteristics, having an infant predicts a substantially
higher probability of becoming attentive to WIC. Thus, the elevated
participation of households with infants appears to operate not only through
their larger benefit packages but also through greater attention to the
program.

\subsubsection{Racial differences}
We also find meaningful racial differences. Compared with the baseline household, a Black household has a predicted attention probability that is 1.54 pp lower, while its predicted choice probability is only 0.30 pp lower. The Black coefficient is negative and statistically significant in the attention stage, whereas the corresponding choice coefficient is not statistically different from zero. In contrast, a Hispanic household has predicted attention and choice probabilities that are 3.03 pp and 1.52 pp higher than those of the baseline household, respectively. Both Hispanic coefficients are statistically significant.

Conditional on the observed characteristics included in the model, these estimates suggest that the lower participation associated with Black households operates primarily through the attention margin rather than the choice margin. This pattern points to outreach and information provision as potentially relevant policy margins. In particular, it provides a rationale for examining targeted outreach designed to increase program awareness and engagement among Black families, complementing existing WIC outreach initiatives such as targeted outreach grant programs conducted in partnership with CinnaMoms.

\subsubsection{Gender differences and SNAP participation status}
Two other covariates---gender and SNAP participation---strongly influence predicted attention. The Female indicator has positive and statistically significant coefficients in both the attention and choice stages. Compared with the baseline household, changing the gender indicator from male to female increases the predicted attention probability by 9.15 pp and the predicted choice probability by 2.31 pp. The difference is particularly large at the attention stage.

SNAP participation exhibits a similar pattern. Changing the baseline household from a SNAP nonparticipant to a SNAP participant raises the predicted attention probability by 6.57 pp, while the predicted choice probability increases by only 0.69 pp. Thus, the substantially higher WIC take-up among SNAP participants documented in the descriptive statistics appears to be associated primarily with greater attention to WIC rather than a large difference in the conditional choice to participate.

\subsection{Hassle costs and attention during recertification} \label{sec recertification attention less than one}

Because the imputed benefit 
$\hat{B}_{it}$ enters both stages in dollar units, ratios of coefficients to the corresponding stage-specific benefit coefficient admit a dollar-equivalent interpretation. For example, the implied (re)certification hassle cost is $-\hat{\beta}^S/\hat{\beta}^B$, which equals approximately \$67.16 under the full model.

Identification of the hassle cost relies on the assumption that households participating in the previous month are fully attentive, including during recertification months. We assess the sensitivity of the estimates to this assumption by mechanically reducing the attention probability of participating households in recertification months. In addition to the full model, which sets this probability to one, we estimate specifications imposing attention probabilities of 0.99, 0.98, 0.97, and 0.96. We report key results for all five specifications in \Cref{table recertification attention robustness}.

The 0.98 specification provides a particularly informative benchmark. The
observed continuation probability is 96.1\% in nonrecertification months and
94.4\% in recertification months, and
$94.4/96.1 \approx 0.982$. Thus, if the entire raw difference in continuation
rates were attributed to temporary inattention, the implied attention
probability during recertification would be approximately 98\%. Even under
this conservative benchmark for the hassle-cost estimate, the implied hassle
cost remains \$38.67, more than one-third of the sample median monthly benefit
of \$100. The 97\% and 96\% specifications provide increasingly stringent
sensitivity checks by assigning an even larger role to temporary inattention.

\begin{table}[htbp]
\centering
\caption{Selected Estimates under Alternative Attention Probabilities during Recertification}
\label{table recertification attention robustness}

\footnotesize
\setlength{\tabcolsep}{7pt}
\renewcommand{\arraystretch}{1.05}

\begin{adjustbox}{max width=\textwidth}
\begin{tabular}{lccccc}
    \toprule
    & Full
    & A99
    & A98
    & A97
    & A96 \\
    & $(100\%)$
    & $(99\%)$
    & $(98\%)$
    & $(97\%)$
    & $(96\%)$ \\
    \midrule

    \multicolumn{6}{l}{\textbf{Panel A: Attention parameters}} \\[2pt]

    Imputed monthly benefit, $\widehat B_{it}$
    & \makecell{$0.0054$ \\ $(0.0004)$}
    & \makecell{$0.0054$ \\ $(0.0004)$}
    & \makecell{$0.0055$ \\ $(0.0004)$}
    & \makecell{$0.0055$ \\ $(0.0004)$}
    & \makecell{$0.0056$ \\ $(0.0004)$} \\

    Has infant
    & \makecell{$0.2899$ \\ $(0.0302)$}
    & \makecell{$0.2888$ \\ $(0.0301)$}
    & \makecell{$0.2878$ \\ $(0.0300)$}
    & \makecell{$0.2868$ \\ $(0.0300)$}
    & \makecell{$0.2858$ \\ $(0.0299)$} \\

    SNAP participant
    & \makecell{$0.3925$ \\ $(0.0254)$}
    & \makecell{$0.3919$ \\ $(0.0253)$}
    & \makecell{$0.3912$ \\ $(0.0253)$}
    & \makecell{$0.3906$ \\ $(0.0252)$}
    & \makecell{$0.3900$ \\ $(0.0252)$} \\

    \addlinespace[6pt]
    \multicolumn{6}{l}{\textbf{Panel B: Choice parameters}} \\[2pt]

    Imputed monthly benefit, $\widehat B_{it}$
    & \makecell{$0.0049$ \\ $(0.0005)$}
    & \makecell{$0.0049$ \\ $(0.0005)$}
    & \makecell{$0.0049$ \\ $(0.0005)$}
    & \makecell{$0.0049$ \\ $(0.0005)$}
    & \makecell{$0.0049$ \\ $(0.0005)$} \\

    Has infant
    & \makecell{$0.0558$ \\ $(0.0314)$}
    & \makecell{$0.0515$ \\ $(0.0316)$}
    & \makecell{$0.0479$ \\ $(0.0317)$}
    & \makecell{$0.0442$ \\ $(0.0319)$}
    & \makecell{$0.0404$ \\ $(0.0320)$} \\

    SNAP participant
    & \makecell{$0.0623$ \\ $(0.0250)$}
    & \makecell{$0.0617$ \\ $(0.0252)$}
    & \makecell{$0.0614$ \\ $(0.0254)$}
    & \makecell{$0.0610$ \\ $(0.0255)$}
    & \makecell{$0.0605$ \\ $(0.0257)$} \\

    Eligibility (re)certification, $S_{it}$
    & \makecell{$-0.3291$ \\ $(0.0394)$}
    & \makecell{$-0.2602$ \\ $(0.0434)$}
    & \makecell{$-0.1895$ \\ $(0.0481)$}
    & \makecell{$-0.1133$ \\ $(0.0538)$}
    & \makecell{$-0.0293$ \\ $(0.0611)$} \\

    \addlinespace[6pt]
    \multicolumn{6}{l}{\textbf{Panel C: Heterogeneity and model fit}} \\[2pt]

    $\sigma_1$
    & \makecell{$0.1553$ \\ $(0.0045)$}
    & \makecell{$0.1560$ \\ $(0.0045)$}
    & \makecell{$0.1566$ \\ $(0.0045)$}
    & \makecell{$0.1571$ \\ $(0.0045)$}
    & \makecell{$0.1577$ \\ $(0.0044)$} \\

    $\sigma_2$
    & \makecell{$0.1012$ \\ $(0.0045)$}
    & \makecell{$0.1015$ \\ $(0.0045)$}
    & \makecell{$0.1016$ \\ $(0.0046)$}
    & \makecell{$0.1017$ \\ $(0.0046)$}
    & \makecell{$0.1016$ \\ $(0.0046)$} \\[2pt]

    Implied hassle cost (\$)
    & $67.16$
    & $53.10$
    & $38.67$
    & $23.12$
    & $5.98$ \\

    Log likelihood
    & $-19707.82$
    & $-19722.21$
    & $-19737.93$
    & $-19754.54$
    & $-19771.89$ \\

    Observations
    & $109{,}591$
    & $109{,}591$
    & $109{,}591$
    & $109{,}591$
    & $109{,}591$ \\

    \bottomrule
\end{tabular}
\end{adjustbox}

\vspace{3pt}

\parbox{\textwidth}{
    \footnotesize
    \textit{Notes:}
    Each column reports estimates from a specification imposing the
    indicated attention probability on households that participated in
    WIC in the preceding month and face eligibility recertification in the
    current month. The full model imposes an attention probability of one.
    Standard errors are reported in parentheses. The implied hassle cost is
    calculated as $-\widehat{\beta}^{S}/\widehat{\beta}^{B}$ and is reported
    as a positive dollar amount.
    \par
}
\end{table}

\begin{figure}[htbp]
    \centering
    \includegraphics[width=\linewidth]{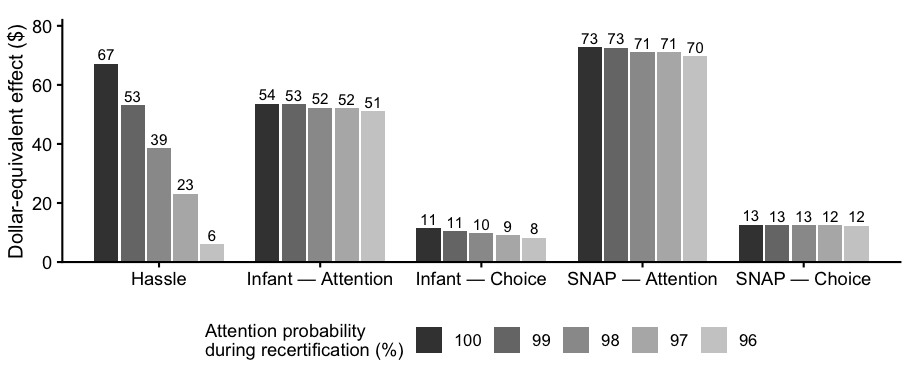}
    \caption{Dollar-Equivalent Effects under Alternative Attention Probabilities during Recertification}
    \label{fig hassle cost comparison}

    \caption*{\footnotesize
    \textit{Notes:} The figure compares dollar-equivalent effects across five
    specifications that set the attention probability of participating
    households during recertification to 100\%, 99\%, 98\%, 97\%, or 96\%.
    Dollar-equivalent effects are obtained by dividing the corresponding
    coefficient by the benefit coefficient within the same stage. The implied
    recertification hassle cost declines sharply as the assumed attention
    probability during recertification decreases, because a larger share of
    nonparticipation is attributed to inattention rather than to the
    choice-stage recertification cost. In contrast, the dollar-equivalent
    effects of infant status and SNAP participation on attention and choice
    remain comparatively stable across specifications.
    }
\end{figure}

As the imposed attention probability during recertification decreases from
100\% to 96\%, the estimated hassle cost falls sharply from \$67.16 to \$5.98.
Importantly, this sensitivity is concentrated in the hassle-cost estimate
rather than appearing broadly across the other parameters.
\Cref{fig hassle cost comparison} illustrates this contrast using
dollar-equivalent effects. Across the five specifications, the
dollar-equivalent effect of having an infant ranges only from approximately
\$51 to \$54 in the attention stage and from \$8 to \$11 in the choice stage.
Similarly, the dollar-equivalent effect of SNAP participation remains between
approximately \$70 and \$73 in the attention stage and between \$12 and \$13
in the choice stage.

The concentration of sensitivity in the hassle-cost estimate is consistent
with the identification argument in \Cref{prop alpha overidentified}.
Identification of $\alpha$ relies specifically on variation in
recertification status: the compensating benefit variation is identified by
comparing otherwise comparable participation sequences that differ in whether
recertification is required. Consequently, changing how program exit during
recertification is allocated between the attention and choice stages directly
affects the identification and estimation of $\alpha$. In contrast, the identifying
variation for the remaining choice parameters $\beta$ and the attention
parameters $\gamma$ is not tied exclusively to recertification status.
Consistent with this distinction, their estimates remain comparatively stable
across the five specifications.


\subsection{Addressing endogeneity from omitted heterogeneity}

Some unobserved characteristics may be correlated with observed covariates
while also entering the attention and choice utility shocks,
$\Xi_{it}$ and $\mathcal{E}_{it}$. We examine the robustness of our results
to such endogeneity concerns using grouped fixed-effect specifications
inspired by \citet{bonhomme2022discretizing}. In the first step, we construct
observable summary statistics that are informative about particular sources
of omitted heterogeneity and use these statistics to classify observations
into a small number of types. In the second step, we take these classifications
as given and include type-specific fixed effects in the parametric
maximum-likelihood estimation. We consider two sources of omitted
heterogeneity: persistent household characteristics related to safety-net
participation and time-varying local economic conditions. The results are summarized in \Cref{table grouped fe robustness}.

\begin{table}[htbp]
    \centering
    \caption{Selected Estimates under Grouped Fixed-Effect Specifications}
    \label{table grouped fe robustness}

    \footnotesize
    \setlength{\tabcolsep}{6pt}
    \renewcommand{\arraystretch}{1.05}

    \begin{adjustbox}{max width=\textwidth}
    \begin{tabular}{lrrrrr}
        \toprule
        & Full
        & \multicolumn{2}{c}{Household-type FE}
        & \multicolumn{2}{c}{Regional economy FE} \\
        \cmidrule(lr){3-4}
        \cmidrule(lr){5-6}
        Variable
        & Estimate
        & Estimate & SE
        & Estimate & SE \\
        \midrule

        \multicolumn{6}{l}{\textbf{Panel A: Attention parameters}} \\[2pt]

        Imputed monthly benefit, $\widehat B_{it}$
        & 0.0054
        & 0.0053 & 0.0004
        & 0.0055 & 0.0004 \\

        Has infant
        & 0.2899
        & 0.2981 & 0.0303
        & 0.2827 & 0.0302 \\

        SNAP participant
        & 0.3925
        & 0.2736 & 0.0300
        & 0.3927 & 0.0253 \\

        Income
        & -0.0150
        & -0.0261 & 0.0183
        & -0.0157 & 0.0180 \\

        Income squared
        & 0.0041
        & 0.0052 & 0.0015
        & 0.0041 & 0.0014 \\[2pt]

        Household type 2 FE, $\delta_2^{A}$
        & --
        & 0.3099 & 0.0388
        & -- & -- \\

        Household type 3 FE, $\delta_3^{A}$
        & --
        & 0.2755 & 0.0428
        & -- & -- \\

        Household type 4 FE, $\delta_4^{A}$
        & --
        & 0.3816 & 0.0445
        & -- & -- \\

        Local downturn FE
        & --
        & -- & --
        & -0.1065 & 0.0296 \\

        Local upturn FE
        & --
        & -- & --
        & -0.0820 & 0.0262 \\

        \addlinespace[6pt]
        \multicolumn{6}{l}{\textbf{Panel B: Choice parameters}} \\[2pt]

        Imputed monthly benefit, $\widehat B_{it}$
        & 0.0049
        & 0.0049 & 0.0004
        & 0.0048 & 0.0004 \\

        Has infant
        & 0.0558
        & 0.0570 & 0.0312
        & 0.0613 & 0.0314 \\

        SNAP participant
        & 0.0623
        & -0.0119 & 0.0323
        & 0.0639 & 0.0251 \\

        Income
        & 0.0188
        & 0.0138 & 0.0156
        & 0.0177 & 0.0295 \\

        Income squared
        & -0.0023
        & -0.0018 & 0.0012
        & -0.0022 & 0.0024 \\

        Eligibility (re)certification, $S_{it}$
        & -0.3291
        & -0.3276 & 0.0394
        & -0.3298 & 0.0395 \\[2pt]

        Household type 2 FE, $\delta_2^{C}$
        & --
        & 0.0670 & 0.0427
        & -- & -- \\

        Household type 3 FE, $\delta_3^{C}$
        & --
        & 0.1403 & 0.0458
        & -- & -- \\

        Household type 4 FE, $\delta_4^{C}$
        & --
        & 0.2218 & 0.0479
        & -- & -- \\

        Local downturn FE
        & --
        & -- & --
        & 0.0670 & 0.0324 \\

        Local upturn FE
        & --
        & -- & --
        & -0.0529 & 0.0279 \\

        \addlinespace[6pt]
        \multicolumn{6}{l}{\textbf{Panel C: Heterogeneity and model fit}} \\[2pt]

        $\sigma_1$
        & 0.1553
        & 0.1541 & 0.0045
        & 0.1554 & 0.0045 \\

        $\sigma_2$
        & 0.1012
        & 0.0995 & 0.0045
        & 0.1017 & 0.0045 \\

        Implied hassle cost (\$)
        & 67.16
        & 66.86 & --
        & 68.71 & -- \\

        Log likelihood
        & -19707.82
        & -19651.87 & --
        & -19692.39 & -- \\

        Observations
        & 109,591
        & 109,591 & --
        & 109,591 & -- \\

        \bottomrule
    \end{tabular}
    \end{adjustbox}

    \vspace{3pt}

    \parbox{\textwidth}{
        \footnotesize
        \textit{Notes:}
        The Full column reports estimates from the baseline full model.
        Standard errors for the full-model estimates are reported in
        \Cref{table preliminary full estimates} and are omitted here to avoid
        repetition. The household-type FE specification classifies households
        according to their long-run SNAP participation histories and retains
        contemporaneous SNAP participation as an observed covariate.
        Household Type 1 is the omitted category. The local-condition FE
        specification classifies region-year observations according to changes
        in median household income, with the intermediate-growth group serving
        as the omitted category. The implied hassle cost is calculated as
        $-\widehat{\beta}^{S}/\widehat{\beta}^{B}$ and is reported as a
        positive dollar amount. Dashes indicate parameters that are not
        included in the corresponding specification.
    }
\end{table}

\subsubsection{Persistent household characteristics}

The full model includes a household-level random effect to capture persistent
unobserved heterogeneity. A restriction of the random-effects specification,
however, is that the random effect is independent of the observed covariates.
This restriction may be violated. For example, households may differ persistently in
their familiarity with safety-net programs, administrative capacity,
information, financial vulnerability, or preferences toward public
assistance. These characteristics may affect both SNAP participation and WIC take-up, generating correlation between observed SNAP participation and the unobserved determinants of attention and choice.

\begin{figure}[htbp]
    \centering

    \subfigure[Distribution of household SNAP participation rates]{
        \includegraphics[width=0.47\linewidth]{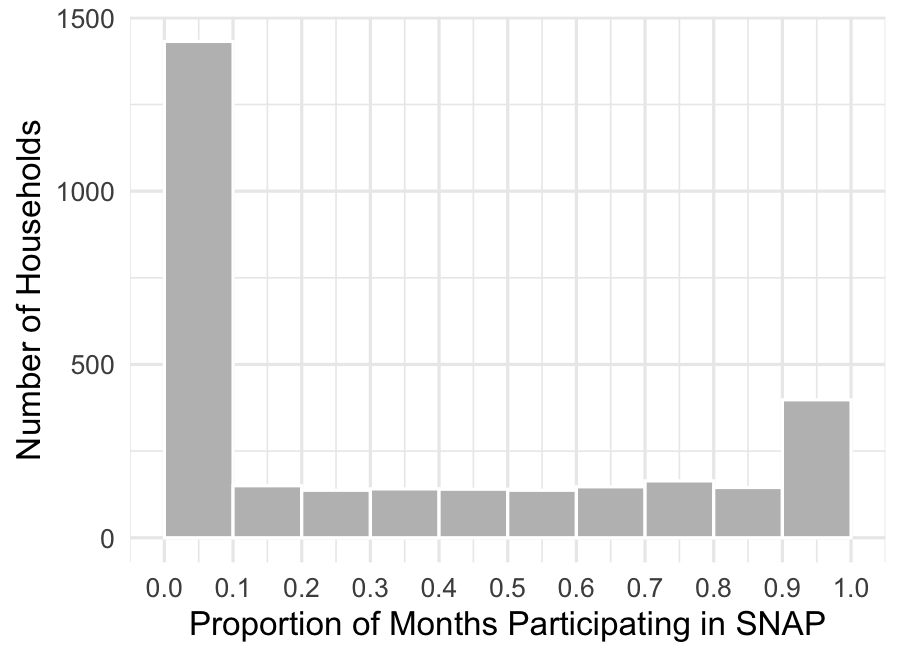}
        \label{fig distribution snap rate}
    }
    \hfill
    \subfigure[WIC participation by household SNAP participation rate]{
        \includegraphics[width=0.47\linewidth]{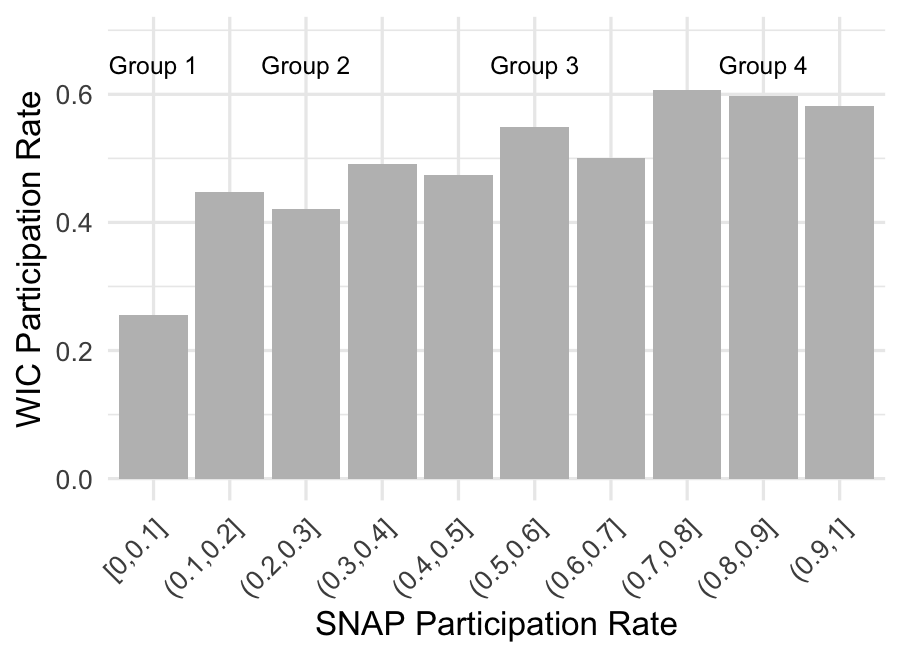}
        \label{fig wic snap}
    }

    \caption{Household SNAP Participation Histories and WIC Participation}
    \label{fig snap household types}

    \caption*{\footnotesize
    \textit{Notes:}
    For each household, the SNAP participation rate is defined as the fraction
    of observed months in which the household participates in SNAP.
    Panel (a) shows the distribution of this household-specific participation
    rate. Panel (b) reports the average WIC participation rate for households
    grouped into 0.1-wide bins of their SNAP participation rate.
    The substantial mass near zero and one, together with variation in WIC
    participation across the SNAP-participation distribution, motivates
    discretizing households into types according to their long-run SNAP
    participation histories.
    }
\end{figure}

To allow for such correlated persistent heterogeneity, we classify households
according to their long-run SNAP participation histories and include
household-type fixed effects in both stages of the model. Let
$SP_i$ denote the fraction of observed months in which household $i$
participates in SNAP. We discretize $SP_i$ into four household types\footnote{\Cref{fig distribution snap rate} shows substantial heterogeneity in households'
SNAP participation histories, with a large mass of households participating
in SNAP in fewer than 10\% of their observed months. This distribution motivates the definition of $g(i)$.}
\begin{align*}
g(i)=
\begin{cases}
1, & SP_i < 0.1,\\
2, & 0.1 \leq SP_i < 0.4,\\
3, & 0.4 \leq SP_i < 0.7,\\
4, & SP_i \geq 0.7.
\end{cases}
\end{align*}

The attention and choice indices are augmented by
type-specific effects $\delta^A_{g(i)}$ and $\delta^C_{g(i)}$, respectively,
while retaining the continuous household random effect from the full model. The augmented utility functions are 
\begin{align*}
U^{A}_{it}
&=
X_{it}'\gamma
+\delta^{A}_{g(i)}
+\sigma_{1}Q_{i}
+\xi_{it},
\\
U^{C}_{it}
&=
X_{it}'\beta
+\alpha S_{it}
+\delta^{C}_{g(i)}
+\sigma_{2}Q_{i}
+\epsilon_{it}.
\end{align*}
We normalize $\delta^A_1=\delta^C_1=0$, so the remaining type-specific
effects are interpreted relative to households with $p_i<0.1$.

\Cref{fig wic snap} shows that households with higher long-run SNAP
participation rates also generally exhibit higher WIC participation rates.
This relationship suggests that the SNAP participation history contains
information about persistent household characteristics that are relevant for
WIC take-up, motivating its use to construct household types.

\Cref{table grouped fe robustness} shows that the central structural
estimates are highly stable after incorporating household-type fixed effects.
The benefit and infant coefficients change little in either stage, as do the
estimated heterogeneity parameters $\sigma_1$ and $\sigma_2$. The
recertification coefficient is also nearly unchanged, leaving the implied
hassle cost at \$66.86 compared with \$67.16 in the full model.

The main change occurs in the coefficients on contemporaneous SNAP
participation. In the attention stage, the SNAP coefficient declines from
0.39 in the full model to 0.27 after the household-type fixed effects are
included. This decline is consistent with the grouped fixed effects absorbing
part of the persistent household heterogeneity that is correlated with
long-run SNAP participation. The remaining SNAP coefficient captures the
conditional association between contemporaneous SNAP participation and WIC
attention among households with similar long-run SNAP participation
histories.

The corresponding SNAP coefficient in the choice stage falls from 0.06 to
$-0.01$ and is no longer statistically different from zero. Thus, after
controlling for persistent household heterogeneity associated with long-run
SNAP participation, we find little evidence of an additional association
between contemporaneous SNAP participation and the conditional choice to
participate in WIC. This result suggests that much of the baseline
choice-stage association between SNAP and WIC participation is absorbed by
persistent household characteristics captured by the grouped fixed effects.

The estimated household-type fixed effects are positive relative to Group 1
in both stages. In the choice stage, the fixed effects increase monotonically
with the household's long-run SNAP participation group, from 0.07 for Group 2
to 0.14 for Group 3 and 0.22 for Group 4. The attention-stage fixed effects
are also positive and sizable, although they are not strictly monotonic
across the four groups. Overall, the estimates indicate substantial
persistent differences in WIC attention and choice across households with
different long-run histories of safety-net participation.

\subsubsection{Regional economic downturns and upturns}

Time-varying regional economic conditions may be correlated with observed
household characteristics such as income while also affecting unobserved
determinants of WIC attention and choice. Economic downturns, for example,
may increase the salience of WIC as safety-net programs become more relevant
and more frequently discussed among friends, neighbors, and coworkers. At
the same time, worsening economic conditions may increase financial stress
and cognitive burden, potentially exacerbating limited mental bandwidth problems.
Economic upturns may operate through the opposite margins. We therefore
allow the attention and choice utilities to vary flexibly across different
regional economic conditions.

We classify each region-year combination, hereafter referred to as a cell,
into one of three types: economic downturn, normal conditions, or economic
upturn. The classification is constructed using the unrestricted NLSY97
sample rather than the WIC-eligible estimation sample. The unrestricted
sample contains 76,031 annual household-income observations during
2000--2009. For each household with valid income observations in consecutive
years, we calculate its annual percentage change in household income and then
compute the median income growth within each region-year cell. We classify
cells with median income growth below 3\% as downturns, cells with growth
between 3\% and 6\% as normal conditions, and cells with growth above 6\%
as upturns.

\Cref{table region year income growth} summarizes the resulting
classification. Downturn cells are concentrated in the South and West,
whereas normal-growth cells occur disproportionately in the Northeast and
Midwest. The downturn classification also exhibits a clear temporal pattern:
all but one downturn cells occur during 2001--2003 or 2008--2009. Thus, the
classification captures variation across both regions and years rather than
simply reproducing either regional or calendar-year fixed effects.
\begin{table}[htbp]
    \centering
    \caption{Median Household Income Growth by Region and Year}
    \label{table region year income growth}

    \begin{tabular}{lrrrr}
        \toprule
        Year & Northeast & Midwest & South & West \\
        \midrule

        2001
        & \cellcolor{gray!31}5.08
        & \cellcolor{gray!31}4.44
        & \cellcolor{gray!71}2.02
        & 8.33 \\

        2002
        & \cellcolor{gray!31}4.00
        & \cellcolor{gray!31}3.03
        & \cellcolor{gray!71}1.34
        & \cellcolor{gray!71}-1.58 \\

        2003
        & \cellcolor{gray!71}1.08
        & \cellcolor{gray!71}-1.70
        & \cellcolor{gray!71}1.15
        & \cellcolor{gray!71}2.08 \\

        2004
        & 11.20
        & \cellcolor{gray!31}3.68
        & 10.60
        & 22.10 \\

        2005
        & \cellcolor{gray!31}5.48
        & 6.23
        & \cellcolor{gray!31}3.71
        & \cellcolor{gray!71}2.51 \\

        2006
        & \cellcolor{gray!31}5.20
        & \cellcolor{gray!31}4.44
        & 6.19
        & 7.86 \\

        2007
        & 6.65
        & 9.09
        & 6.22
        & 8.09 \\

        2008
        & \cellcolor{gray!71}2.86
        & 8.09
        & \cellcolor{gray!31}4.80
        & 7.06 \\

        2009
        & 6.67
        & \cellcolor{gray!31}5.25
        & \cellcolor{gray!71}2.69
        & \cellcolor{gray!31}5.00 \\

        \bottomrule
    \end{tabular}

    \caption*{\footnotesize
    \textit{Notes:}
    Each entry reports the median annual percentage change in household income
    among households in the unrestricted sample for the indicated region-year
    cell. Darker gray cells indicate downturns, defined as growth below 3\%.
    Lighter gray cells indicate normal conditions, defined as growth between
    3\% and 6\%. Unshaded cells indicate upturns, defined as growth above 6\%.
    }
\end{table}

\Cref{table grouped fe robustness} shows that the central structural
estimates remain highly stable after incorporating region-year
economic-condition fixed effects. The benefit and infant coefficients change
little in either stage, as do the estimated heterogeneity parameters
$\sigma_1$ and $\sigma_2$. The recertification coefficient is also nearly
unchanged, leaving the implied hassle cost at \$68.71, compared with \$67.16
in the full model. Thus, allowing the unobserved determinants of attention
and choice to vary with regional economic conditions has little effect on the
full model's main structural estimates.

Relative to cells experiencing
normal economic conditions, upturn cells have a lower attention utility. The
choice-stage upturn coefficient is also negative; its dollar-equivalent
magnitude is
$-0.0529/0.0048 \approx -\$11$.
Thus, households in upturn cells exhibit lower predicted attention to WIC
and, conditional on attention, a lower estimated utility from participation.

Downturn cells exhibit a different pattern across the two stages. Relative
to normal-condition cells, the estimated downturn fixed effect is negative
in the attention stage but positive in the choice stage. The former is
consistent with the bandwidth tax argument in \cite{mullainathan2013scarcity}. The latter implies a dollar-equivalent increase in the value
of participation of
$0.0670/0.0048 \approx \$14$.
This is consistent with safety-net benefits becoming more valuable when
economic conditions deteriorate. 

\section{Policy comparison} \label{sec policy comparison}

The estimation results reveal a striking asymmetry between the two stages. 
For a baseline household that did not participate in the preceding month, 
the predicted attention probability is only 6.60\%, whereas conditional on 
attention, the predicted choice probability is 94.11\%. Viewed 
\textit{statically}, this large gap suggests that attention is the more 
promising policy margin. Yet evidence from the WIC2Five intervention, which will be presented in
\Cref{sec WIC2Five}, points in the opposite direction: messages designed to 
nudge households' choice are substantially more effective in increasing takeup than 
messages designed to boost households' attention. 
The \textit{dynamic} structure of our model provides an explanation for this 
puzzle.

We proceed in three steps. First, we introduce the attention-boosting message (ABM) and 
choice-nudging message (CNM) interventions implemented in WIC2Five. Using the WIC2Five quasi-experiment data, we compare these policies' effects on retention rate. Second, we use the two-stage structural model to simulate ABM and CNM and examine the dynamic mechanism underlying the greater effectiveness of CNM. Finally, we compare the two policies using the marginal value of public funds (MVPF), incorporating both 
their welfare effects and implementation costs. We find that the dynamic 
advantage of CNM does \textit{not} extend beyond participation effects, and it has a lower MVPF than ABM.

\subsection{CNM and ABM in WIC2Five} \label{sec WIC2Five}
The WIC2Five pilot implemented in Vermont provides real-world examples of
the two policy margins considered in our model. The program sent two types
of text messages to WIC participants. CNM
encouraged currently participating households to remain enrolled by
emphasizing the benefits of continued WIC participation. ABM, in contrast, were sent around missed appointments and
reminded households to take the steps necessary to maintain or restore
their benefits. 
Below are examples of each intervention.

\begin{table}[ht]
\centering
\begin{tabular}{>{\raggedright\arraybackslash}p{0.45\textwidth}>{\raggedright\arraybackslash}p{0.45\textwidth}}

\textbf{CNM Example} & \textbf{ABM Example} \\
\justifying Welcome to WIC2Five! Get texts on family nutrition, health, activities and more. Kids who stick with WIC for the first 5 years grow healthy, happy and smart!
& 
\justifying Hi Lynne! We missed seeing you at WIC today. Keep your benefits active. Call us at 802-388-4644 to reschedule. Thanks, the Middlebury Health Department. \\ 
\end{tabular}
\label{table examples of CNM and ABM}
\end{table}
The two interventions were implemented at the site level. Panel A of
\Cref{table WIC2Five evidence} documents treatment assignment and reports
the 36 site-year retention rates underlying our analysis. We estimate the
effects of CNM and ABM using the following difference-in-differences (DiD)
and event-study (ES) specifications:
\begin{align*}
\text{DiD} \qquad
RR_{it}
=&~
\theta_i+\phi_t
+\beta_{DiD}\mathbbm{1}\{t=2017\}\times \mathbbm{d}_i
+\epsilon_{it},
\\
\text{ES} \qquad
RR_{it}
=&~
\theta_i+\phi_t
+\beta_{pre}\mathbbm{1}\{t=2015\}\times\mathbbm{d}_i
+\beta_{post}\mathbbm{1}\{t=2017\}\times\mathbbm{d}_i
+\epsilon_{it},
\end{align*}
where $RR_{it}$ denotes the retention rate at site $i$ in year $t$,
$\theta_i$ and $\phi_t$ denote site and year fixed effects, respectively,
and $\mathbbm{d}_i$ indicates whether site $i$ received the intervention
being evaluated. In the event-study specification, 2016 is the omitted
year. We estimate the two specifications separately for CNM and ABM. For
each intervention, sites that did not receive that intervention form the
control group, including sites assigned only to the other intervention.
Panel B of \Cref{table WIC2Five evidence} reports the estimates.

CNM increased child retention by approximately 8.7 pp in the DiD
specification and 9.7 pp in the event-study specification. In contrast,
the estimated effects of ABM are negative and statistically insignificant.
Because treatment assignment in WIC2Five was not randomized and the analysis is
based on only 12 sites, we interpret these estimates as
suggestive evidence rather than as the primary source of causal
identification in this paper. Nevertheless, we provide
additional robustness checks tailored to causal identification and small sample inference in \Cref{appendix WIC2Five}.

\begin{table}[htbp]
\centering
\footnotesize
\caption{WIC2Five Retention Rates and Treatment Effect Estimates}
\label{table WIC2Five evidence}

\vspace{2em}

\textit{Panel A. Site-level treatment assignment and retention rates}

\smallskip

\begin{tabular}{lccrrr}
\toprule
\multirow{2}{*}{Location}
& \text{CNM}
& \text{ABM}
& \multicolumn{3}{c}{\text{Retention Rate}} \\
\cmidrule(lr){4-6}
& \text{Treatment}
& \text{Treatment}
& \text{2015}
& \text{2016}
& \text{2017} \\
\midrule

Rutland       &              &              & 69.9\% & 65.9\% & 59.2\% \\
Springfield   & $\checkmark$ & $\checkmark$ & 66.9\% & 62.5\% & 64.7\% \\
Bennington    &              &              & 74.8\% & 73.7\% & 63.3\% \\
White River   & $\checkmark$ &              & 70.8\% & 63.3\% & 66.3\% \\
Brattleboro   & $\checkmark$ &              & 72.6\% & 70.3\% & 67.6\% \\
St. Johnsbury &              & $\checkmark$ & 79.9\% & 77.3\% & 70.0\% \\
Newport       &              &              & 81.5\% & 73.7\% & 59.4\% \\
Morrisville   &              & $\checkmark$ & 83.8\% & 84.3\% & 77.9\% \\
St. Albans    &              &              & 73.2\% & 70.4\% & 59.0\% \\
Burlington    & $\checkmark$ &              & 68.1\% & 62.6\% & 62.3\% \\
Middlebury    &              & $\checkmark$ & 84.0\% & 79.7\% & 72.4\% \\
Barre         & $\checkmark$ &              & 66.8\% & 61.4\% & 61.9\% \\

\bottomrule
\end{tabular}

\vspace{2em}

\textit{Panel B. Difference-in-differences and event-study estimates}

\smallskip

\begin{tabular}{lcccc}
\toprule
& \multicolumn{2}{c}{\text{CNM}}
& \multicolumn{2}{c}{\text{ABM}} \\
\cmidrule(lr){2-3}
\cmidrule(lr){4-5}
& DiD & ES & DiD & ES \\
\midrule

$\beta_{DiD}$
& 8.723$^{***}$
&
& $-4.713$
& \\

&
(1.646)
&
& (3.485)
& \\[0.2em]

$\beta_{pre}$
&
& 1.863
&
& 1.075 \\

&
& (1.323)
&
& (1.597) \\[0.2em]

$\beta_{post}$
&
& 9.654$^{***}$
&
& $-4.175$ \\

&
& (1.527)
&
& (3.546) \\

\midrule
Observations & 36 & 36 & 36 & 36 \\
Sites        & 12 & 12 & 12 & 12 \\
\bottomrule
\end{tabular}

\vspace{2em}

\caption*{\footnotesize
\textit{Notes:}
Panel A reports child retention rates for the 12 Vermont WIC sites in
2015--2017 and indicates participation in the choice-nudging message (CNM)
and attention-boosting message (ABM) interventions.
Panel B reports difference-in-differences (DiD) and event-study (ES)
estimates using the 36 site-year observations. Standard errors clustered
at the site level are reported in parentheses.
$^{***}p<0.01$.
}
\end{table}

The finding that CNM promotes ABM takeup more effectively is notable in light of our
structural estimates. Although baseline attention probability
is much lower than the baseline choice probability, CNM, a policy that targets choice mechanism, produces substantially larger
retention gains than the ABM, a policy that targets attention mechanism. In \Cref{sec counterfactual takeup dynamic mechanism}, we use the structural model to show how the dynamic link
between current participation and future attention can rationalize these seemingly contradictory patterns.

\subsection{Policy counterfactuals for takeup} 
\label{sec counterfactual takeup dynamic mechanism}

In this section, our main counterfactual outcome is the average expected
participation rate across all eligible household-months.
Let $\mathcal{EH}$ denote the set of eligible household-month observations
in the counterfactual sample. The average predicted takeup rate is
\begin{align} \label{eq average predicted takeup}
\bar{p}
=
\frac{1}{\lvert \mathcal{EH} \rvert}
\sum_{(i,t)\in\mathcal{EH}}
\int
P\left(
D_{it}=1
\mid
X_{i,1:T},Z_{i,1:T},Q_i=q
\right)
d\Phi(q).
\end{align}

To compute this object, we first obtain the participation probability
recursively conditional on $Q_i=q$. Define
$p_{it}(q)
\equiv
P\left(
D_{it}=1
\mid
X_{i,1:T},Z_{i,1:T},Q_i=q
\right)$. For the first period of each eligibility spell, the household enters as a
nonparticipant. It must therefore first become attentive and then choose to
participate. The first-period takeup probability conditional on $Q_i=q$ is
\begin{align*}
p_{i1}(q)
&=
\Phi\left(
X_{i1}\gamma+\sigma_1 q
\right)
\Phi\left(
X_{i1}\beta+\alpha+\sigma_2 q
\right).
\end{align*}

In the next three subsubsections, we derive the recursive participation
probabilities under three scenarios: the baseline with no intervention,
ABM, and CNM. We
then use Guassian-Hermite quadrature to integrate $p_{it}(q)$ over $q$ and compute the corresponding average predicted
takeup rate $\bar p$. \Cref{table takeup counterfactual} report $\bar p$ under different interventions.

\begin{table}[htbp]
\centering
\footnotesize
\caption{Counterfactual WIC takeup and Dynamic Mechanisms}
\label{table takeup counterfactual}

\begin{tabular}{lrrrr}
\toprule
& \shortstack{$\bar{p}$\\(\%)}
& \shortstack{$\Delta \bar{p}$ \\(pp)}
& \shortstack{Choice\\Contribution\\(pp)}
& \shortstack{Indirect Attention\\Contribution\\(pp)}
\\
\midrule

Observed
& 47.29 & -- & -- & -- \\

Model baseline
& 49.28 & 0.00 & -- & -- \\

\addlinespace
\multicolumn{5}{l}{\textit{Panel A: Attention-boosting messages (ABM)}} \\

ABM: attention floor = 25\%
& 50.72 & 1.45 & -- & -- \\

ABM: attention floor = 50\%
& 53.34 & 4.06 & -- & -- \\

ABM: attention floor = 75\%
& 56.42 & 7.14 & -- & -- \\

ABM: attention floor = 100\%
& 60.05 & 10.77 & -- & -- \\

\addlinespace
\multicolumn{5}{l}{\textit{Panel B: Choice-nudging messages (CNM)}} \\

CNM: choice increase = 1 pp
& 51.77 & 2.49 & 0.58 & 1.91 \\

CNM: choice increase = 2 pp
& 54.40 & 5.13 & 1.19 & 3.94 \\

CNM: choice increase = 3 pp
& 56.90 & 7.63 & 1.75 & 5.88 \\

CNM: choice increase = 4 pp
& 59.09 & 9.81 & 2.22 & 7.59 \\

CNM: choice increase = 5 pp
& 60.86 & 11.58 & 2.61 & 8.97 \\

CNM: choice increase = 6 pp
& 62.17 & 12.90 & 2.90 & 9.99 \\

CNM: choice increase = 7 pp
& 63.07 & 13.80 & 3.10 & 10.70 \\

CNM: choice increase = 8 pp
& 63.67 & 14.40 & 3.24 & 11.16 \\

CNM: choice increase = 9 pp
& 64.07 & 14.79 & 3.33 & 11.46 \\

CNM: choice increase = 10 pp
& 64.32 & 15.05 & 3.39 & 11.65 \\

\bottomrule
\end{tabular}

\vspace{0.5em}
\begin{minipage}{0.95\textwidth}
\footnotesize
\textit{Notes:}
takeup is the average predicted WIC participation rate across eligible
household-months. $\Delta$ takeup is measured relative to the model
baseline. For CNM, the total effect is decomposed into a choice contribution
and an indirect attention contribution. The choice contribution combines the
direct increase in retention induced by CNM with the smaller indirect
choice-state effect generated by additional households remaining participants
in subsequent periods. The indirect attention contribution captures the
increase in future participation arising because households retained by CNM
remain fully attentive in subsequent periods. The two components sum to the
total change in takeup. The decomposition is specific to CNM and is therefore
not reported for ABM.
\end{minipage}

\end{table}

\subsubsection{Baseline}

We first compute the baseline takeup rate in the absence of any
intervention. Given $p_{i,t-1}(q)$, the participation probability in period
$t$ can be computed recursively. By the law of iterated expectations,

\begin{equation}
\label{eq recursion}
\begin{aligned}
p_{it}(q)
={}&
p_{i,t-1}(q)
\Phi\left(
X_{it}\beta+Z_{it}\alpha+\sigma_2 q
\right)
\\
&+
\left[1-p_{i,t-1}(q)\right]
\Phi\left(
X_{it}\gamma+\sigma_1 q
\right)
\Phi\left(
X_{it}\beta+\alpha+\sigma_2 q
\right).
\end{aligned}
\end{equation}

A derivation of \Cref{eq recursion} is provided in
\Cref{proof for eq recursion}. Iterating \Cref{eq recursion} over each
eligibility spell gives $p_{it}(q)$ for every eligible household-month.
We then integrate over $q$ using Gauss--Hermite quadrature and average
across eligible household-months to obtain $\bar p$ as in \Cref{eq average predicted takeup}. \Cref{table takeup counterfactual} reports a baseline predicted takeup rate
of 49.28\%, compared with an observed average takeup rate of 47.29\% in the
data. Thus, the estimated model reproduces the aggregate participation rate
in the sample reasonably well.

\subsubsection{ABM}
We consider four ABM intensities, with attention floors $\nu \in \{25\%,50\%,75\%,100\%\}$. To simulate ABM, we must keep track of participation over the previous two periods because the intervention is targeted specifically to households that have just exited WIC. 
A household receives the attention boost in period $t$ only if $(D_{i,t-2},D_{i,t-1})=(1,0)$. For these recently exited households, ABM raises the attention probability to at least $\nu$,
$\max\left\{
\Phi(X_{it}\gamma+\sigma_1q),\nu
\right\}$.
For households that were nonparticipants in both preceding periods, the attention probability remains at its baseline value
$\Phi(X_{it}\gamma+\sigma_1q)$. Households participating in period $t-1$ remain fully attentive.
\begin{equation}
\begin{aligned}
p^{ABM}_{it}(q)
={}&
p^{ABM}_{i,t-1}(q)
\Phi\left(
X_{it}\beta+Z_{it}\alpha+\sigma_2 q
\right)
\\
&+
r^{ABM}_{i,t-1}(q)
\max\left\{
\Phi\left(X_{it}\gamma+\sigma_1 q\right),\nu
\right\}
\Phi\left(
X_{it}\beta+\alpha+\sigma_2 q
\right)
\\
&+
\left[
1-p^{ABM}_{i,t-1}(q)-r^{ABM}_{i,t-1}(q)
\right]
\Phi\left(
X_{it}\gamma+\sigma_1 q
\right)
\Phi\left(
X_{it}\beta+\alpha+\sigma_2 q
\right).
\end{aligned}
\label{eq abm recursion}
\end{equation}
where $
r^{ABM}_{it-1}(q)
=
p^{ABM}_{i,t-2}(q)
\left[
1-
\Phi\left(
X_{it-1}\beta+Z_{it-1}\alpha+\sigma_2 q
\right)
\right]$. Similarly, we integrate the conditional participation probabilities over
$Q_i$ using Gauss--Hermite quadrature to obtain $\bar p$. The derivation of
the ABM recursion follows the same law-of-iterated-expectations argument as
\Cref{eq recursion} and is therefore omitted.

\Cref{table takeup counterfactual} shows that predicted takeup increases
monotonically with the ABM attention floor. Raising $\nu$ from 25\% to
100\% increases the predicted takeup rate from 50.72\% to 60.05\%.
Nevertheless, even relatively aggressive attention interventions generate
moderate changes in aggregate participation. The median predicted attention probability, integrating over the distribution of household heterogeneity, is only approximately 9\%, so imposing an
attention floor of $\nu=50\%$ represents a substantial increase in attention
for many households. Yet this intervention raises aggregate takeup by only
4.06 pp relative to the baseline.

\subsubsection{CNM}
We consider ten CNM intensities corresponding to increases in the choice
probability of 1 to 10 percentage points. Let
$\delta \in \{0.01,0.02,\ldots,0.10\}$ denote the increase in the choice
probability. Under CNM, a household that participated in WIC in period
$t-1$ receives the choice nudge in period $t$. Thus, its choice probability
is increased by $\delta$, subject to an upper bound of one. The behavior of
previous nonparticipants is unchanged.

The recursion for $p_{it}^{CNM}(q)$ therefore takes the same form as the
baseline recursion in \Cref{eq recursion}, except that the continuation
probability for previous participants is replaced by its CNM counterpart:
\begin{equation}
\label{eq cnm recursion}
\begin{aligned}
p_{it}^{CNM}(q)
={}&
p_{i,t-1}^{CNM}(q)
\min\left\{
\Phi\left(
X_{it}\beta+Z_{it}\alpha+\sigma_2 q
\right)+\delta,\,1
\right\}
\\
&+
\left[1-p_{i,t-1}^{CNM}(q)\right]
\Phi\left(
X_{it}\gamma+\sigma_1 q
\right)
\Phi\left(
X_{it}\beta+\alpha+\sigma_2 q
\right).
\end{aligned}
\end{equation}

\Cref{table takeup counterfactual} shows that as $\delta$ increases from 0.01 to 0.10, the predicted takeup rate rises from 51.77\% to 64.32\%. The increase is monotonic but concave. This pattern reflects an important limitation of CNM: the intervention affects only households that have already entered WIC and therefore does not directly alter the participation decision of previous nonparticipants. As $\delta$ increases, the continuation probability of current participants approaches one, leaving progressively less scope for additional retention gains. Setting $\delta=1$, so that all current participants remain enrolled in the following period, yields a model-implied upper bound of 64.74\%. Thus, at $\delta=0.10$, predicted takeup is already close to the maximum attainable under CNM.

\begin{figure}[htbp]
    \centering

    \includegraphics[width=0.78\textwidth]{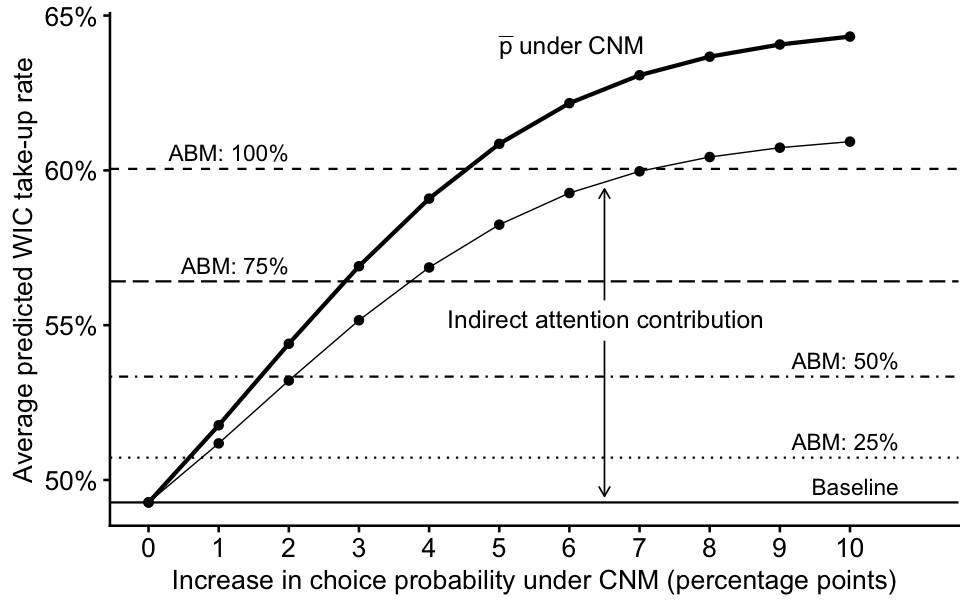}

    \caption{WIC takeup under ABM and CNM}
    \label{fig policy compare takeup}

    \caption*{\footnotesize
    \textit{Notes:}
    The thick solid line with markers reports the average predicted WIC takeup rate under choice-nudging messages (CNM), as the increase in the choice probability varies from 0 to 10 percentage points. The vertical distance between the baseline and the thin solid line represents the indirect attention contribution of CNM; the distance between the thin and thick solid lines represents the choice contribution.
    The horizontal lines report the average predicted takeup rates under attention-boosting messages (ABM) that impose attention floors of 25, 50, 75, and 100 percent.
    The bottom horizontal line reports the baseline predicted takeup rate.
    }
\end{figure}

\subsubsection{Comparison between ABM and CNM}
\Cref{fig policy compare takeup} shows that relatively small CNM interventions can generate larger participation effects than substantially more aggressive ABM interventions. A one-pp choice nudge produces a higher takeup rate than an ABM attention floor of 25\%; a two-pp nudge outperforms a 50\% attention floor; a three-pp nudge outperforms a 75\% attention floor; and a five-pp nudge outperforms even a 100\% attention floor. This pattern aligns with the results from WIC2Five in \Cref{table WIC2Five evidence}: CNM promotes takeup more effectively than ABM.

However, when viewed in light of the model estimates, this pattern is surprising because estimated attention probabilities are generally much lower than choice probabilities, suggesting substantial scope for attention interventions. We reconcile this pattern by decomposing the effectiveness of CNM into choice contribution and indirect attention contribution. CNM nudges household $i$'s choice in period $t$, increasing their retention in month $t$. Once household $i$ is retained, it is fully attentive in period $t+1$. We call this dynamic effect on attention the indirect attention effect. The following decomposition isolates the indirect attention effect.
\begin{align}
\Delta p_{it}^{CNM}(q)
:= & p_{it}^{CNM}(q)-p_{it}(q)
\nonumber\\
=\;&
p_{i,t-1}^{CNM}(q)
\left[
\pi_{it}^{1,\delta}(q)-\pi_{it}^{1}(q)
\right]
+
\Delta p_{i,t-1}^{CNM}(q)
\left[
\pi_{it}^{1}(q)-a_{it}(q)\pi_{it}^{0}(q)
\right]
\nonumber\\
=\;&
\underbrace{
p_{i,t-1}^{CNM}(q)
\left[
\pi_{it}^{1,\delta}(q)-\pi_{it}^{1}(q)
\right]
}_{\text{Direct retention effect}}
+
\underbrace{
\Delta p_{i,t-1}^{CNM}(q)
\left[
\pi_{it}^{1}(q)-\pi_{it}^{0}(q)
\right]
}_{\text{Indirect choice-state effect}}
\nonumber\\
&+
\underbrace{
\Delta p_{i,t-1}^{CNM}(q)
\left[
1-a_{it}(q)
\right]
\pi_{it}^{0}(q)
}_{\text{Indirect attention effect}}.
\label{eq cnm decomposition}
\end{align}
where
\begin{align*}
&a_{it}(q)
\equiv
\Phi\left(
X_{it}\gamma+\sigma_1 q
\right),
&\pi_{it}^{0}(q)
\equiv
\Phi\left(
X_{it}\beta+\alpha+\sigma_2 q
\right),\\
&\pi_{it}^{1}(q)
\equiv
\Phi\left(
X_{it}\beta+Z_{it}\alpha+\sigma_2 q
\right),
&\pi_{it}^{1,\delta}(q)
\equiv
\min\left\{
\pi_{it}^{1}(q)+\delta,\,1
\right\}.
\end{align*}
Using \Cref{eq cnm decomposition}, we compute the contribution of the
indirect attention channel to the overall CNM effect by integrating over
$q$ using Gauss--Hermite quadrature and averaging across $(i,t)$. We combine the other two channels related to choice and refer to their sum as the choice contribution.
\Cref{table takeup counterfactual} summarizes the decomposition results for
all $\delta$ values. \Cref{fig policy compare takeup} shows that, across all
CNM intensities, the indirect attention channel accounts for approximately
77\% of the total increase in takeup. The dynamic link between current
participation and future attention explains why CNM, an intervention that
targets the choice mechanism, is so effective when inattention, rather than
active choice, is the primary barrier to WIC participation.

\subsection{Policy counterfactuals for MVPF}
\label{sec MVPF}

In this section, we evaluate the welfare implications of the counterfactual
policies considered in \Cref{sec counterfactual takeup dynamic mechanism}. Our main welfare object
is the marginal value of public
funds (MVPF) from \cite{hendren2020unified}.

Using model estimates, we back out total expected willingness to pay (WTP) generated by WIC participation across eligible household-months.  Using these WTP estimates and policy cost data, we compare the MVPF (defined as $\frac{\text{marginal WTP}}{\text{marginal cost}}$ in our policy counterfactual context) of ABM and CNM across the policy intensities considered above. The
results show that ABM has a substantially higher MVPF than CNM throughout
the range of $\nu$ and $\delta$ we consider. 

\subsubsection{WTP construction}
We are interested in total expected willingness to pay (WTP) generated by WIC participation
across eligible household-months
\begin{align*}
   \sum_{(i,t) \in \mathcal{EH}} \mathbbm{E}\left[\frac{U^c_{it} - \xi_{it}}{\beta^B} D_{it} \middle| X_{i,1:T}, Z_{i,1:T}\right].  
\end{align*}
We normalize the choice stage utility by the benefit parameter $\beta^B$ so that the WTP is measured in dollar amount. 
\begin{align*}
    \mathbbm{E}\left[\frac{U^c_{it} - \xi_{it}}{\beta^B} D_{it} \middle| X_{i,1:T}, Z_{i,1:T}\right] =&~ \int \mathbbm{E}\left[\frac{U^c_{it} - \xi_{it}}{\beta^B} D_{it} \middle| X_{i,1:T}, Z_{i,1:T}, Q_i = q\right] d\Phi(q)
\end{align*}
For the baseline, we characterize the conditional expectation as a recursive equation
\begin{align} \label{eq recursion for wtp}
    &\mathbbm{E}\left[(U^c_{it} - \xi_{it}) D_{it} \middle| X_{i,1:T}, Z_{i,1:T}, Q_i = q\right]  \nonumber \\
    =~& \underbrace{\psi(V_{it}^1(q))p_{it-1}(q)}_{\text{from previous participant}} + 
    \underbrace{\psi(V_{it}^0(q))\Phi(X_{it}\gamma + \sigma_1 q)(1 - p_{it-1}(q))}_{\text{from previous nonparticipant}}
\end{align}
where $\psi(x) = x\Phi(x) + \phi(x)$, $V_{it}^1(q) = X_{it}\beta + Z_{it}\alpha + \sigma_2 q$, $V_{it}^0(q) = X_{it}\beta + \alpha + \sigma_2 q$.
The proof for \Cref{eq recursion for wtp} can be found in \Cref{appendix proof for eq recursion for wtp}. \Cref{eq recursion for wtp} has a clean interpretation. $\psi(V_{it}^1(q))$ is the expected choice-stage surplus for a previous participant ($D_{it-1} = 1$), incorporating both the probability of continuing WIC participation and the realized surplus conditional on continuation. It is weighted by $p_{it-1}$, the conditional probability that household $i$ participated in the previous period.
$\psi(V_{it}^0(q))$ is the expected choice-stage surplus for a previous nonparticipant ($D_{it-1} = 0$). It is multiplied by the additional attention probability $\Phi(X_{it}\gamma + \sigma_1 q)$ as previous nonparticipants need to pay attention before moving on to the choice stage. It is also weighted by $1 - p_{it-1}$.



Under ABM, the choice-stage utility and choice rule are unchanged. The policy affects WTP only by increasing the probability that a recently exited household reaches the choice stage. Using the recent-exit probability $r_{i,t-1}^{ABM}(q)$ defined above, the conditional expected WTP under ABM can therefore be written as
\begin{align}
&\mathbbm{E}\left[
(U^c_{it}-\xi_{it})D_{it}
\middle|
X_{i,1:T},Z_{i,1:T},Q_i=q;\mathrm{ABM}
\right]
\nonumber\\
={}&
p_{i,t-1}^{ABM}(q)
\psi\left(V_{it}^1(q)\right)
+
r_{i,t-1}^{ABM}(q)
\max\left\{
\Phi(X_{it}\gamma+\sigma_1q),\nu
\right\}
\psi\left(V_{it}^0(q)\right)
\nonumber\\
&+
\left[
1-p_{i,t-1}^{ABM}(q)-r_{i,t-1}^{ABM}(q)
\right]
\Phi(X_{it}\gamma+\sigma_1q)
\psi\left(V_{it}^0(q)\right).
\label{eq abm wtp recursion}
\end{align}

Under CNM, the previous-nonparticipant branch is unchanged, but the welfare calculation for previous participants requires an additional adjustment. CNM raises their choice probability without changing the systematic choice utility $V_{it}^1(q)$. Let
\begin{align*}
\pi_{it}^{1,\delta}(q)
\equiv
\min\left\{
\Phi\left(V_{it}^1(q)\right)+\delta,,1
\right\},
c_{it}^{1,\delta}(q)
\equiv
\Phi^{-1}\left(
\pi_{it}^{1,\delta}(q)
\right).
\end{align*}
The CNM choice rule can then be represented as choosing WIC whenever
$\xi_{it}<c_{it}^{1,\delta}(q)$. Holding $V_{it}^1(q)$ fixed, expected
choice-stage surplus for a previous participant becomes
\begin{align*}
\psi^{CNM}\left(V_{it}^1(q),\delta\right)
\equiv
\int_{-\infty}^{c_{it}^{1,\delta}(q)}
\left(V_{it}^1(q)-c\right)d\Phi(c)
=
V_{it}^1(q)\pi_{it}^{1,\delta}(q)
+
\phi\left(c_{it}^{1,\delta}(q)\right).
\end{align*}
When $\delta=0$, $c_{it}^{1,\delta}(q)=V_{it}^1(q)$ and
$\psi^{CNM}(V_{it}^1(q),0)=\psi(V_{it}^1(q))$, recovering the baseline
expression. The conditional expected WTP under CNM is therefore
\begin{align}
&\mathbbm{E}\left[
(U^c_{it}-\xi_{it})D_{it}
\middle|
X_{i,1:T},Z_{i,1:T},Q_i=q;\mathrm{CNM}
\right]
\nonumber\\
=~&
p_{i,t-1}^{CNM}(q)
\psi^{CNM}\left(V_{it}^1(q),\delta\right)
+
\left[
1-p_{i,t-1}^{CNM}(q)
\right]
\Phi(X_{it}\gamma+\sigma_1q)
\psi\left(V_{it}^0(q)\right).
\label{eq cnm wtp recursion}
\end{align}
Proof for \Cref{eq abm wtp recursion} and \Cref{eq cnm wtp recursion} are similar to \Cref{eq recursion for wtp} and therefore, is omitted. For all policies, we integrate the conditional expected WTP over $Q_i$
using Gauss--Hermite quadrature, divide by $\beta^B$ to express utility in
dollar-equivalent units, and sum across eligible household-months to obtain
total WTP.

Since we evaluate policies over a range of intensities, we report marginal
changes in WTP between adjacent policy intensities as ``$\Delta$WTP including $\beta^0$'' in \Cref{table mvpf counterfactual}. For ABM, marginal $\Delta$WTP generally
increases with the attention floor. For CNM, marginal $\Delta$WTP initially
changes little but subsequently declines sharply as the choice nudge becomes
more intensive. These patterns are broadly consistent with the takeup
results in \Cref{sec counterfactual takeup dynamic mechanism}: the marginal
participation gains from ABM increase with the attention floor, whereas the
additional takeup generated by CNM eventually diminishes as retention
approaches its upper bound.

The WTP measure defined above includes the choice-stage
intercept $\beta^0$ in welfare monetization. Following \cite{wagner2023can}, we consider an alternative
net-of-intercept WTP measure,
$\sum_{(i,t)\in\mathcal{EH}}
\mathbbm{E}\left[
\frac{U^c_{it}-\beta^0-\xi_{it}}{\beta^B}D_{it}
\middle|
X_{i,1:T},Z_{i,1:T}
\right]$, which excludes the contribution of $\beta^0$ from
welfare monetization while leaving counterfactual participation behavior
unchanged. We report the corresponding estimates in the
``$\Delta$WTP Net of $\beta^0$'' column of
\Cref{table mvpf counterfactual}. The qualitative patterns across ABM and
CNM are similar under the two WTP measures. The key difference is that,
under the net-of-intercept measure, marginal $\Delta$WTP from CNM becomes
negative for increases in the choice nudge beyond 5 percentage points. The sign switch indicates that, as CNM becomes more intensive, the marginal households induced to continue participating have progressively lower WTP. At sufficiently high intensities, the welfare loss from these marginal participants outweighs the welfare gain from additional retention under the net-of-intercept measure. As shown in the next section, this pattern translates into a sharply declining MVPF for CNM.

\begin{table}[!htbp]
\centering
\caption{Welfare Effects and Marginal Value of Public Funds}
\label{table mvpf counterfactual}

\resizebox{0.9\textwidth}{!}{
\begin{tabular}{lrrrrr}
\toprule
& \shortstack{$\Delta$WTP\\including $\beta^0$\\(\$ million)}
& \shortstack{$\Delta$WTP\\Net of $\beta^0$\\(\$ million)}
& \shortstack{$\Delta$Government\\Cost\\(\$ million)}
& \shortstack{MVPF\\including $\beta^0$}
& \shortstack{MVPF\\Net of $\beta^0$}
\\
\midrule

\multicolumn{6}{l}{\textit{Panel A: Attention-boosting messages (ABM)}} \\

Baseline $\rightarrow$ 25\% floor
& 0.541 & 0.125 & 0.0620 & 8.73 & 2.02 \\

25\% $\rightarrow$ 50\% floor
& 1.004 & 0.250 & 0.1074 & 9.35 & 2.33 \\

50\% $\rightarrow$ 75\% floor
& 1.175 & 0.289 & 0.1262 & 9.31 & 2.29 \\

75\% $\rightarrow$ 100\% floor
& 1.376 & 0.331 & 0.1490 & 9.24 & 2.22 \\

\addlinespace
\multicolumn{6}{l}{\textit{Panel B: Choice-nudging messages (CNM)}} \\

0 $\rightarrow$ 1 pp
& 0.815 & 0.098 & 0.1049 & 7.77 & 0.94 \\

1 $\rightarrow$ 2 pp
& 0.838 & 0.081 & 0.1080 & 7.76 & 0.75 \\

2 $\rightarrow$ 3 pp
& 0.775 & 0.055 & 0.1027 & 7.55 & 0.53 \\

3 $\rightarrow$ 4 pp
& 0.657 & 0.028 & 0.0896 & 7.33 & 0.31 \\

4 $\rightarrow$ 5 pp
& 0.515 & 0.006 & 0.0725 & 7.10 & 0.09 \\

5 $\rightarrow$ 6 pp
& 0.370 & -0.008 & 0.0538 & 6.87 & $-\infty$ \\

6 $\rightarrow$ 7 pp
& 0.247 & -0.013 & 0.0371 & 6.65 & $-\infty$ \\

7 $\rightarrow$ 8 pp
& 0.159 & -0.013 & 0.0245 & 6.47 & $-\infty$ \\

8 $\rightarrow$ 9 pp
& 0.102 & -0.011 & 0.0161 & 6.31 & $-\infty$ \\

9 $\rightarrow$ 10 pp
& 0.065 & -0.009 & 0.0105 & 6.17 & $-\infty$ \\

\bottomrule
\end{tabular}
}

\vspace{0.5em}
\begin{minipage}{0.95\textwidth}
\footnotesize
\textit{Notes:}
WTP is calculated under the random-utility interpretation of the choice-stage
shock. The net-of-intercept specification excludes the choice-stage
alternative-specific constant from welfare monetization while leaving
counterfactual participation behavior unchanged. Each row reports the
incremental welfare and fiscal effects of increasing policy intensity from
the preceding level. Government cost equals direct messaging costs plus the
fiscal cost of additional WIC participation. Direct messaging costs are
\$0.30 per eligible household-year (\$0.025 per household-month), and each
additional participant-month costs \$37.42 in WIC benefits. MVPF is defined
as the marginal change in household WTP divided by the corresponding marginal
change in government cost when marginal WTP is nonnegative. When a marginal
increase in policy intensity raises government cost but reduces household WTP,
we assign MVPF equal to $-\infty$, reflecting that the marginal policy
expansion is welfare-reducing.
\end{minipage}

\end{table}

\subsubsection{MVPF comparison} \label{sec mvpf comparison}
The WIC2Five report states that the total implementation cost of the messaging intervention is \$4,000, corresponding to approximately 30 cents per household per year, or 2.5 cents per eligible household-month. Because the messaging intervention has a fixed-cost structure that does not vary with the number of message recipients according to the report, we assume that the initial program spending of either ABM or CNM is 0.025×109,591=\$2,739.775. In addition, each additional month of WIC participation generates \$37.42 in food-package costs according to the WIC Food Package Costs and Rebates Summary: Fiscal Year 2005. Thus, for policy $k$, the fiscal externality (FE) generated by additional participation relative to baseline is
$$
FE^{k}
=37.42 \times
\sum_{(i,t) \in \mathcal{EH}} (p_{it}^k - p_{it}^0).
$$
Total government cost is the sum of the fixed implementation cost and this FE. Because \Cref{table mvpf counterfactual} reports \textit{marginal} changes in government cost between adjacent policy intensities, the fixed implementation cost enters only when moving from no intervention to the first positive policy intensity, i.e., the first rows in Panel A and Panel B. For subsequent increases in either $\nu$ or $\delta$, the fixed cost cancels, so the reported marginal government cost reflects only the additional FE generated by higher WIC participation. This distinction between rows should influence our ultimate MVPF calculations minimally because the initial program spending is a magnitude smaller than change in FE across all policy intensities.

\begin{figure}[htbp]
    \centering

    \includegraphics[width=.7\textwidth]{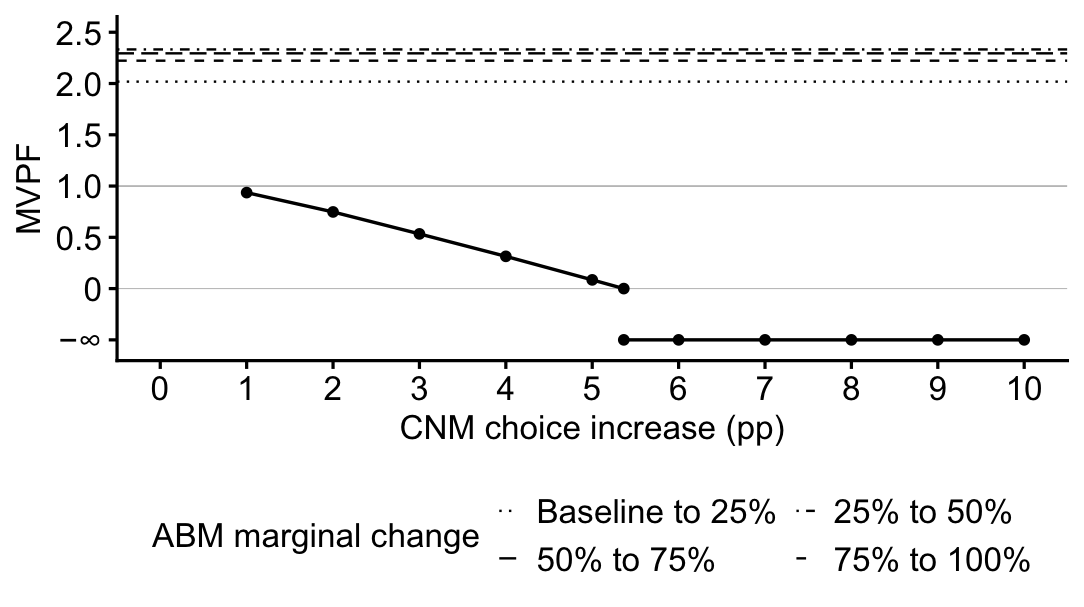}

    \caption{Marginal Value of Public Funds under Alternative Messaging Policies}
    \label{fig mvpf counterfactual}

    \caption*{\footnotesize
    \textit{Notes:}
    The solid line with markers reports the marginal value of public funds (MVPF)
    for choice-nudging messages (CNM). Each point gives the change in household
    willingness to pay from a one-percentage-point increase in the choice
    probability divided by the corresponding change in government cost.
    The horizontal lines report the MVPF for successive increases in the
    attention floor under attention-boosting messages (ABM): baseline to
    25 percent, 25 to 50 percent, 50 to 75 percent, and 75 to 100 percent.
    The horizontal gray line denotes an MVPF of one.
    Welfare is monetized net of the choice-stage alternative-specific intercept,
    while counterfactual participation behavior is unchanged.
    }
\end{figure}

With the two WTP measures, we report two MVPFs in \Cref{table mvpf counterfactual},
\begin{align*}
     \text{MVPF including } \beta^0 =&~ \frac{\Delta \text{WTP including } \beta^0}{\Delta \text{Government Cost}} \quad \text{and}\\
     \text{MVPF Net of } \beta^0 =&~ \frac{\Delta \text{WTP Net of } \beta^0}{\Delta \text{Government Cost}}.
\end{align*}
Both measures show substantially higher MVPF for ABM than for CNM across the policy intensities we consider. Including $\beta^0$, the MVPF of ABM remains around 9 across all four marginal increases in the attention floor, whereas the MVPF of CNM declines from 7.77 for the first one-pp increase to 6.17 for the increase from 9 to 10 pp. The contrast is substantially stronger under the net-of-$\beta^0$ welfare measure. ABM maintains an MVPF above 2 at every intensity, whereas CNM's MVPF declines from 0.94 to 0.09 as $\delta$ increases from 0.01 to 0.05. For increases beyond 5 pp, $\Delta$WTP becomes negative while $\Delta$government cost remains positive. We therefore assign these marginal CNM expansions an MVPF of $-\infty$ since they are Pareto dominated by lower intensity CNM. We depict this MVPF comparison in \Cref{fig mvpf counterfactual}. 

The difference in MVPF between ABM and CNM reflects how the two policies
select households into participation. ABM affects whether a household reaches
the choice stage but leaves the choice rule unchanged. Consequently, among
households whose attention is increased by ABM, only those who prefer WIC
under the baseline choice rule ultimately participate. The additional
participants induced by ABM are therefore positively selected on
choice-stage surplus, i.e., $WTP \times \beta^B$. CNM instead acts directly on the choice margin by
relaxing the effective threshold for participation. As $\delta$ increases,
CNM induces participation among households with progressively lower
choice-stage surplus. This explains why CNM can generate larger increases in
takeup while producing smaller marginal WTP per additional participant and,
therefore, a lower MVPF than ABM.

This comparison depends on how WTP is calculated for the CNM intervention. In \Cref{eq cnm wtp recursion}, CNM changes the
probability of choosing WIC but does not directly enter the household's
choice-stage utility. In other words, the choice nudge itself
is not internalized as part of the choice utility or equivalently, WTP. This
welfare evaluation of CNM is conservative. If the CNM instead generates a
welfare-relevant utility gain---for example, by increasing the perceived value of the WIC food package---and this gain is
included in household WTP, the estimated WTP and MVPF of CNM would be higher; see \cite{handel2013adverse} for an example.
\section{Conclusion} \label{sec conclusion}

This paper develops a new econometric framework that separately identifies two latent behavioral
mechanisms in incomplete social assistance takeup: inattention and active choice. Although these mechanisms are
observationally indistinguishable, we show
that they can be semiparametrically identified using standard panel data by
exploiting institutional features common to many public programs.

Applied to WIC, our framework yields three main findings. First, inattention
is the primary barrier to WIC participation among previously nonenrolled
households. Second, CNM increases takeup substantially more than ABM because
of the dynamic mechanism linking current participation to future
attention: by preventing exit today, CNM also prevents households from
becoming inattentive in subsequent periods. Third, despite generating smaller
takeup effects, ABM has substantially higher MVPF than CNM across the policy
intensities we consider. The reversal in policy ranking arises from the selection on marginal participants' WTP.

More broadly, this paper demonstrates the value of distinguishing inattention
from active choice when studying incomplete program takeup. By separating these
two mechanisms, the proposed framework provides policymakers with a richer
basis for diagnosing the sources of incomplete takeup and evaluating policies
that target them. 
Beyond WIC, the framework can be applied to other public programs in which eligible agents underutilize available resources.
\bibliographystyle{aea}
\bibliography{JMP_WIC}
\clearpage

\pagenumbering{arabic}
\clearpage
\appendix
\counterwithin{figure}{section}
\renewcommand{\thefigure}{\thesection.\arabic{figure}}

\begin{center}
    \Large Online Appendix
\end{center}

\section{Comparison with other technical literature} \label{appendix related literature and contribution}

\subsection{Inattention models} 
Our model introduces two innovations relative to existing two-stage inattention models in the marketing literature. First, our model does not require an exclusive shifter for the attention stage. This contrasts with the typical approach in the inattention model literature, which relies on two sets of exclusive shifters for both attention and choice stages \citep{heiss2021inattention,agarwal2025demand}. Second, we allow unobserved household-level heterogeneity to influence both attention and choice probabilities simultaneously. In contrast, most inattention models rule out unobserved heterogeneity \citep{honka2017advertising,crawford2021survey}. 
We compare in detail our contribution and two leadings papers in this field.

\subsubsection{\citet{abaluck2021consumers}}
\citet{abaluck2021consumers} exploits the Slutsky asymmetry when \textit{prices} of goods vary. Importantly, the price variation of the default good (the good that inattentive consumers choose by default) has a different impact from the price variation of non-default goods. In the welfare take-up context, the default option is nonparticipation, which \textit{does not have a price}. Hence, our context calls for a different identification strategy.


The strength of our identification argument over \cite{abaluck2021consumers} is that we allow for unobserved heterogeneity. To accommodate household-level unobserved heterogeneity, we look for two \textit{sequences} of take-up decisions whose ratio identifies the ratio between attention parameters. 

There are two prices we pay 
for allowing unobserved heterogeneity. First,  
we specify linear utility functions, whereas \cite{abaluck2021consumers} does not impose a functional form constraint on the utility functions. Second, we require individual-level participation data, whereas \cite{abaluck2021consumers} requires market share data, not individual consumer data.

\subsubsection{\citet{agarwal2025demand}}   
\citet{agarwal2025demand} assumes two sets of exclusive shifters, one for the attention stage and the other for the choice stage. In contrast, our key identifying assumption is the existence of an observed fully attentive subpopulation. 
In our identification theorems, the attention shifter is not required to be exclusive but has to generate an observed fully attentive subgroup. Moreover, the exclusive attention shifter in \citet{agarwal2025demand} is assumed to have large support. In our model, the non-exclusive attention shifter is the last period \textit{binary} participation status. 

    
Admittedly, though our theorems have a lower data requirement, our model setup, which requires the presence of a full attention group and a recertification process, is much more restrictive. Hence, we anticipate that readers find our work a better fit for specific empirical setups, such as welfare take-up or some other program participation contexts, whereas \cite{agarwal2025demand}'s theoretical results apply to a wider range of topics but may impose unrealistic data assumptions in the empirical contexts that this paper specializes in.

\subsection{Double hurdle model}
Some readers may find our model similar to the classical double hurdle model \citep{cragg1971some}. We clarify that our setup is substantially different from the classical double hurdle model, as our outcome variable is binary, whereas \cite{cragg1971some} explicitly states that part of the observed outcome is a ``continuous and positive random variable''.

\subsection{Dynamic panel discrete choice with random effect} 
The full model introduces a dynamic panel (AR1) covariate into a generalized linear mixed-effect model (GLMM). Though such a design has been used by other works from econometrics, biometrics, and psychometrics \citep{wooldridge2005simple,escaramis2008detection,cho2018autoregressive}, the full model allows the AR1 term (i.e., last period take-up decision) and the random-effect component to \textit{simultaneously} enter the attention and choice stages. This is a special feature of the full model that, to the best of our knowledge, has only been used by one paper: \cite{heiss2021inattention}. However, \cite{heiss2021inattention} argues that the identifiability of their model comes from the presence of two sets of large-support exclusive shifters; we show that certain aspects of the model identifiability can be established semiparametrically when a fully attentive subgroup is created by the \textit{non-exclusive} attention shifter. Furthermore, the attention shifter in our paper is the binary-support lagged outcome, imposing virtually no data requirements if the data is longitudinal.

\section{Mathematical appendix for preliminary model} 

\subsection{Proof of \Cref{eq decomposing observed prob}} \label{proof of eq decomposing observed prob}
\begin{proof}
\begin{align*}
    &~ P(D_{it} = 1 \mid D_{it-1},X_{it},Z_{it}) \\
    =&~ P(A_{it}=1, C_{it}=1 \mid D_{it-1},X_{it},Z_{it}) \\
    =&~ P(A_{it}=1\mid D_{it-1},X_{it}) P(C_{it}=1\mid D_{it-1},X_{it},Z_{it}) \nonumber \quad \text{\Cref{assumption on shocks preliminary model} (c)}\\ 
    =&~ G(X_{it}\gamma)^{1-D_{it-1}} H(X_{it}\beta + S_{it}\alpha) \quad \text{\Cref{assumption on shocks preliminary model} (a) and (b)}
\end{align*}
\end{proof}

\subsection{An extension to binary exclusive attention shifter} \label{appendix extension to binary exclusive attention shifter}
In many empirical cases, economists observe binary exclusive attention shifters that make subgroups fully attentive. Examples include:  households that recently moved are fully attentive to choosing an electricity billing company \citep{hortaccsu2017power}; Medicare Part D consumers whose plan choice is dropped out of the Medicare marketplace in year $t$ are fully attentive in year $t$ \citep{heiss2021inattention}; subscription service users are fully attentive if their credit card expires when they need to renew their subscription \citep{einav2025selling}.

We adapt \Cref{prop alpha overidentified} to these cases, demonstrating the broad applicability of our identification strategy. We use a slightly different set of notations. $D_{it}$ denotes the outcome (e.g., whether a user renews a subscription service or not), $X_{it}$ refers to consumer or alternative characteristics, $Z_{it}$ is the exclusive attention shifter that makes the subgroup experiencing it fully attentive.
\begin{align*}
    P(D_{it} = 1 \mid X_{it}, Z_{it}) =&~ P(A_{it} = 1, C_{it} = 1 \mid X_{it}, Z_{it}) \\
    =&~ P(A_{it} = 1 \mid X_{it}, Z_{it})P(C_{it} = 1 \mid X_{it}, Z_{it}) \\
    =&~ P(A_{it} = 1 \mid X_{it})^{1-Z_{it}} P(C_{it} = 1 \mid X_{it})
\end{align*}
When $Z_{it} = 1$, consumers are fully attentive. Hence, $P(D_{it} = 1 \mid X_{it}, Z_{it} = 1) = P(C_{it} = 1 \mid X_{it})$. We use these observations to study choice parameters. When $Z_{it} = 0$, $P(D_{it} = 1 \mid X_{it}, Z_{it} = 0) = P(A_{it} = 1 \mid X_{it}) P(C_{it} = 1 \mid X_{it})$. We already identified $P(C_{it} = 1 \mid X_{it})$, so $P(A_{it} = 1 \mid X_{it})$ is identified as the ratio $\frac{P(D_{it} = 1 \mid X_{it}, Z_{it} = 0)}{P(C_{it} = 1 \mid X_{it})}$. This identification argument extends trivially to the multinomial outcome setup like the two-stage structural model in \cite{hortaccsu2017power}. It also extends to models that incorporate switching costs as long as there are two fully attentive groups: one does not experiences switching costs and one experiences switching costs. For example, in \cite{heiss2021inattention}, the former is the new Medicare participants, the latter is existing participants whose plan choice in year $t-1$ is dropped from the choice menu in year $t$.

\section{Mathematical appendix for full model} 
\subsection{Full Model vs Preliminary Model} \label{appendix strength of the full model}
The preliminary model treats the lagged decision $D_{it-1}$ as exogenous, whereas the full model allows $D_{it-1}$ to be endogenous. 
The preliminary model requires both unobserved terms, $\mathcal{E}_{it}$ and $\Xi_{it}$, to be uncorrelated with $D_{it-1}$. In contrast, the full model allows persistent household-level heterogeneity $Q_i$ to influence both the attention and choice utilities in every period, thereby affecting takeup decisions across \textit{all} periods. Consequently, $D_{it-1}$ correlate with $\mathcal{E}_{it}$ and $\Xi_{it}$, for all $t > 1$. 

A second difference concerns the relation between attention and choice.  
Under \Cref{assumption on shocks preliminary model}, $A_{it}$ and $C_{it}$ are conditionally independent given $X_{it}$ and $D_{it-1}$, since their shocks are independent.  
This independence can be unrealistic: thriftier households (higher $Q_i$) may both pay more attention to the program and be more inclined to choose participation.  
The full model captures this by letting $Q_i$ enter both $U_{it}^a$ and $U_{it}^c$, generating correlation between $A_{it}$ and $C_{it}$ even after conditioning on observables.  

Together, these modifications relax the strong exogeneity and independence restrictions of the preliminary model, yielding a more realistic framework in which persistent unobserved heterogeneity drives both dynamics across time and dependence across stages.

\subsection{Proof of \Cref{lemma multiple period likelihood}} \label{proof for multiple period likelihood}
\begin{proof}
    \begin{align}
    &~P(D_{i,1:T} \mid  W_{i,1:T}, Q_i) \nonumber\\
    =&~ P(D_{i,1:T} \mid  X_{i,1:T}, Z_{i,1:T}, Q_i) \nonumber\\
    =&~ P(D_{iT},D_{i,1:(T-1)} \mid  X_{i,1:T}, Z_{i,1:T}, Q_i) \nonumber\\
    =&~ P(D_{iT}\mid D_{i,1:(T-1)},X_{i,1:T}, Z_{i,1:T}, Q_i) P(D_{i,1:(T-1)} \mid  X_{i,1:T}, Z_{i,1:T}, Q_i) \label{eq def of conditional prob} \\ 
    =&~ P(D_{iT}\mid D_{iT-1},X_{iT}, Z_{iT}, Q_i) P(D_{i,1:(T-1)} \mid  X_{i,1:(T-1)}, Z_{i,1:(T-1)}, Q_i) \label{eq markov property}\\ 
    =&~ \prod_{t=1}^{T} P(D_{it}\mid D_{it-1},X_{it}, Z_{it}, Q_i) \label{eq recursive}\\
    =&~ \prod_{t=1}^{T} P(D_{it}\mid R_{it}, Q_i) \nonumber
\end{align}

\Cref{eq def of conditional prob} is by the definition of conditional probability. 

The first half of \Cref{eq markov property} requires \Cref{assumption on shocks full model}   (c1). The second half of \Cref{eq markov property} requires \Cref{assumption on shocks full model} (c2).

\Cref{eq recursive} is by recursively applying the previous reasoning to the last term in \Cref{eq markov property}.
\end{proof}

\subsection{Proof of \Cref{lemma decomposing fod}} \label{proof decomposing fod}
\begin{proof}
\begin{align*}
    &~ \frac{\partial}{\partial X_{i\tau}^\omega} P(D_{i,1:T} \mid W_{i,1:T}) \\
    =&~ \frac{\partial}{\partial X_{i\tau}^\omega} \int P(D_{i,1:T} \mid W_{i,1:T},Q_i =q) dF(q) \\
    =&~ \frac{\partial}{\partial X_{i\tau}^\omega} \int \prod_{t=1}^{T} P(D_{it}\mid R_{it}, Q_i = q) dF(q) ~\text{by \Cref{lemma multiple period likelihood}}\\
    =&~ \int \frac{\prod_{t=1}^{T} P(D_{it}\mid R_{it},Q_i = q)}{P(P(D_{i\tau}\mid R_{i\tau}, Q_i = q)}  \frac{\partial}{\partial X_{i\tau}^\omega} P(D_{i\tau}\mid R_{i\tau}, Q_i = q) dF(q) \\
    =&~ \int \prod_{t=1}^{T} P(D_{it}\mid R_{it},Q_i = q)\frac{ \frac{\partial}{\partial X_{i\tau}^\omega} P(D_{i\tau}\mid R_{i\tau}, Q_i = q)}{P(D_{i\tau}\mid R_{i\tau}, Q_i = q)} dF(q). 
\end{align*}
\end{proof}

\subsection{Theorems}
\subsubsection{\Cref{theorem identifying beta}} \label{appendix theorem identifying beta}
We first aggregate these identifying equations for the same $(d,w)$ across all continuation periods by defining
\begin{align*}
SFOD^\omega(d,w)
:=
\sum_{\tau=2}^T
\mathbbm{1}\{d_{\tau-1}=d_\tau=1\}
FOD_\tau^\omega(d,w).
\end{align*}
Summing \Cref{eq beta overidentified} over all continuation
periods gives
\begin{align}
SFOD^\omega(d,w)
=
\beta^\omega SFOD^B(d,w)
\label{eq beta overidentified SFOD}
\end{align}
Aggregating \Cref{eq beta overidentified SFOD} across all relevant combinations of $(d,w)$ yields the following theorem.



\begin{theorem}\label{theorem identifying beta}
Let $\widetilde{\mathcal D}
:=
\bigcup_{\tau=2}^T
\widetilde{\mathcal D}_\tau$
denote the set of takeup sequences containing at least one continuation
period. Under \Cref{assumption on shocks full model}, normalize $\beta^B=1$.
For any continuous covariate $X^\omega$, suppose
$P(D_{i,1:T}\in\widetilde{\mathcal D})>0$ and
$E\left[
SFOD^B(D_{i,1:T},W_{i,1:T})
\mid
D_{i,1:T}\in\widetilde{\mathcal D}
\right]
\neq0$.
Then $\beta^\omega$ is constructively identified as
\begin{align*}
\beta^\omega
=
\frac{
E\left[
SFOD^\omega(D_{i,1:T},W_{i,1:T})
\mid
D_{i,1:T}\in\widetilde{\mathcal D}
\right]
}{
E\left[
SFOD^B(D_{i,1:T},W_{i,1:T})
\mid
D_{i,1:T}\in\widetilde{\mathcal D}
\right]
}.
\end{align*}
\end{theorem}
\begin{proof}
  By \Cref{eq beta overidentified SFOD}, for every
$d\in\widetilde{\mathcal D}$ and every admissible covariate sequence $w$,
\begin{align*}
SFOD^\omega(d,w)
=
\beta^\omega SFOD^B(d,w).
\end{align*}

Integrating this pointwise identifying equation over all admissible
covariate sequences and summing over all
$d\in\widetilde{\mathcal D}$ gives
\begin{align*}
&\sum_{d\in\widetilde{\mathcal D}}
\int
SFOD^\omega(d,w) P(D_{i,1:T} = d \mid W_{it} = w)
\,dF_{W_{i,1:T}}(w)
\\
=&~
\beta^\omega
\sum_{d\in\widetilde{\mathcal D}}
\int
SFOD^B(d,w) P(D_{i,1:T} = d \mid W_{it} = w)
\,dF_{W_{i,1:T}}(w).
\end{align*}

Dividing both sides by
$P(D_{i,1:T}\in\widetilde{\mathcal D})>0$ yields
\begin{align*}
&E\left[
SFOD^\omega(D_{i,1:T},W_{i,1:T})
\mid
D_{i,1:T}\in\widetilde{\mathcal D}
\right]
=
\beta^\omega
E\left[
SFOD^B(D_{i,1:T},W_{i,1:T})
\mid
D_{i,1:T}\in\widetilde{\mathcal D}
\right].
\end{align*}

Since the expectation multiplying $\beta^\omega$ is nonzero by
assumption, dividing both sides by it gives
\begin{align*}
\beta^\omega
=
\frac{
E\left[
SFOD^\omega(D_{i,1:T},W_{i,1:T}) 
\mid
D_{i,1:T}\in\widetilde{\mathcal D}
\right]
}{
E\left[
SFOD^B(D_{i,1:T},W_{i,1:T})
\mid
D_{i,1:T}\in\widetilde{\mathcal D}
\right]
}.
\end{align*}  
\end{proof}

\subsubsection{\Cref{theorem identifying alpha}} 
\label{appendix theorem identifying alpha}

Let $K_{i,1:T}=(D_{i,1:T},W_{i,1:T})$ denote an observed sequence, and let
$k=(d,w)$ denote a realization. Let $p(k)$ denote the joint density/mass
of $K_{i,1:T}$. Denote by $\mathcal K$ a set of admissible matched sequence pairs
$(\tilde k,\ddot k)$. For each matched pair, let
$\mathcal J(\tilde k,\ddot k)\subseteq\{2,\ldots,T\}$ denote the set of
periods at which the pair satisfies the relevant matching conditions.
For $\alpha$ identification, $|\mathcal J(\tilde k,\ddot k)|=1$ by
construction. For $\gamma$ identification, we allow
$|\mathcal J(\tilde k,\ddot k)|>1$, as discussed in
\Cref{appendix theorem identifying gamma}.

For any integrable period-specific function
$m(\tilde k,\ddot k,\tau)$, define its pair-level aggregate as
\begin{align*}
\check m(\tilde k,\ddot k)
:=
\sum_{\tau\in\mathcal J(\tilde k,\ddot k)}
m(\tilde k,\ddot k,\tau).
\end{align*}
We define the pair-weighted expectation as
\begin{align}
E_{\mathcal K}
\left[
\check m(\tilde K_{i,1:T},\ddot K_{i,1:T})
\right]
:=
\frac{
\displaystyle
\iint
\mathbbm{1}
\left\{
(\tilde k,\ddot k)\in\mathcal K
\right\}
\check m(\tilde k,\ddot k)
p(\tilde k)p(\ddot k)
\,d\ddot k\,d\tilde k
}{
\displaystyle
\iint
\mathbbm{1}
\left\{
(\tilde k,\ddot k)\in\mathcal K
\right\}
|\mathcal J(\tilde k,\ddot k)|
p(\tilde k)p(\ddot k)
\,d\ddot k\,d\tilde k
}.
\label{eq pair weighted expectation}
\end{align}

\begin{theorem}
\label{theorem identifying alpha}

Let
$\pi(d,w)
:=
P(D_{i,1:T}=d\mid W_{i,1:T}=w)$.
For any sequence pair
$\tilde k=(\tilde d,\tilde w)$ and
$\ddot k=(\ddot d,\ddot w)$, define
$\mathcal J_\alpha(\tilde k,\ddot k)$ as the set of periods
$\tau\in\{2,\ldots,T\}$ satisfying
\begin{align*}
&\tilde d=\ddot d,\qquad
\tilde d_{\tau-1}=\tilde d_\tau=1,\\
&\tilde z_\tau=0,\qquad
\ddot z_\tau=1,\\
&\tilde w_t=\ddot w_t
\qquad \forall~t\neq\tau,
\end{align*}
such that all remaining components of $\tilde w_\tau$ and
$\ddot w_\tau$ are identical except for the benefit amount,
and
\begin{align*}
\pi(\tilde d,\tilde w)
=
\pi(\ddot d,\ddot w).
\end{align*}
We further require that $\ddot b_\tau$ is the unique admissible benefit
amount satisfying this equality.

Define the set of admissible matched sequence pairs as
\begin{align*}
\mathcal K_\alpha
:=
\left\{
(\tilde k,\ddot k):
|\mathcal J_\alpha(\tilde k,\ddot k)|=1
\right\},
\end{align*}
and, for
$\tau\in\mathcal J_\alpha(\tilde k,\ddot k)$, define
\begin{align*}
m_\alpha(\tilde k,\ddot k,\tau)
:=
\tilde b_\tau-\ddot b_\tau.
\end{align*}
Accordingly,
\begin{align*}
\check m_\alpha(\tilde k,\ddot k)
:=
\sum_{\tau\in\mathcal J_\alpha(\tilde k,\ddot k)}
m_\alpha(\tilde k,\ddot k,\tau).
\end{align*}

Under \Cref{assumption on shocks full model}, normalize $\beta^B=1$.
Suppose $\mathcal K_\alpha$ has positive matched-pair weight so that
$E_{\mathcal K_\alpha}[\cdot]$ is well defined. Then $\alpha$ is
constructively identified as
\begin{align*}
\alpha
=
E_{\mathcal K_\alpha}
\left[
\check m_\alpha(\tilde K_{i,1:T},\ddot K_{i,1:T})
\right].
\end{align*}

\end{theorem}

\subsubsection{\Cref{theorem identifying gamma}}
\label{appendix theorem identifying gamma}

\begin{theorem}
\label{theorem identifying gamma}

For any sequence pair
$\tilde k=(\tilde d,\tilde w)$ and
$\ddot k=(\ddot d,\ddot w)$, define
$\mathcal J_\gamma(\tilde k,\ddot k)$ as the set of periods
$\tau\in\{2,\ldots,T\}$ satisfying
\begin{align*}
\tilde d_{\tau-1}
&=
\tilde d_\tau
=
1,\\
\ddot d_{\tau-1}
&=
0,
\ddot d_\tau
=
1,
\end{align*}
for which the base and comparison sequences satisfy matching conditions
H1 and H2 at period $\tau$.

Define the set of admissible matched sequence pairs as
\begin{align*}
\mathcal K_\gamma
:=
\left\{
(\tilde k,\ddot k):
\mathcal J_\gamma(\tilde k,\ddot k)\neq\varnothing
\right\}.
\end{align*}

For any continuous covariate $X^\omega$ and
$\tau\in\mathcal J_\gamma(\tilde k,\ddot k)$, define the period-specific
FOD difference
\begin{align*}
m_\gamma^\omega(\tilde k,\ddot k,\tau)
:=
FOD_\tau^\omega(\ddot d,\ddot w)
-
FOD_\tau^\omega(\tilde d,\tilde w),
\end{align*}
and its pair-level aggregate
\begin{align*}
\check m_\gamma^\omega(\tilde k,\ddot k)
:=
\sum_{\tau\in\mathcal J_\gamma(\tilde k,\ddot k)}
m_\gamma^\omega(\tilde k,\ddot k,\tau).
\end{align*}

Under \Cref{assumption on shocks full model}, normalize $\gamma^B=1$.
Suppose $\mathcal K_\gamma$ has positive matched-pair weight so that
$E_{\mathcal K_\gamma}[\cdot]$ is well defined, and
\begin{align*}
E_{\mathcal K_\gamma}
\left[
\check m_\gamma^B(\tilde K_{i,1:T},\ddot K_{i,1:T})
\right]
\neq0.
\end{align*}
Then, for every continuous covariate $X^\omega$,
$\gamma^\omega$ is constructively identified as
\begin{align*}
\gamma^\omega
=
\frac{
E_{\mathcal K_\gamma}
\left[
\check m_\gamma^\omega
(\tilde K_{i,1:T},\ddot K_{i,1:T})
\right]
}{
E_{\mathcal K_\gamma}
\left[
\check m_\gamma^B
(\tilde K_{i,1:T},\ddot K_{i,1:T})
\right]
}.
\end{align*}

\end{theorem}

\subsection{Fully nonparametric model} \label{proof for fully nonparametric model}
We show that our FOD-based identification strategy extends to a fully nonparametric setup. For exposition, we assume that three are three observed covariates: $B_{it},X^{-B}_{it},Z_{it}$ with unobserved time-invariant household random effect $Q_i$. The choice utility function for those who already enrolled boils down to
\begin{equation} \label{eq new choice utility}
    U^c(1,B_{it},X^{-B}_{it},Z_{it},Q_i) = B_{it} + X^{-B}_{it}\beta(B_{it}) + Z_{it}\alpha + \sigma_2 Q_i
\end{equation}
where $\beta$ can vary with benefit amount $B_{it}$. Since $B_{it}$ is a continuous variable, the utility function's deterministic part are characterized by an infinite number of parameters.

Substituting \Cref{eq new choice utility} into \Cref{lemma decomposing fod}, we obtain the FODs
\begin{align*}
&FOD_{\tau}^\omega(\tilde{d},B_{i,1:T},X^{-B}_{i,1:T},Z_{i,1:T}) \\
=~& \int CPL(\tilde{d},B_{i,1:T},X^{-B}_{i,1:T},Z_{i,1:T},q) \frac{h(U^c(1,B_{i\tau},X^{-B}_{i\tau},Z_{i\tau},Q_i))}{H(U^c(1,B_{i\tau},X^{-B}_{i\tau},Z_{i\tau},Q_i))} dF(q) \beta(B_{i\tau})\\
&FOD_{\tau}^B(\tilde{d},B_{i,1:T},X^{-B}_{i,1:T},Z_{i,1:T}) \\
=~& \int CPL(\tilde{d},B_{i,1:T},X^{-B}_{i,1:T},Z_{i,1:T},q) \frac{h(U^c(1,B_{i\tau},X^{-B}_{i\tau},Z_{i\tau},Q_i))}{H(U^c(1,B_{i\tau},X^{-B}_{i\tau},Z_{i\tau},Q_i))}  dF(q) (1+X^{-B}_{i\tau}\beta'(B_{i\tau}))
\end{align*}

Dividing these two FODs, we obtain
\begin{align*}
\frac{FOD_{\tau}^B(\tilde{d},B_{i,1:T},X^{-B}_{i,1:T},Z_{i,1:T})}{FOD_{\tau}^\omega(\tilde{d},B_{i,1:T},X^{-B}_{i,1:T},Z_{i,1:T})} = \frac{1+X^{-B}_{i\tau}\beta'(B_{i\tau})}{\beta(B_{i\tau})} = \frac{1}{\beta(B_{i\tau})} + \frac{X^{-B}_{i\tau}\beta'(B_{i\tau})}{\beta(B_{i\tau})}
\end{align*}
Since the LHS is observed data, we can regress it on $X^{-B}_{i\tau}$ to obtain the intercept, which is the inverse of $\beta(B_{i\tau})$. With sufficient variations in $B_{i\tau}$ and sufficient variations in $X^{-B}_{i\tau}$ for each $B_{i\tau}$ value, the entire $\beta$ function can be recovered for all possible benefit values. 

\subsection{Model extension} \label{sec model extension}
\textbf{Inattention among enrollees}: In both models, enrollees are assumed to be fully attentive in the following period. This assumption might be too strong for those who are at the recertification period. Lack of awareness of eligibility or program continuation guidelines can lead existing enrollees to miss benefits in the recertification period. We show that such a concern is not severe in our empirical application with extremely high continuation rate during recertification period, though the continuation probability at recertification period is lower than other periods. To further test the sensitivity of our model to this the full attention setup, we mechanical decreasing the attention probability for those who need to recertify their eligibility and present the estimation results in \Cref{sec estimation results}.

\textbf{$Q_i^a$ and $Q_i^c$}: In our full model, we assume an identical random effect for the attention and choice stages. This assumption can be relaxed. Consider bivariate iid $\{Q_i^a,Q_i^c\}$ and that they are independent from $X_{1,1:T}$ and $Z_{i,1:T}$, though $Q_i^a$ and $Q_i^c$ do not be mutually independent from each other. This relaxation of identical $Q_i$ for the two stages changes \Cref{eq decomposing CPL and SE} from a single integral to a double integral like in \cite{heiss2021inattention}, but does not undermine our identification proof.

\textbf{Different sign-up and recertification hassle}: Our model assumes that recertification and sign-up share the same hassle; therefore, variation in $Z_{it}$ identifies $\alpha$. This assumption is more likely to hold for programs with administratively identical recertification and sign-up. In some programs, recertification is administratively easier than sign-up. In this case, having a sign-up/recertification complexity measure as proposed in \cite{kleven2011transfer} would be helpful for identifying $\alpha$. For example, we can use the average number of trips households need to take for recertification and sign-up in a given region and a given year. As long as there is overlap between the distributions of the recertification complexity and that of sign-up complexity, then we can condition on the same value for the complexity measures to identify $\alpha$.

\textbf{Household type fixed effect}: Economists may find the orthogonality assumption between the household-level unobserved heterogeneity and observed characteristics too strong. One way to relax this assumption is to allow household-type fixed effects \citep{bonhomme2022discretizing}. Many panel datasets contain households' participation status for \textit{multiple} programs. Suppose the economists are interested in one particular program. In that case, they can use other programs' participation status information to cluster households into types and augment our model with households' latent types fixed effects. This augmentation decomposes the household-level unobserved heterogeneity into a household-type fixed effect that can arbitrarily correlate with observed characteristics and a household-level random effect that is orthogonal to the observed characteristics. 





\section{Mathematical appendix for welfare calculation}
\subsection{Proof for \Cref{eq recursion}} \label{proof for eq recursion}
\begin{equation*}
\begin{split}
p_{it}(q)
&=
\mathbbm{E}\left[
D_{it}
\mid
X_{i,1:T},Z_{i,1:T},Q_i=q
\right]
\\
&=
\mathbbm{E}\left[
\mathbbm{E}\left[
D_{it}
\mid
D_{i,t-1},
X_{i,1:T},Z_{i,1:T},Q_i=q
\right]
\middle|
X_{i,1:T},Z_{i,1:T},Q_i=q
\right]
\\
&=
\mathbbm{E}\Big[
D_{i,t-1}
\Phi\left(
X_{it}\beta+Z_{it}\alpha+\sigma_2 q
\right)
\\
&\qquad\qquad
+(1-D_{i,t-1})
\Phi\left(
X_{it}\gamma+\sigma_1 q
\right)
\Phi\left(
X_{it}\beta+\alpha+\sigma_2 q
\right)
\Big|
X_{i,1:T},Z_{i,1:T},Q_i=q
\Big]
\\
&=
p_{i,t-1}(q)
\Phi\left(
X_{it}\beta+Z_{it}\alpha+\sigma_2 q
\right)
\\
&\qquad\qquad + \left[1-p_{i,t-1}(q)\right]
\Phi\left(
X_{it}\gamma+\sigma_1 q
\right)
\Phi\left(
X_{it}\beta+\alpha+\sigma_2 q
\right).
\end{split}
\end{equation*}
The first term captures continuation among households participating in the previous period. Because these households are assumed to be fully attentive, their current participation probability is determined solely by the choice stage. The second term captures entry or re-entry among previous nonparticipants, who must first become attentive and then choose to participate. Iterating this recursion over each eligibility spell yields $p_{it}(q)$ for every eligible household-month. We then integrate over the distribution of $Q_i$ and average across eligible household-months to obtain $\bar p$.

\subsection{Proof for \Cref{eq recursion for wtp}} \label{appendix proof for eq recursion for wtp}
For a fixed $q$, we can decompose the conditional expectation into two cases: $D_{it-1} = 1$ and $D_{it-1} = 0$.
\begin{align*}
    &\mathbbm{E}\left[(U^c_{it} - \xi_{it}) D_{it} \middle| X_{i,1:T}, Z_{i,1:T}, Q_i = q\right] \\
    =~& \mathbbm{E}\left[ (X_{it}\beta + S_{it}\alpha + \sigma_2 q - \xi_{it}) D_{it} \middle| X_{i,1:T}, Z_{i,1:T}, Q_i = q\right]\\
    =~& \mathbbm{E}\left[ (V_{it}^1(q) - \xi_{it}) D_{it} D_{it-1} \middle| X_{i,1:T}, Z_{i,1:T}, Q_i = q\right] + \\
    & \qquad \qquad \mathbbm{E}\left[ (V_{it}^0(q) - \xi_{it}) D_{it} (1-D_{it-1}) \middle| X_{i,1:T}, Z_{i,1:T}, Q_i = q\right]
\end{align*}

The first term can be expressed recursively as
\begin{align*}
    &\mathbbm{E}\left[ (V_{it}^1(q) - \xi_{it}) D_{it} D_{it-1} \middle| X_{i,1:T}, Z_{i,1:T}, Q_i = q\right] \\
    =~& \mathbbm{E}\left[ (V_{it}^1(q) - \xi_{it}) C_{it} D_{it-1} \middle| X_{i,1:T}, Z_{i,1:T}, Q_i = q\right] \\
    =~& \mathbbm{E}\left[ (V_{it}^1(q) - \xi_{it}) \mathbbm{1}\{(V_{it}^1(q) > \xi_{it} \} D_{it-1} \middle| X_{i,1:T}, Z_{i,1:T}, Q_i = q\right]\\
    =~& \mathbbm{E}\left[ (V_{it}^1(q) - \xi_{it}) \mathbbm{1}\{(V_{it}^1(q) > \xi_{it} \} \middle| X_{i,1:T}, Z_{i,1:T}, Q_i = q\right] p_{it-1} \\
    =~& \int (V_{it}^1(q) - c) \mathbbm{1}\{(V_{it}^1(q) > c \} d\Phi(c) p_{it-1} \\
    =~& \int_{-\infty}^{V_{it}^1(q)} (V_{it}^1(q) - c)  d\Phi(c) p_{it-1} \\
    =~& ((V_{it}^1(q))\Phi(V_{it}^1(q)) + \phi(V_{it}^1(q)))p_{it-1}
\end{align*}

The second term can be expressed recursively as
\begin{align*}
    & \mathbbm{E}\left[ (V_{it}^0(q) - \xi_{it}) D_{it} (1-D_{it-1}) \middle| X_{i,1:T}, Z_{i,1:T}, Q_i = q\right] \\
    =~& \mathbbm{E}\left[ (V_{it}^0(q) - \xi_{it}) A_{it} C_{it} (1-D_{it-1}) \middle| X_{i,1:T}, Z_{i,1:T}, Q_i = q \right]\\
    =~& \mathbbm{E}\left[ (V_{it}^0(q) - \xi_{it}) C_{it} \middle| X_{i,1:T}, Z_{i,1:T}, Q_i = q  \right] \\
    & \qquad \qquad \qquad P(A_{it} = 1, D_{it-1} = 0 \mid X_{i,1:T}, Z_{i,1:T}, Q_i = q)\\
    =~& \mathbbm{E}\left[ (V_{it}^0(q) - \xi_{it}) C_{it} \middle| X_{i,1:T}, Z_{i,1:T}, Q_i = q  \right] \\
    & \qquad \qquad \qquad P(A_{it} = 1 \mid D_{it-1} = 0, X_{i,1:T}, Z_{i,1:T}, Q_i = q) (1 - p_{it-1})\\
    =~& (V_{it}^0(q)\Phi(V_{it}^0(q)) + \phi(V_{it}^0(q)))\Phi(X_{it}\gamma + \sigma_1 q)(1 - p_{it-1})
\end{align*}

\section{Data details} \label{appendix data details}
\subsection{Benefit imputation variable details}
\label{appendix Benefit imputation variable details}

\begin{figure}[htbp]
    \centering
    \includegraphics[width=0.82\linewidth]{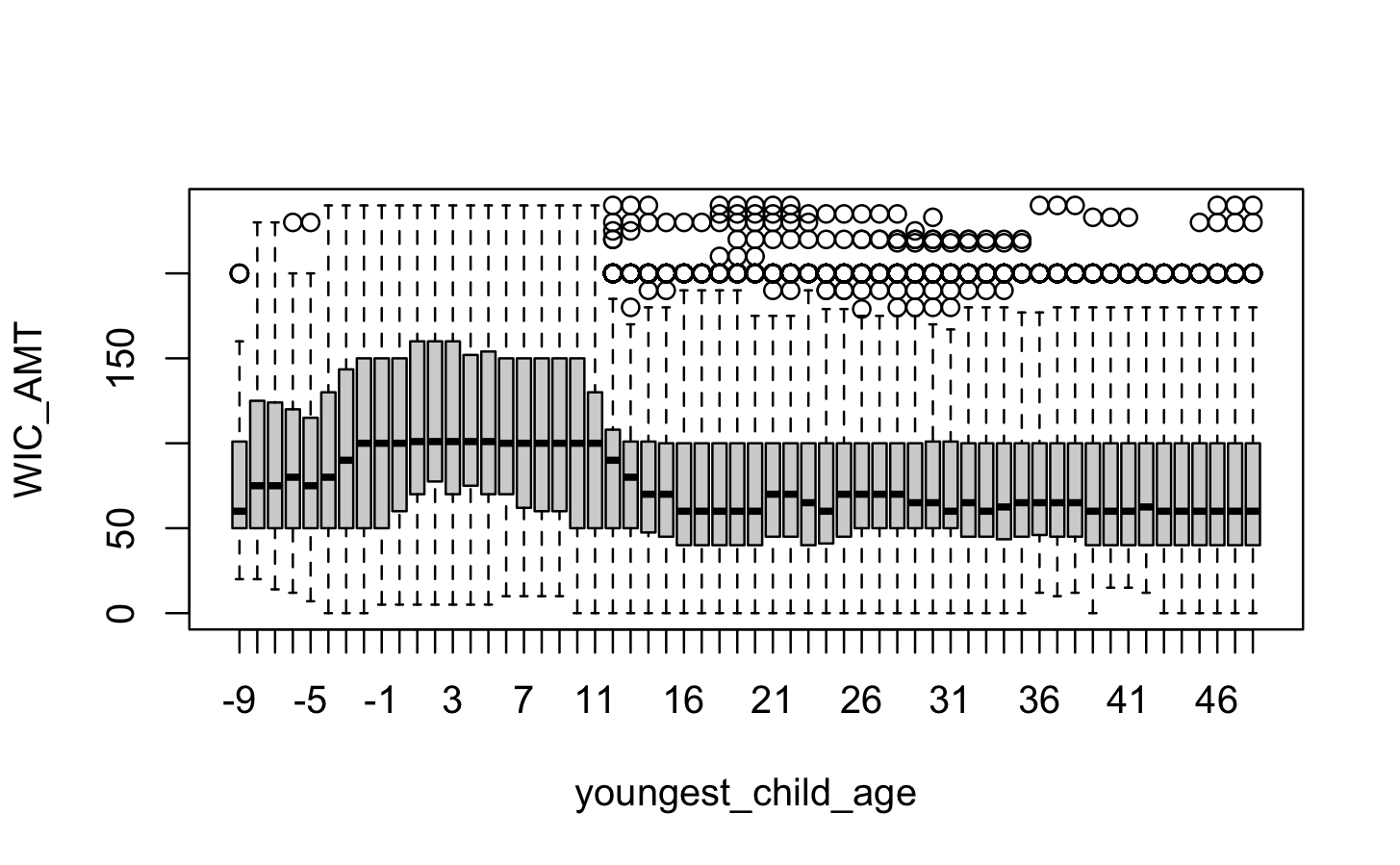}
    \caption{Reported Monthly WIC Benefits over Pregnancy and Early Childhood}
    \label{fig:benefit_vs_age}

    \caption*{\footnotesize
    \textit{Notes:} The sample is restricted to households with only one
    WIC-eligible member: one pregnant woman, one infant, or one
    preschool-aged child. The horizontal axis reports the pregnancy-and-child-age
    index $YA_{it}$. Negative values denote months before birth, values
    1--12 denote infancy, and values 13--48 denote the preschool-aged period.
    The thick horizontal line in each box denotes the median, the lower and
    upper boundaries denote the first and third quartiles, and the circles
    denote observations beyond the boxplot whiskers.
    }
\end{figure}

Reported WIC benefit amounts vary systematically with household composition.
To illustrate this relationship, we restrict the sample to households with
only one WIC-eligible member: one pregnant woman, one infant, or one
preschool-aged child. We then plot the distribution of reported monthly
benefits against the pregnancy-and-child-age index $YA_{it}$ in
\Cref{fig:benefit_vs_age}.

The distribution of reported benefits closely follows the age categories used
to determine WIC food packages. Median benefits are highest when the eligible
child is an infant aged 1--12 months and decline when the child transitions
into the preschool-aged category at 13 months. These patterns motivate the
inclusion of household-composition variables $HC_{it}$ in
\Cref{eq benefit imputation}.

\begin{sidewaysfigure}[p]
    \centering

    \subfigure[One pregnant woman
    \label{fig:benefit_preg_year}]{
        \includegraphics[width=0.47\linewidth]
        {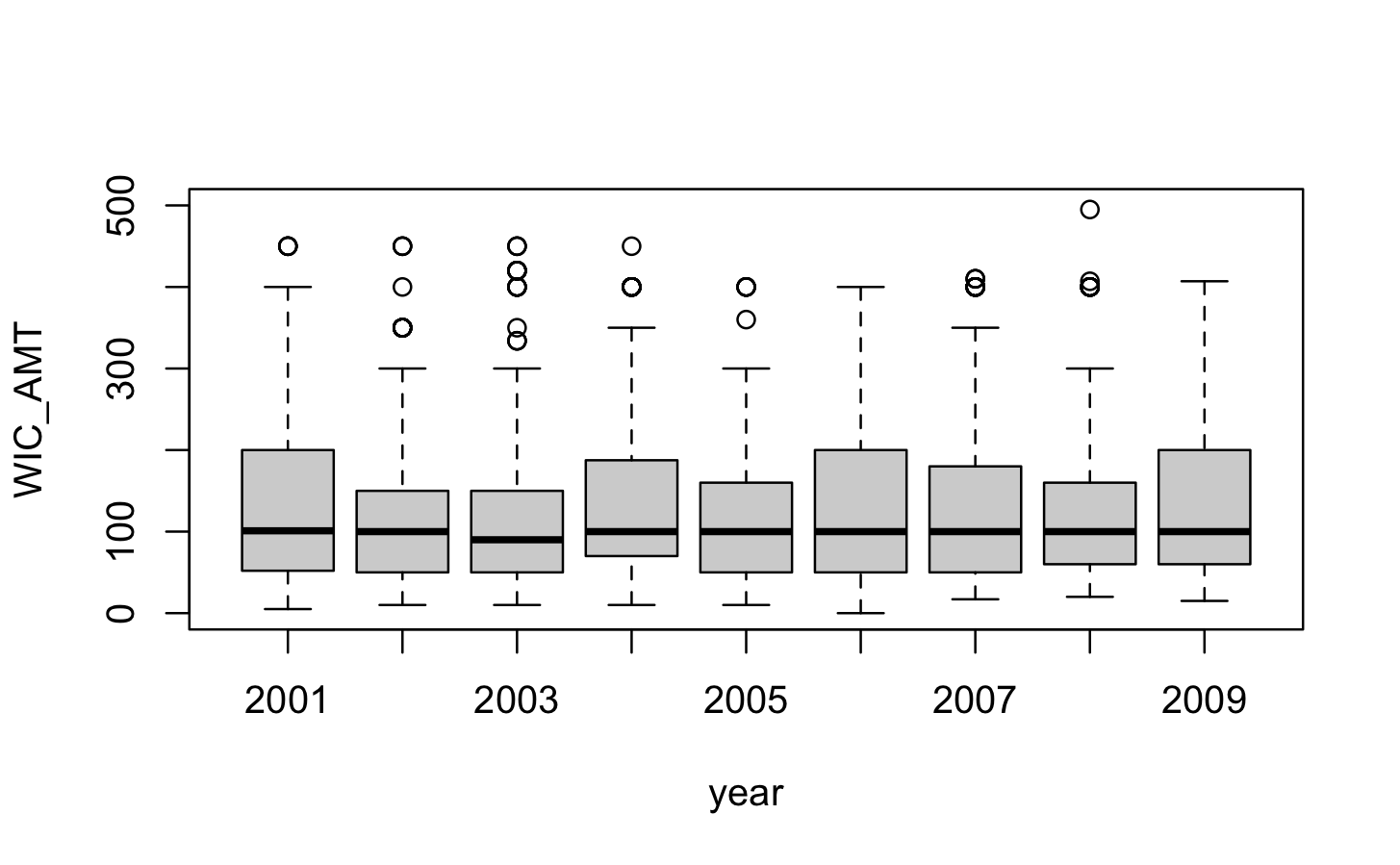}
    }
    \hfill
    \subfigure[One infant
    \label{fig:benefit_infant_year}]{
        \includegraphics[width=0.47\linewidth]
        {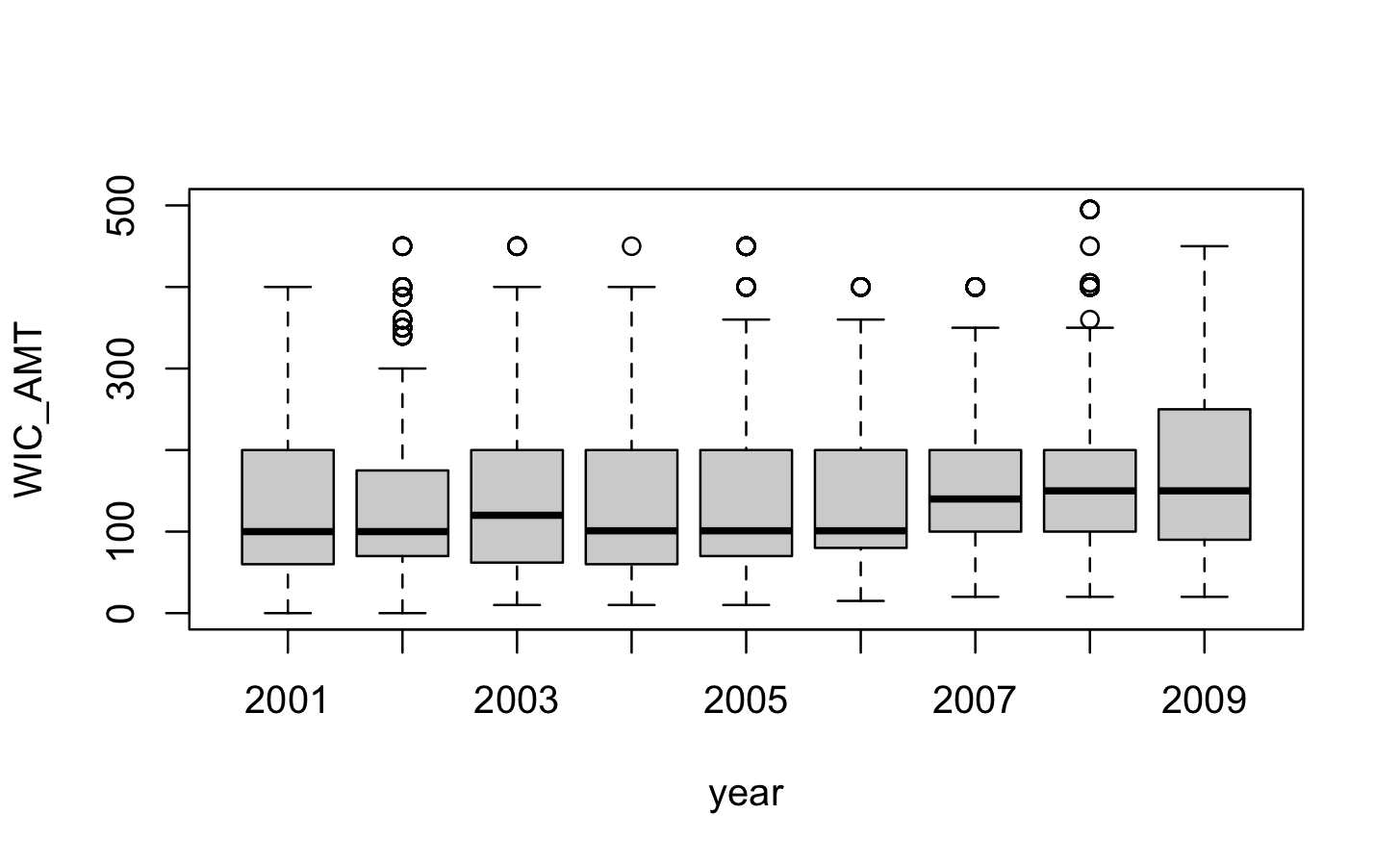}
    }

    \par\medskip

    \subfigure[One preschooler aged 13--30 months
    \label{fig:benefit_ps1_year}]{
        \includegraphics[width=0.47\linewidth]
        {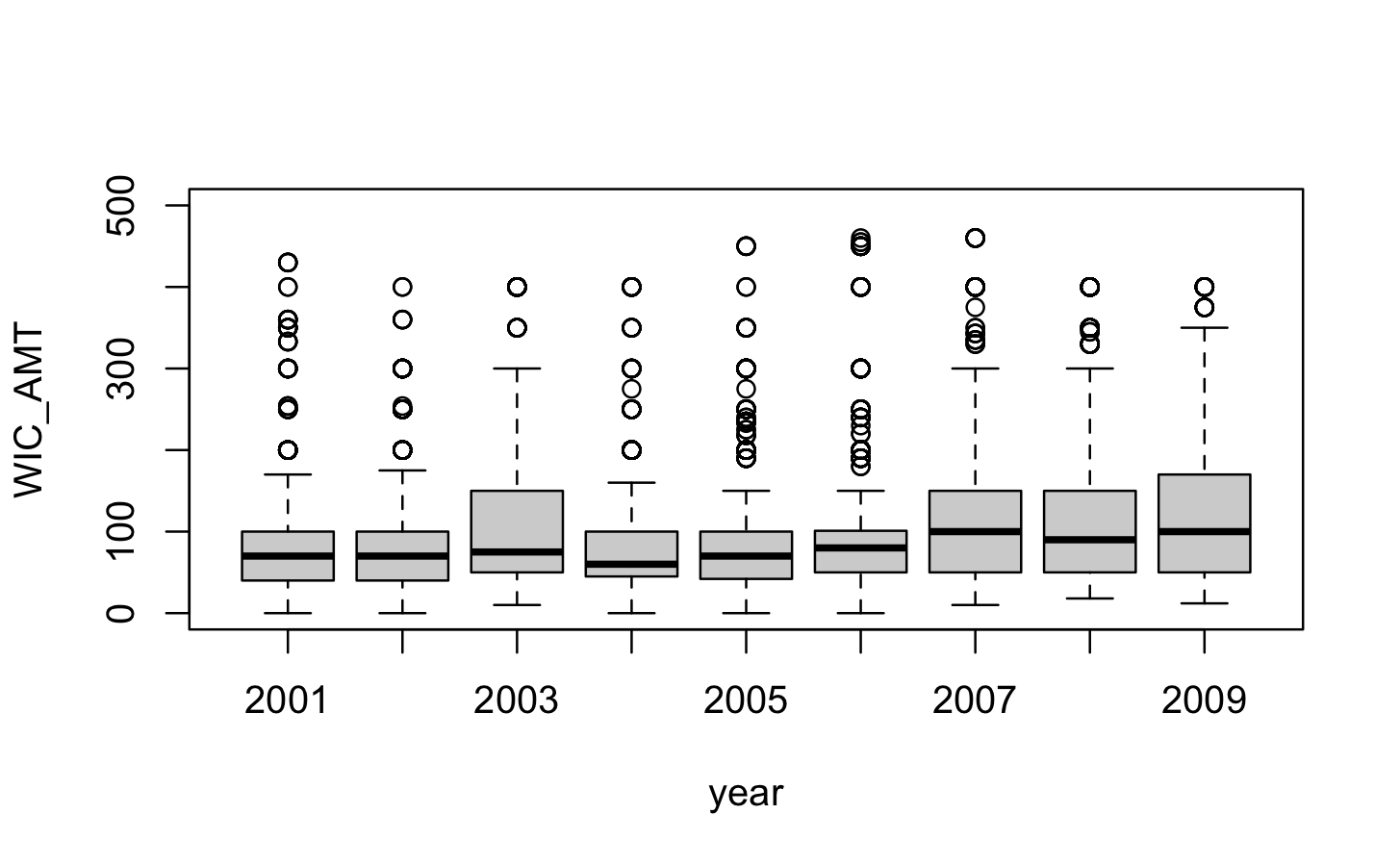}
    }
    \hfill
    \subfigure[One preschooler aged 31--48 months
    \label{fig:benefit_ps2_year}]{
        \includegraphics[width=0.47\linewidth]
        {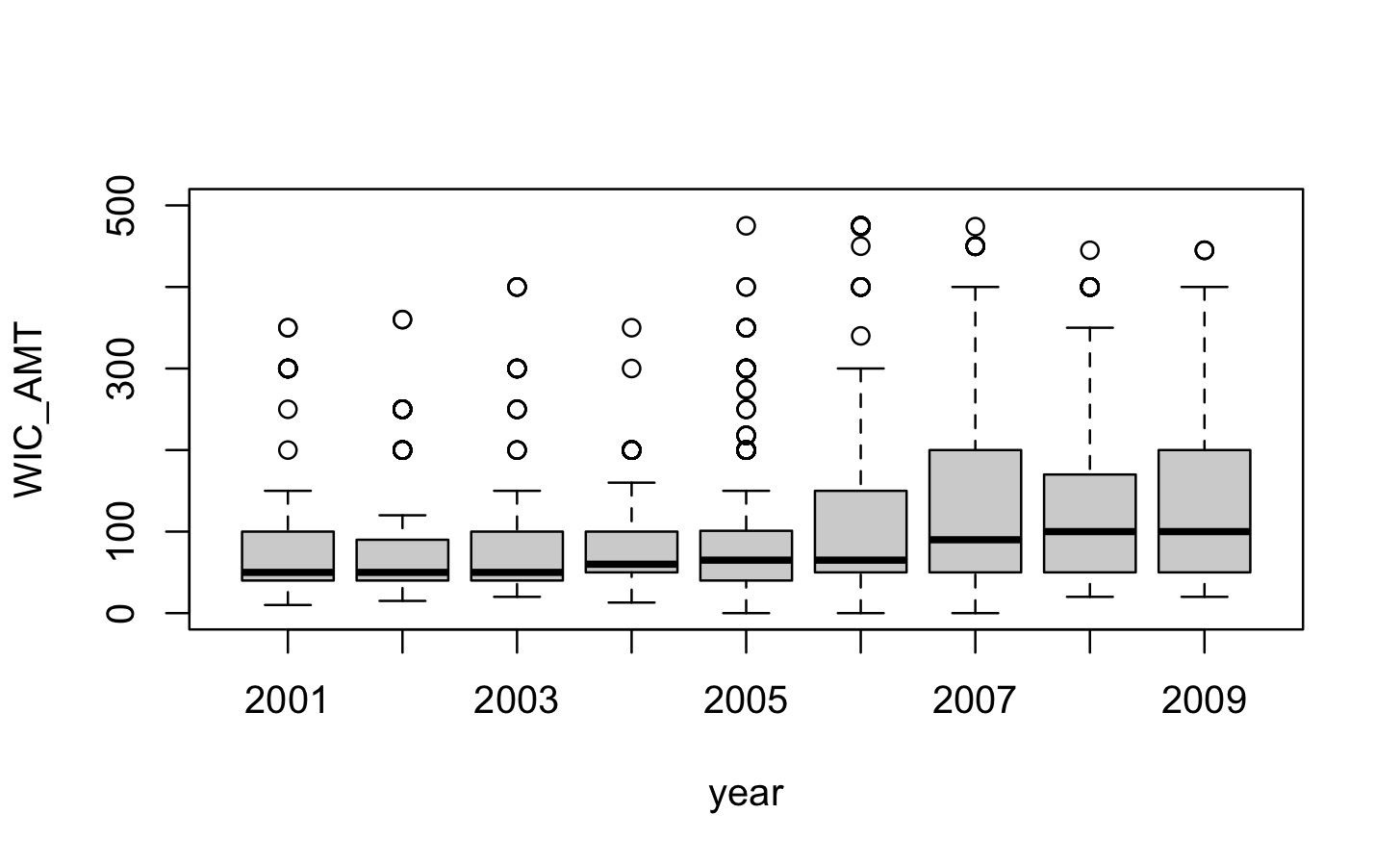}
    }

    \caption{Reported Monthly WIC Benefits by Year and Household Composition}
    \label{fig benefit across years}

    \caption*{\footnotesize
    \textit{Notes:} Each panel reports the annual distribution of reported
    monthly WIC benefits for households with exactly one member in the
    indicated eligibility category and no members in the other eligibility
    categories. The thick horizontal line denotes the median, the lower and
    upper boundaries of each box denote the first and third quartiles, and
    the circles denote observations beyond the boxplot whiskers. The common
    vertical scale facilitates comparisons across household-composition
    groups and years.
    }
\end{sidewaysfigure}

We next examine how reported benefit amounts vary over time. 
\Cref{fig benefit across years} presents annual distributions separately for
households with one pregnant woman, one infant, one preschool-aged child aged
13--30 months, and one preschool-aged child aged 31--48 months. Median
benefits for households with an infant and for both preschool-aged groups are
generally higher during 2007--2009 than during the preceding years, whereas
the median for households with one pregnant woman remains comparatively
stable.

The observed break occurs during a period of major changes to the national WIC
food packages. The Institute of Medicine published its review in 2006, USDA
issued an interim rule revising the food packages in 2007, and state agencies
were required to implement the revised packages by October 2009. 
These observed patterns and events motivate allowing the effects of household
composition to differ between the pre-2007 and 2007--2009 periods through
$CH_{it}\times Post_{it}$ in \Cref{eq benefit imputation}.

Though WIC is funded federally, it is administered by state government. Therefore, we further include $HC_{it} \times Region_{it}$ and $HC_{it} \times Region_{it} \times Post_{it}$ in \Cref{eq benefit imputation}. We report the regression results of \Cref{eq benefit imputation} in \Cref{table benefit imputation regression}.

\begin{table}[htbp]
    \centering
    \caption{Median Regression for Reported Monthly WIC Benefits}
    \label{table benefit imputation regression}

    \begin{threeparttable}
    \scriptsize
    \setlength{\tabcolsep}{5pt}
    \renewcommand{\arraystretch}{0.92}

    \begin{adjustbox}{max width=\textwidth}
    \begin{tabular}{lccc}
        \toprule


        & Number of pregnant women
        & Number of infants
        & Number of preschoolers \\
        \midrule

        \addlinespace[2pt]
        \multicolumn{4}{l}{
            \textit{Household-composition effects}
        } \\

        Pre-2007 Northeast effect
        & \makecell{$50.00^{***}$ \\ $(2.61)$}
        & \makecell{$51.00^{***}$ \\ $(2.62)$}
        & \makecell{$25.00^{***}$ \\ $(3.24)$} \\

        Post-2007 adjustment
        & \makecell{$0.00$ \\ $(13.87)$}
        & \makecell{$19.00$ \\ $(13.93)$}
        & \makecell{$25.00^{***}$ \\ $(3.31)$} \\

        \addlinespace[2pt]
        \multicolumn{4}{l}{
            \textit{Regional adjustments relative to the Northeast}
        } \\

        Midwest
        & \makecell{$0.00$ \\ $(2.51)$}
        & \makecell{$7.00$ \\ $(4.45)$}
        & \makecell{$0.00$ \\ $(3.30)$} \\

        South
        & \makecell{$-14.00^{***}$ \\ $(3.66)$}
        & \makecell{$19.00^{***}$ \\ $(3.86)$}
        & \makecell{$-10.00^{***}$ \\ $(3.47)$} \\

        West
        & \makecell{$20.00^{***}$ \\ $(5.08)$}
        & \makecell{$29.00^{***}$ \\ $(6.23)$}
        & \makecell{$5.00$ \\ $(4.10)$} \\

        \addlinespace[2pt]
        \multicolumn{4}{l}{
            \textit{Additional post-2007 regional adjustments}
        } \\

        Midwest
        & \makecell{$0.00$ \\ $(14.11)$}
        & \makecell{$23.00$ \\ $(14.54)$}
        & \makecell{$-16.67^{***}$ \\ $(5.91)$} \\

        South
        & \makecell{$31.50^{**}$ \\ $(14.94)$}
        & \makecell{$18.50$ \\ $(14.36)$}
        & \makecell{$2.50$ \\ $(4.21)$} \\

        West
        & \makecell{$-60.00^{***}$ \\ $(15.04)$}
        & \makecell{$-39.00^{**}$ \\ $(16.15)$}
        & \makecell{$-15.00^{***}$ \\ $(5.70)$} \\

        \bottomrule
    \end{tabular}
    \end{adjustbox}

    \begin{tablenotes}
        \footnotesize
        \item \textit{Notes:}
        The dependent variable is the reported monthly WIC benefit amount,
        measured in dollars. The table reports estimates from the median
        regression in \Cref{eq benefit imputation}; standard errors are shown
        in parentheses. The omitted period is before 2007, and the omitted
        Census region is the Northeast. The first row of household-composition
        effects therefore reports the effects of an additional pregnant woman,
        infant, or preschooler for a household in the Northeast before 2007.
        The regional rows report interactions between household composition
        and region, while the final three rows report the corresponding
        household-composition-by-region adjustments for 2007--2009. The intercept is estimated to be 50.00 with a standard error of 1.87.
        $^{*}p<0.10$, $^{**}p<0.05$, and $^{***}p<0.01$.
    \end{tablenotes}

    \end{threeparttable}
\end{table}

\subsection{Takeup comparison across subgroups}
\Cref{fig takeup by race} shows the takeup rate comparison across different racial groups.
\Cref{fig takeup by SNAP} shows the takeup rate comparison across different SNAP participation status. \Cref{fig takeup 1314} and \Cref{fig takeup 2526} shows the takeup rate comparison across different age of the youngest kid in the household.

\begin{figure}[htbp]
    \centering

    \subfigure[By race\label{fig takeup by race}]{
        \includegraphics[width=0.75\textwidth]{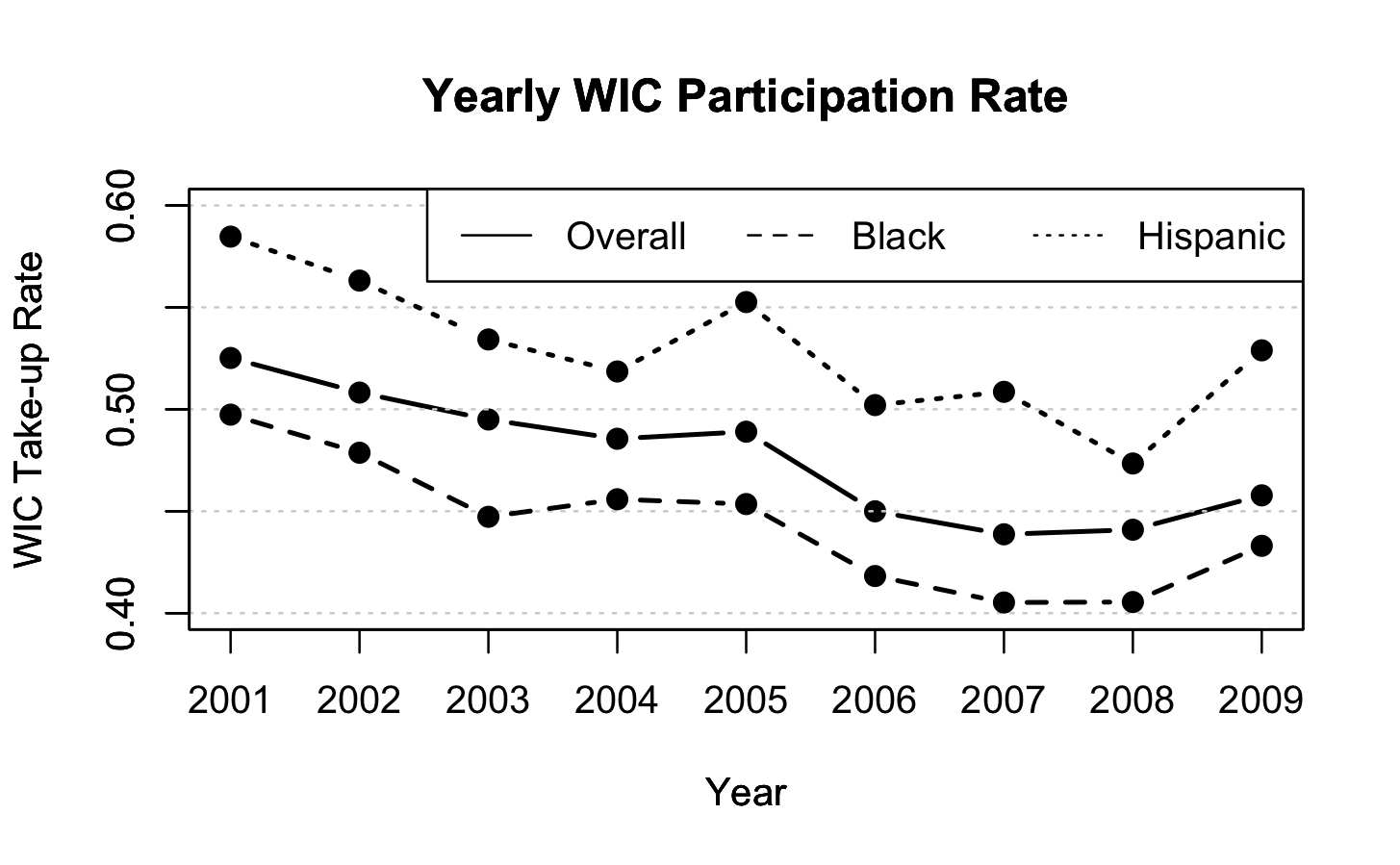}
    }
    \hfill
    \subfigure[By SNAP participation status\label{fig takeup by SNAP}]{
        \includegraphics[width=0.75\textwidth]{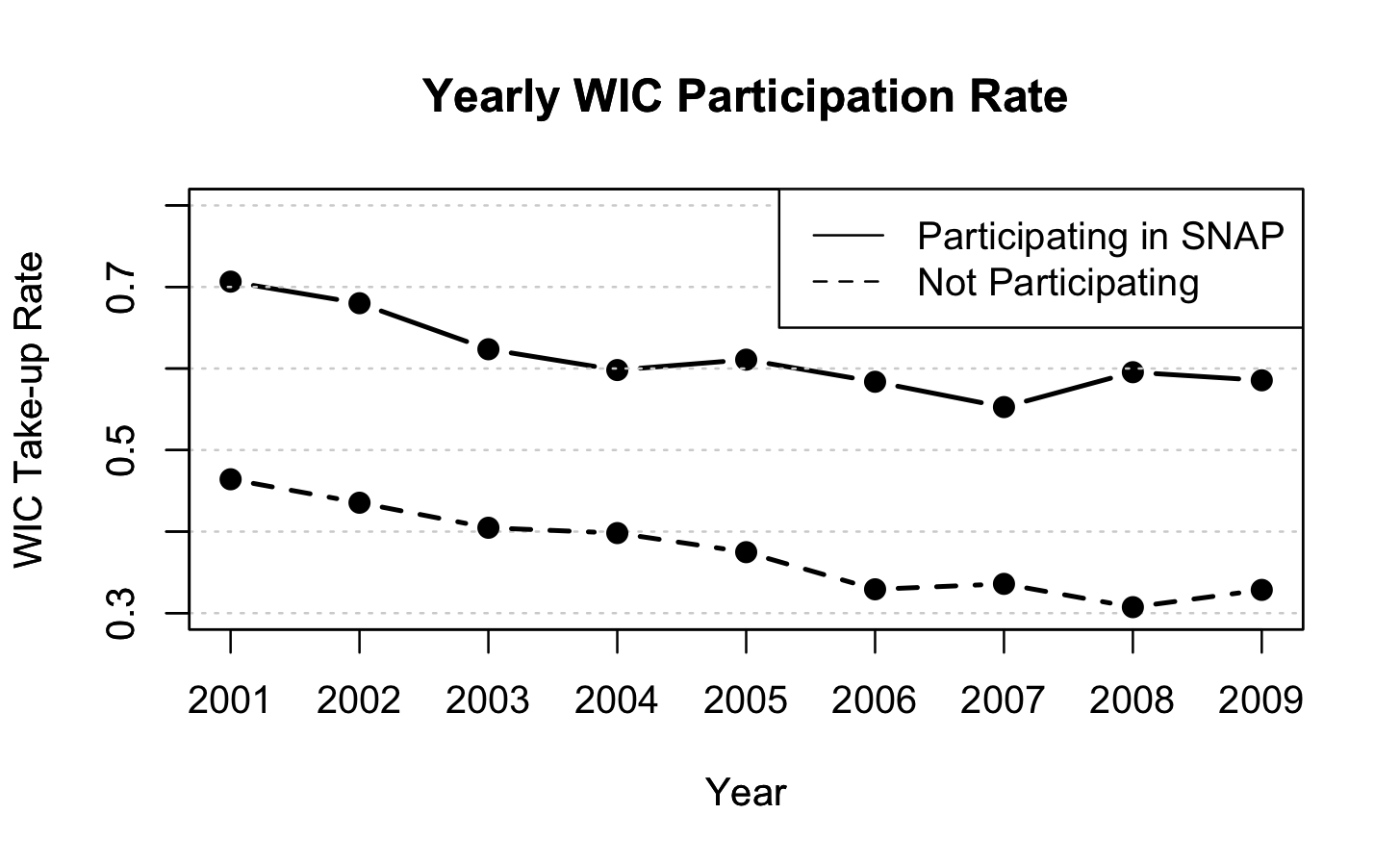}
    }

    \caption{Annual WIC Take-Up Rates across Selected Subgroups, 2001--2009}
    \label{fig:takeup_selected_subgroups}

    \caption*{\footnotesize
    \textit{Notes:} Each panel compares subgroup-specific annual WIC take-up
    rates with the corresponding overall rate. The figures illustrate the
    final column of \Cref{table sample summary}. Black and Hispanic households
    exceed the overall rate in 0 and 9 of the 9 sample years, respectively;
    SNAP participants and nonparticipants do so in 9 and 0 years, respectively.
    }
\end{figure}

\begin{figure}[htbp]
    \centering

    \subfigure[13 month-old vs 14 month-old\label{fig takeup 1314}]{
        \includegraphics[width=0.75\textwidth]{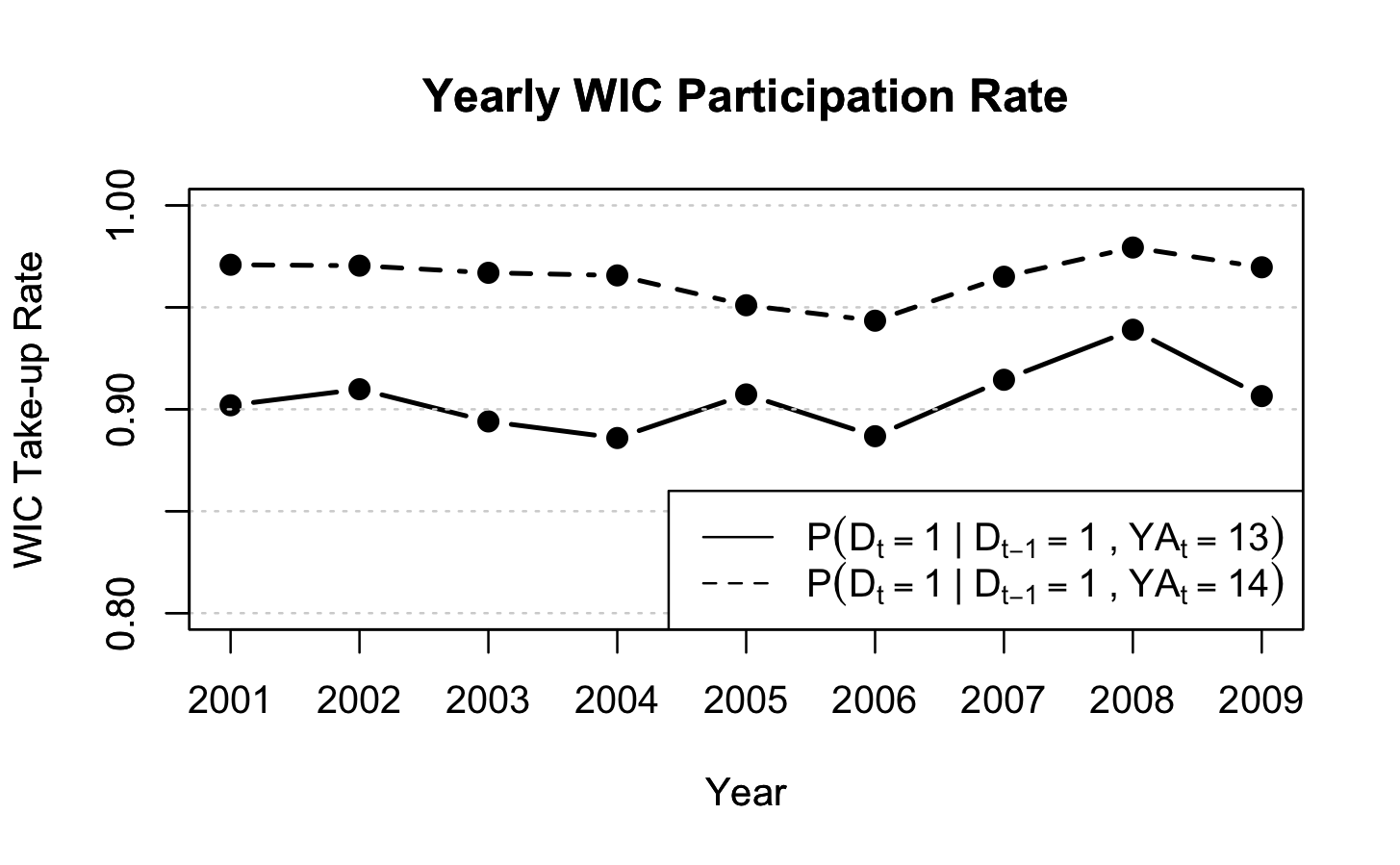}
    }
    \hfill
    \subfigure[25 month-old vs 26 month-old\label{fig takeup 2526}]{
        \includegraphics[width=0.75\textwidth]{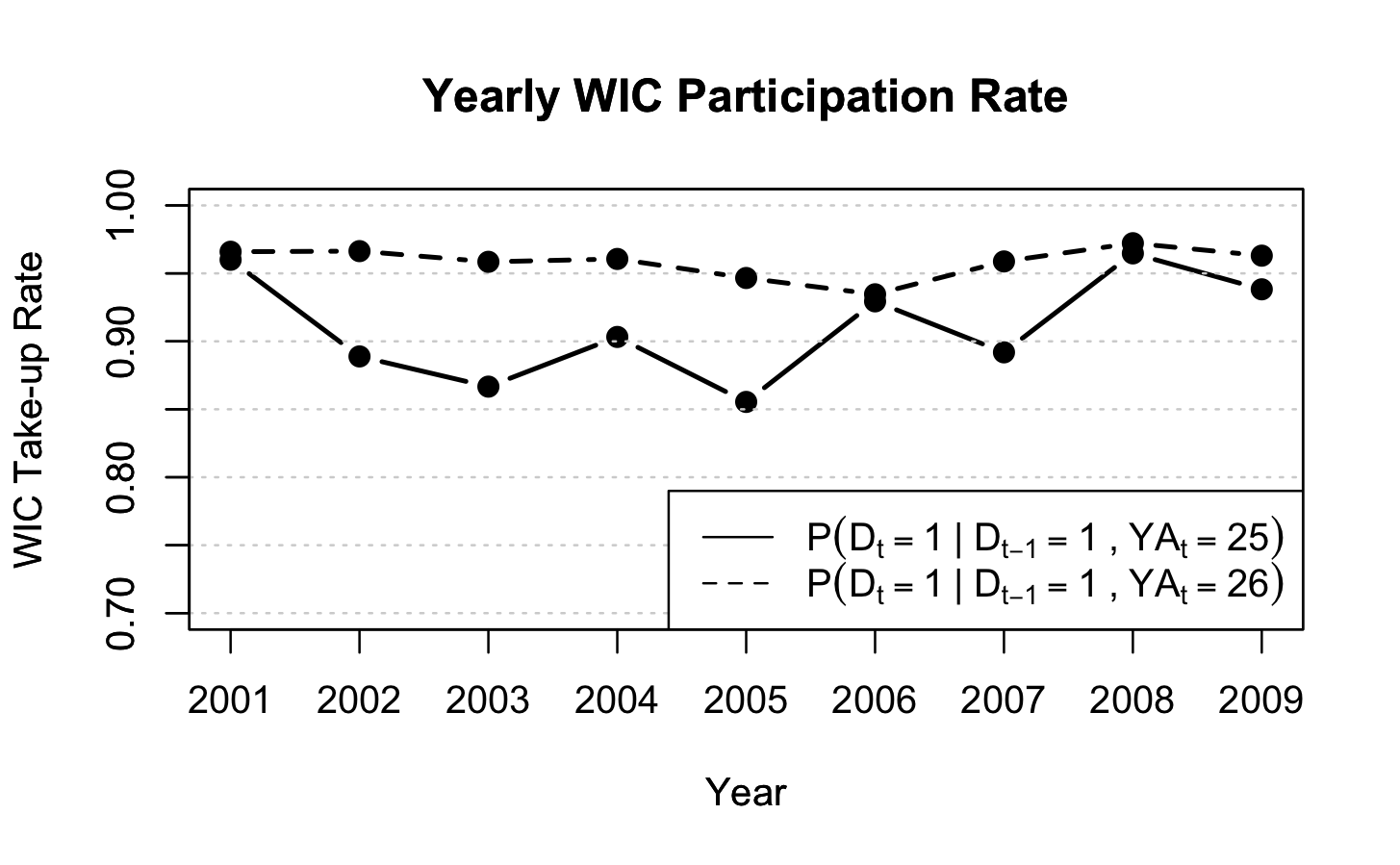}
    }

    \caption{Continuation Probabilities in and after Predicted Recertification Months}
    \label{fig recertification}

    \caption*{\footnotesize
\textit{Notes:} The figure reports annual continuation probabilities conditional
on participation in the preceding month. Panel (a) compares households whose
youngest eligible child is 13 and 14 months old, while Panel (b) compares ages
25 and 26 months. Ages 13 and 25 are classified as recertification months under
the construction of $Z_{it}$; the immediately following months are not. The
benefit category is unchanged within each age pair.
}
\end{figure}

\section{WIC2Five} \label{appendix WIC2Five}
In 2016 and 2017, the Vermont Department of Health WIC Program State Office implemented a pilot program aiming to increase the WIC participation duration among households who are already in the program. The pilot program sent out two types of text messages to WIC participants: one was a choice-inducing message from Aug 2015 to Aug 2016, and the other was an attention-raising message from Mar 2017 to May 2017. Here, we quote from the message track file and final report of the pilot program one choice-nudging message (CNM) and one attention-boosting message (ABM) in the main text. More examples of CNM and ABM can be found in Appendix \ref{appendix text messages}.


CNM increases households' perceived value of the program. It informs them that WIC participation is good for the future development of the kids.
On the other hand, ABM is a straightforward attention trigger that reminds participants of the necessary steps that need to be taken in order to continue participation. Under our two-stage model, ABM essentially ask the household who choose to exit WIC to reconsider their choices. Our estimation and counterfactual analyses in \Cref{sec estimation results} suggest that CNM is more effective than ABM. 

We anticipate the treatment effect of CNM to kick in \textit{later} than the end of the intervention period, Aug 2016. This is due to two reasons. One,  the vast majority of households start participation at a very young age of the infant, many even during pregnancy; two, households tend to participate until their youngest baby ages until 13 months or 25 months old. In contrast, if there is any treatment effect of ABM, we expect the treatment effect to kick in \textit{immediately} during the year of intervention, 2017. This is because the ABM is sent to households that are about to recertify or have already missed their recertification appointment. 

The outcome variable that is used by the Vermont WIC office to measure the policy effectiveness of ABM and CNM is the retention rate. We contacted the Vermont WIC office to check which month the retention rate is measured. 
The retention rates were measured in July 2015, July 2016, and July 2017. 

\subsection{Site selection}
In this section, we discuss the treatment assignment of CNM and ABM. Table \ref{table retention rates} shows which sites are selected for which treatment. In total, five sites are selected for CNM treatment and four sites are selected for ABM treatment.


\begin{table}[htb]
\centering
\begin{tabular}{lcccrrr}
\toprule
\multirow{2}{*}{\textbf{Location}} & \textbf{CNM} & \textbf{ABM} & \multirow{2}{*}{\textbf{EBT Timing}} & \multicolumn{3}{c}{\textbf{Retention Rates}} \\
\cmidrule(lr){5-7}
 & \textbf{Treatment} & \textbf{Treatment} & & \textbf{2015} & \textbf{2016} & \textbf{2017} \\
\midrule
Rutland & & & Jun 2015 & 69.9\% & 65.9\% & 59.2\% \\
Springfield & $\checkmark$ & $\checkmark$ & Oct 2015 & 66.9\% & 62.5\% & 64.7\% \\
Bennington & & & Oct 2015 & 74.8\% & 73.7\% & 63.3\% \\
White River & $\checkmark$ & & Nov 2015 & 70.8\% & 63.3\% & 66.3\% \\
Brattleboro & $\checkmark$ & & Nov 2015 & 72.6\% & 70.3\% & 67.6\% \\
St. Johnsbury & & $\checkmark$ & Dec 2015 & 79.9\% & 77.3\% & 70.0\% \\
Newport & & & Jan 2016 & 81.5\% & 73.7\% & 59.4\% \\
Morrisville & & $\checkmark$ & Jan 2016 & 83.8\% & 84.3\% & 77.9\% \\
St. Albans & & & Feb 2016 & 73.2\% & 70.4\% & 59.0\% \\
Burlington & $\checkmark$ & & Feb 2016 & 68.1\% & 62.6\% & 62.3\% \\

Middlebury & & $\checkmark$ & Mar 2016 & 84.0\% & 79.7\% & 72.4\% \\
Barre & $\checkmark$ & & Mar 2016 & 66.8\% & 61.4\% & 61.9\% \\
\bottomrule
\end{tabular}
\caption{WIC Child Retention Rates and Intervention Timeline (2015-2017), Ordered by EBT Rollout Date}
\label{table retention rates}
\end{table}

We argue that the CNM-treated sites are negatively selected, hence, our estimates are conservative. On the other hand, ABM-treated sites are self-selected and hence, we are less confident about our claims regarding this policy. In addition, we show that the staggered EBT rollout during the intervention period (June 2015 to March 2016) is unlikely to have an impact on our treatment effect estimates.

\subsubsection{Negative selection on CNM treatment} \label{sec negative selection on CNM treatment}
The pilot program selected Barre, Brattleboro, Springfield, White River districts for CNM intervention because these four sites ``reported the largest decreases in caseload in 2014, and were also the offices with the largest
difference between the number of WIC-eligible children receiving Medicaid and the number of children
who were enrolled in WIC'' \citep{vermont2017wic2five}.\footnote{Burlington was asked to test the CNM intervention because it was the ``largest and most culturally diverse'' site.} The selection process was based on the outcome variable retention rate, implying that selected sites for the CNM treatment tend to have more downward-sloping trends in the retention rate of participating households. This coincides with our permutation test and three-period event study plot in Section \ref{sec CNM treatment evaluation}. Hence, our estimate for CNM's treatment effect is potentially conservative and \textit{understates} the effectiveness of CNM. Nevertheless, we show that the treatment effect is extremely salient and statistically positive in \Cref{table WIC2Five evidence}.

\subsubsection{Self-enrollment on ABM treatment} \label{sec Self-enrollment on ABM treatment}
All 12 districts were asked to take part in the ABM treatment.
``Four offices, Middlebury, Morrisville, Springfield, and St. Johnsbury, quickly stepped up and began sending reminder
texts one to three days before a scheduled appointment.'' Unlike the CNM treatment, it is difficult to determine whether the selection is positive or negative for the ABM treatment. The counties may volunteer because they have more responsible WIC staff, implying positive selection; the counties may also have experienced or were expecting to experience a large reduction in retention rate and self-selected into the ABM treatment, implying negative selection. 

\subsubsection{Staggered rollout of EBT during the intervention period} \label{sec Staggered rollout of EBT}
\cite{vermont2017wic2five} states that it is hard to estimate the true effectiveness of the pilot program because there was a staggered rollout of EBT during the intervention period from Jun 2015 to Mar 2016. Using the EBT rollout timing from \cite{vermontwic2015minutes} meeting minutes, we show that the two treated groups are not different from the control groups in terms of their EBT implementation timing. 

We conduct the Wilcoxon rank-sum test for the EBT rollout timings for both treatments. The test results are summarized in \Cref{table rank sum test for EBT rollout timing}. The ranks of the rollout timings in \Cref{table rank sum test for EBT rollout timing} can be computed based on Table \Cref{table retention rates}. For example, Rutland is the first district to implement EBT and has a rank of 1. The next districts are Springfield and Bennington, both of which implemented EBT in the same month, so their ranks are (2+3)/2 = 2.5. The tests fail to reject the hypothesis that the treated and control groups for each treatment differ by their treatment timing. Hence, it is unlikely that the staggered rollout will contaminate the evaluation of the policy effectiveness for CNM and ABM.

\begin{table}[htb]
    \centering
    \footnotesize
    \begin{tabular}{l c c c}
        \toprule
         & Ranks of the treated group & Ranks of the control group & Two-sided test p-value \\
         \midrule
         CNM & $\{2.5,4.5,4.5,9.5,11.5\}$ & $\{1,2.5,6,7,9.5,11.5\}$ & 1  \\
         ABM & $\{2.5,6,7,11.5\}$ & $\{1,2.5,4.5,4.5,7,9.5,9.5,11.5\}$ & 0.5861\\
         \bottomrule
    \end{tabular}
    \caption{The Wilcoxon rank-sum test shows that we do \textbf{not} have enough evidence to reject that the treated sites and control sites have the same EBT rollout timing distribution.}
    \label{table rank sum test for EBT rollout timing}
\end{table}

\subsection{CNM treatment evaluation} \label{sec CNM treatment evaluation}
We combine the permutation test and DiD estimand to test whether there is a \textit{positive} treatment effect of CNM and whether there is a \textit{negative} pretrend, both of which are \textit{one-sided} hypothesis tests. We supplement our causal analysis with the canonical DiD estimate and an event-study plot to support our findings further.

\begin{figure}[htb]
    \centering
    \subfigure[The black line indicates the actual $DiD_\mathbbm{d}$.]{
    \includegraphics[width=0.47\textwidth]{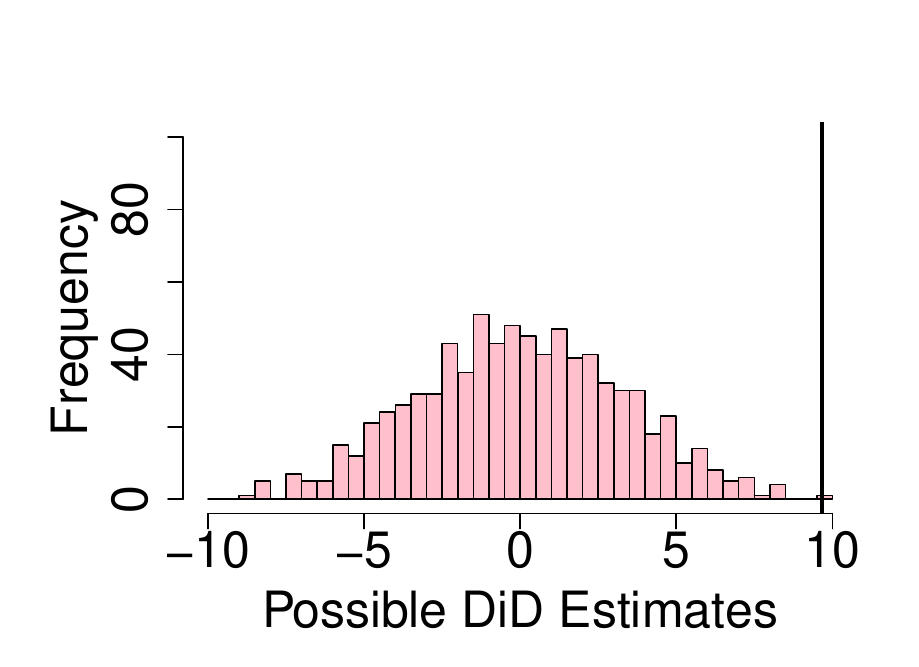}
        \label{fig CNM permutation test treatment effect}
    }
    \hfill
    \subfigure[The black line indicates the actual $PreTrend_{\mathbbm{d}}$.]{
    \includegraphics[width=0.47\textwidth]{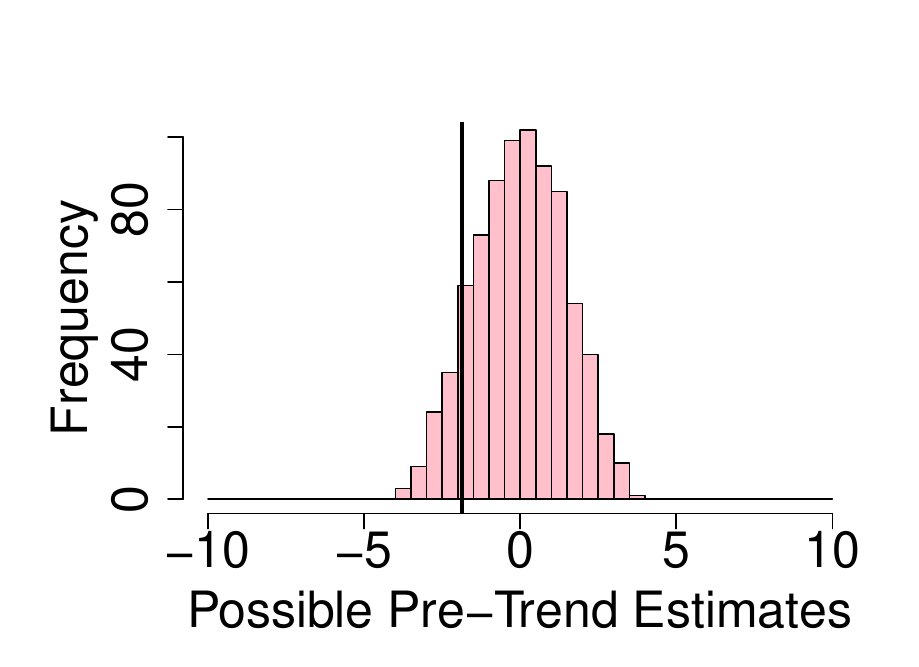}
        \label{fig CNM permutation test pretrend}
    }
    \caption{Permutation test results for the CNM treatment effect evaluation}
    \label{fig CNM permutation test results}
\end{figure}

For the permutation test, we exhaust all possible $\begin{pmatrix}
    12 \\ 
    5
\end{pmatrix} = 792$ treatment assignments. For each treatment assignment, we compute the average of the treated sites' differences between 2017 and 2015 retention rates. We perform the same computation for the control sites. Then, we take the difference of the two differences. To be precise, denote the set of all 792 possible treatment assignments as $\mathcal{D}$, each possible treatment as $\mathbbm{d}$, and the retention rate of site $i$ in year $t$ as $RR_{it}$. Since there are 5 sites treated out of 12, each $\mathbbm{d}$ is a vector with length 12 with 5 entries equal to 1 and 7 entries equal to 0. The DiD estimand that we compute is
\begin{equation*}
    DiD_{\mathbbm{d}} = \frac{1}{5} \sum_{i = 1}^{12} \mathbbm{d}_i (RR_{i,2017} - RR_{i,2015}) - \frac{1}{7} \sum_{i = 1}^{12} (1 - \mathbbm{d}_i) (RR_{i,2017} - RR_{i,2015}).
\end{equation*}
We compute $DiD_{\mathbbm{d}}$ for all $\mathbbm{d} \in \mathcal{D}$ and plot the the empirical distribution of $DiD_{\mathcal{D}}$ in Figure \ref{fig CNM permutation test treatment effect}. 
Out of all the 792 possible assignments, the $DiD_{\mathbbm{d}}$ for the actual treatment assignment is ranked \textbf{first}, indicating strongly that the treatment effect is positive with a p-value of $\frac{1}{792}$. As a robustness check, We replace $RR_{i,2017} - RR_{i,2015}$ with $RR_{i,2017} - RR_{i,2016}$ and perform the same permutation test, the actual DiD estimate is still ranked first and the p-value is still $\frac{1}{792}$.

Just like the canonical DiD setup, one might be concerned about the parallel trend assumption and test the pretrend (as a proxy test for the parallel trend assumption). We perform the same test with permutation. We replace $DiD_{\mathbbm{d}}$ with 
\begin{equation*}
    PreTrend_{\mathbbm{d}} = \frac{1}{5} \sum_{i = 1}^{12} \mathbbm{d}_i (RR_{2016} - RR_{2015}) - \frac{1}{7} \sum_{i = 1}^{12} (1 - \mathbbm{d}_i) (RR_{2016} - RR_{2015}).
\end{equation*}
We plot the empirical distribution of $PreTrend_{\mathcal{D}}$ in \Cref{fig ABM permutation test pretrend}. As suggested by \Cref{sec negative selection on CNM treatment}, there is a negative selection on the treated sites. There is a downward trend among the sites selected for the CNM treatment. The negative pretrend is marginally significant with a p-value of 10.1\%. Hence, our treatment effect estimate will likely understate its true value due to the downward-sloping trend.

We conduct canonical DiD and event-study (ES) to support our permutation test results: first, CNM's treatment effect is positive and statistically significant; second, its pretrend is negative and marginally significant. The estimation equations are as follows:
\begin{align*}
    \text{DiD} \qquad RR_{it} =&~ \theta_i + \phi_t + \beta_{DiD} \mathbbm{1}\{t = 2017\} \times \mathbbm{d}_{i} + \epsilon_{it} 
    \\
    \text{ES} \qquad RR_{it} =&~ \theta_i + \phi_t + \beta_{pre} \mathbbm{1}\{t = 2015\} \times \mathbbm{d}_{i} + \beta_{post} \mathbbm{1}\{t = 2017\} \times \mathbbm{d}_{i} + \epsilon_{it} 
\end{align*}

\Cref{table WIC2Five evidence} shows that the estimation results for DiD and ES align with the permutation test results. DiD estimates that the CNM intervention increases the child retention rate by 8.723\%, whereas ES estimates the treatment effect to be even higher at 9.654\%. Both methods indicate strong statistical significance for the positive treatment effect of the CNM intervention. 



\subsection{ABM treatment evaluation}
\begin{figure}[htb]
    \centering
    \subfigure[]{
    \includegraphics[width=0.47\textwidth]{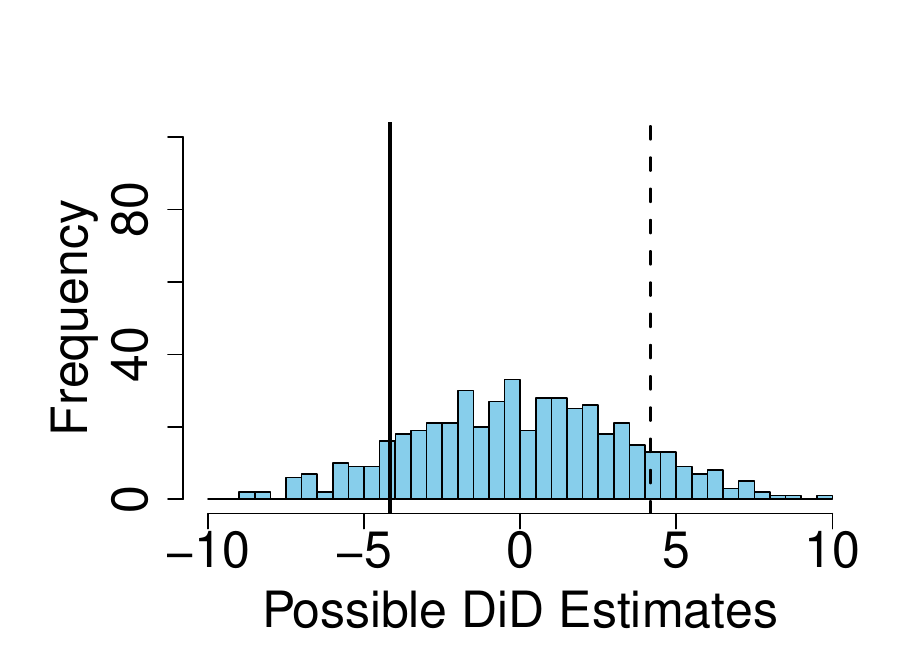}
        \label{fig ABM permutation test treatment effect}
    }
    \hfill
    \subfigure[]{
    \includegraphics[width=0.47\textwidth]{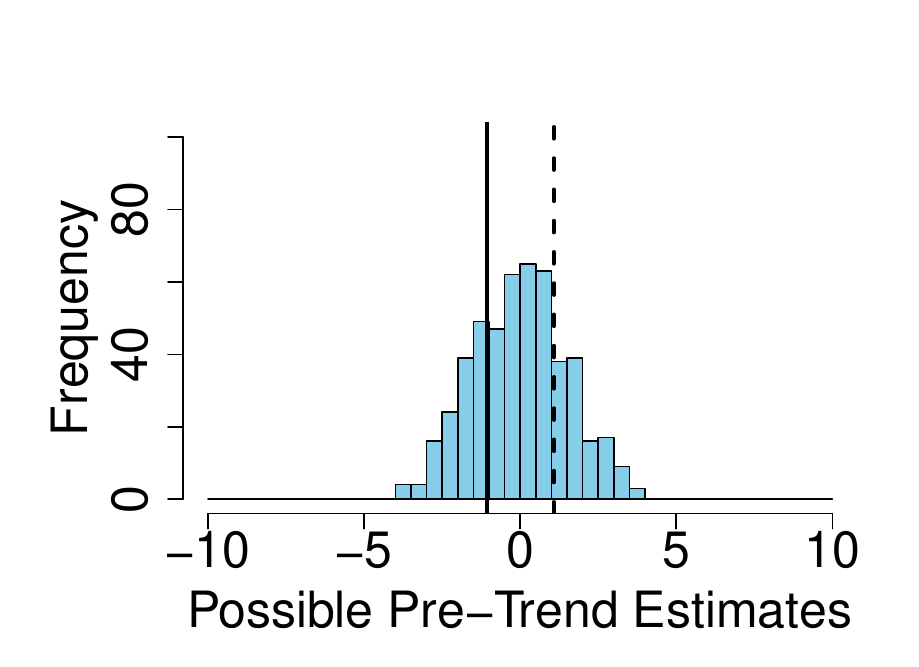}
        \label{fig ABM permutation test pretrend}
    }
    \caption{Permutation test results for the ABM treatment effect evaluation}
    \label{fig ABM permutation test results}
\end{figure}

We test whether there is \textit{any} treatment and whether there is \textit{any} pretrend, both of which are \textit{two-sided} hypothesis tests. We use the same methodologies to evaluate the effectiveness of the ABM treatment: permutation with a DiD estimand, canonical DiD, and event study. The results are reported in \Cref{fig ABM permutation test results} and \Cref{table WIC2Five evidence}. All three sets of results agree that there is neither a significant pretrend nor a significant treatment effect.

Since we expect that the treatment effect of ABM kicks in immediately, we use $RR_{i,2017} - RR_{i,2016}$ instead of $RR_{i,2017} - RR_{i,2015}$. \Cref{fig ABM permutation test treatment effect} shows that the actual DiD estimate is negative with a p-value $119/495 \approx 24.02\%$; \Cref{fig ABM permutation test pretrend} shows that the actual pretrend estimate has a p-value $245/495 \approx 49.50\%$. Our permutation tests show that both the treatment effect and the pretrend are not statistically significant.

\Cref{table WIC2Five evidence} also show that both the pretrend and treatment effect are statistically insignificant. We plot the event study estimates and their 90\% confidence intervals on \Cref{fig ABM permutation test pretrend}. Both confidence intervals cover 0.

Though the treatment effect is estimated to be insignificant, we do not claim that the ABM intervention is ineffective. \cite{vermont2017wic2five} shows that ABM is useful in decreasing the rate of rescheduling and canceling appointments, decreasing workload from WIC staff. This allows the staff to focus on better serving other aspects of the program, potentially having positive long-term impacts, which are out of the scope of our study.

\subsection{Addressing concerns on mean reversion}
\begin{figure}[htb]
    \centering
    \subfigure[]{
    \includegraphics[width=0.47\textwidth]{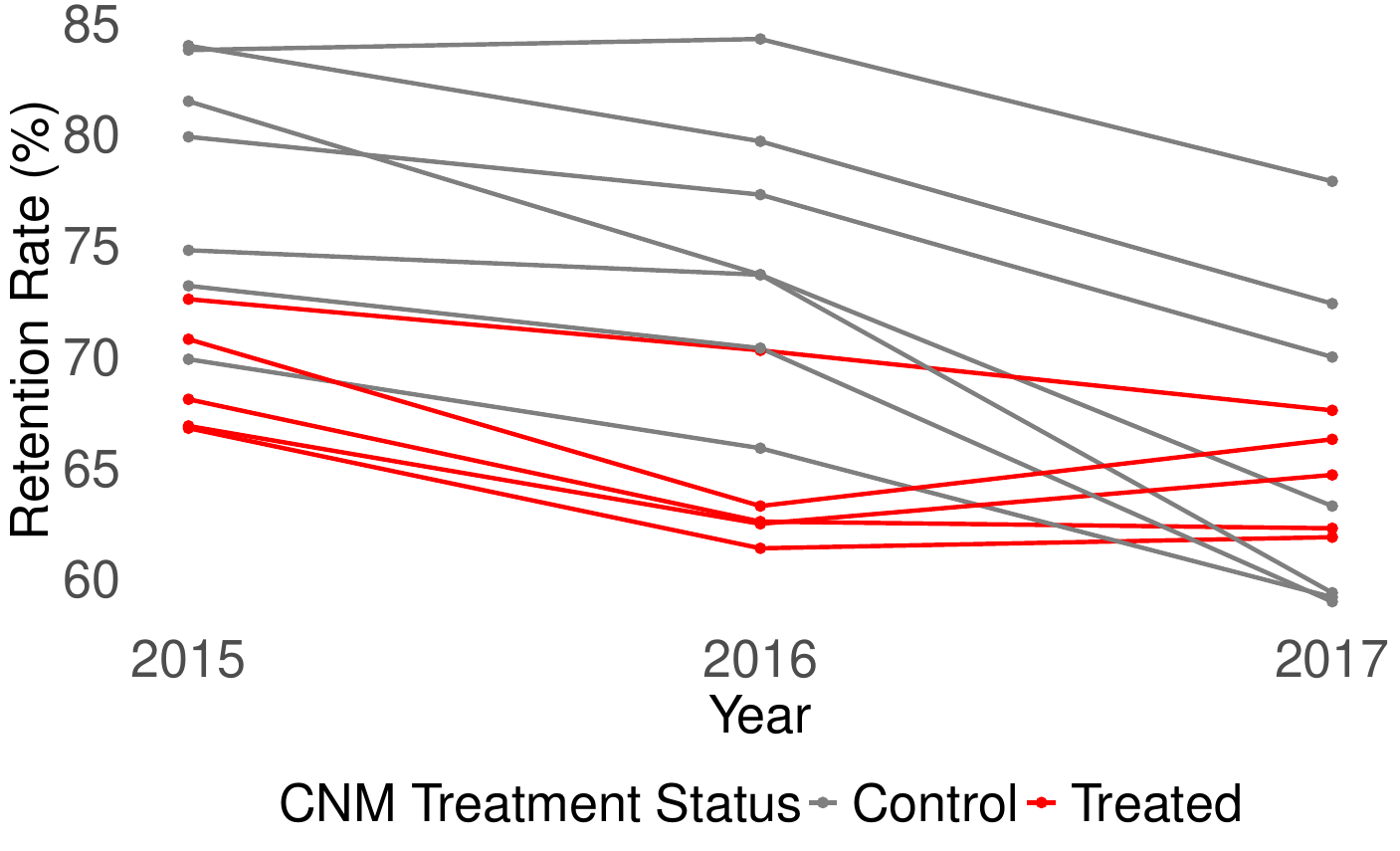}
        \label{fig CNM 12 plots}
    }
    \hfill
    \subfigure[]{
    \includegraphics[width=0.47\textwidth]{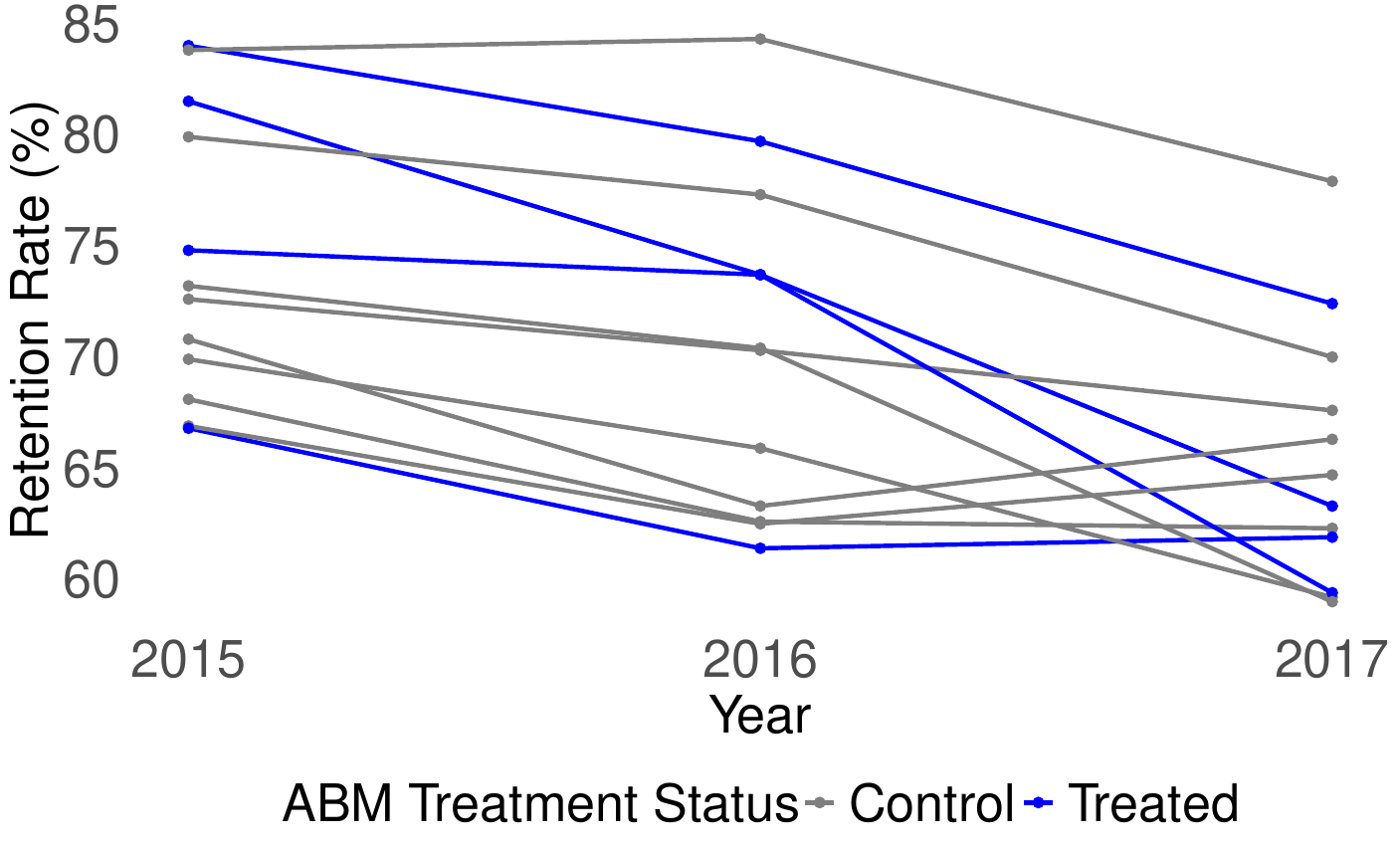}
        \label{fig ABM 12 plots}
    }
    \caption{12 sites' retention rates over the three years. Lines are colored by treatment statuses.}
    \label{fig 12 plots}
\end{figure}

In \Cref{sec negative selection on CNM treatment}, we claim that the sites are negatively selected, hence, our estimate is likely an understatement of the true treatment effect. One potential threat to the correctness of this statement is the mean reversion phenomenon. If sites that previously experienced a large decrease in retention rate are more likely to bounce back up to their mean retention rate, then the negatively selected sites will experience an increment in retention rate in the absence of the treatment. In this case, the estimated positive treatment effect does not have a causal interpretation. We address this concern by plotting \Cref{fig CNM 12 plots} and taking a closer look at the retention rate changes over 2015 to 2017 for both the treated and control sites. For completeness, we make the same plot for the ABM treatment (See \Cref{fig ABM 12 plots}).

Assuming a common mean for all 12 sites, if mean reversion is true, then the slower decline (from 2016 to 2017) in retention rate for orange lines would not necessarily indicate a positive treatment effect, but simply that those which previous decline more (from 2015-2016) bounce back to the mean of the 12 sites' retention rates. However, if this were to be true, the lower gray lines should also exhibit similar ``bounce back'' behavior which we do not observe in \Cref{fig CNM 12 plots}. We observe the opposite, lower gray lines decline more than upper gray lines. Hence, it is unlikely that mean reversion plays a big role in our treatment effect evaluation.

\subsection{Example of text messages from \cite{vermont2017wic2five}} \label{appendix text messages}
In \Cref{table choice-inducing messages}, we present additional examples for the CNM sent out by the WIC office during the WIC2Five pilot program.

\begin{table}[htb]
\centering
\begin{tabular}{>{\RaggedRight}p{3.5cm}>{\RaggedRight}p{3.5cm}>{\RaggedRight}p{3.5cm}>{\RaggedRight}p{3.5cm}}
\toprule
\textbf{1 Year Old} & \textbf{2 Year Old} & \textbf{3 Year Old} & \textbf{4 Year Old} \\
\midrule
Active play helps your toddler build more than muscle. Build her brain with activities like stomping, waddling and marching. Run and jump every day! 
& Active play helps your child build more than muscle. Build her brain with activities like scurrying, chasing and trudging. Run and jump every day! 
& Active play helps your preschooler build more than muscle. Build her brain with activities like hopping, leaping and dashing. Run and jump every day! 
& Active play helps your preschooler build more than muscle. Build her brain with activities like skipping, prancing, and galloping. Run and jump every day! \\
\addlinespace[0.5em]
Get your free copy of \textit{Playing with Your Toddler}. Text Fitwic now and we will send you one. 
& Get your free copy of the \textit{Fit WIC Activity Pyramid}. Text Fitwic now and we will send you one. 
& Get your free copy of the \textit{Fit WIC Activities Book}. Text Fitwic now and we will send you one. 
& Get your free copy of the \textit{Fit WIC Activities Book}. Text Fitwic now and we will send you one. \\
\bottomrule
\end{tabular}
\caption{Example of CNM sent to different age groups}
\label{table choice-inducing messages}
\end{table}

One other example of ABM is
\begin{itemize}
    \item Hi Lynne! Reminding you to complete your Nutrition Education before March 31, 2017 to keep your benefits current. It's easy! Complete a lesson online at wichealth.org. Your WIC household ID is 123456. Or call the Middlebury Health Department, 802-388-4644 for more options
\end{itemize}


\end{document}